\documentclass[nofootinbib,twocolumn,prd]{revtex4}
\usepackage{amsmath}
\usepackage{graphicx}
\usepackage{longtable}
\usepackage{color}
\usepackage{ifpdf}

\newcommand\editremark[1]{ {\color{red} #1}}

\newcommand\hidetosubmit[1]{}
\newcommand\optional[1]{}
\newcommand\ForInternalReference[1]{}

\newcommand\unit[1]{\, {\rm #1}}

\newcommand\Y[1]{Y^{(#1)}}

\newcommand\avL{\left< {\cal L}_{(a} {\cal L}_{b)} \right>}
\newcommand\WeylScalar{{\psi_4}}
\newcommand\WeylScalarFourier{{\tilde{\psi}_4{}}}
\newcommand\FourierWeylScalar{{\tilde{\psi}_4}}
\newcommand\WeylScalarCorot{R^{-1}{\psi_4}}

\newcommand\qmstate[1]{\left|#1\right>}

\newcommand\qmstateproduct[2]{\left<#1|#2\right>}
\newcommand\qmoperatorelement[3]{\left<#1\left|#2\right|#3\right>}

\newcommand\nSimulationsTotal{224}   %
\newcommand\nSimulationsNonprecessing{62}  %
\newcommand\ReferenceMassLow{100}      %
\newcommand\ReferenceMassMiddle{300}   %
\newcommand\ReferenceMassHigh{1000}   %

\def\bbh#1{binary black hole#1 (BBH#1)\gdef\bbh{BBH}}
\def\bh#1{black hole#1 (BH#1)\gdef\bh{BH}}

\begin{document}
\title{Comparing gravitational waves from nonprecessing and precessing black hole binaries in the corotating frame}
\author{L. Pekowsky$^1$}
\email{larne.pekowsky@physics.gatech.edu}
\author{R.\ O'Shaughnessy$^2$}
\email{oshaughn@gravity.phys.uwm.edu}
 \author{ J. Healy$^1$}
 \author{D. Shoemaker$^1$}
 \affiliation{$^1$Center for Relativistic Astrophysics,
 Georgia Tech, Atlanta, GA 30332, USA}
\affiliation{$^2$Center for Gravitation and Cosmology, University of Wisconsin-Milwaukee,
Milwaukee, WI 53211, USA}

\begin{abstract}
Previous analytic and numerical calculations suggest that, at each instant, the emission from a precessing black hole binary closely
resembles the emission from a nonprecessing analog.  
In this paper we quantitatively explore the validity and limitations of that correspondence, extracting the radiation from a large  collection of \textbf{\nSimulationsTotal}  generic black
hole binary merger simulations  both in the simulation frame and in a corotating frame that tracks precession.  
To a first approximation, the corotating-frame waveforms resemble nonprecessing analogs, based on similarity over a
band-limited frequency interval defined using a fiducial detector (here, advanced LIGO) and the source's total mass
$M$.      
By restricting attention to masses $M\in \ReferenceMassLow{}, \ReferenceMassHigh
M_\odot$, we insure our comparisons are sensitive only to our
simulated late-time inspiral, merger, and ringdown signals.  
In this mass region,  every one of our precessing simulations can be fit by some physically similar member of the
\texttt{IMRPhenomB} phenomenological waveform family to  better than 95\%; most fit significantly better.  
The best-fit parameters at low and high mass correspond to natural physical limits: the pre-merger orbit and post-merger
perturbed black hole.   
 Our results  suggest that  physically-motivated synthetic signals can be derived by viewing radiation
from suitable nonprecessing binaries in a suitable nonintertial reference frame.  
While a good first approximation,    precessing systems have  degrees of freedom (i.e., the transverse spins) which a
nonprecessing simulation cannot reproduce.   We quantify the extent to which these missing degrees of
freedom limit the utility of  synthetic precessing signals for detection and parameter estimation.  

\end{abstract}

\maketitle

\ForInternalReference{

PENDING TASKS FOR ROS

- EOB model for effective smaller spin?

- Note: we have a poor fit during the ringdown because we're going to a corotating frame...this shifts the effective
frequency a bit

- 'pre-merger fit' : try the *guessed best fit parameters* using specific M,eta,chi from known simulation parameters

-- try test case : SEOBNR on one simulation (e.g., the resolution-test simulation Sq(4,0.6,270,9)

ACTION ITEMS:ROS

- implement mathematica lal code (lalsimulation)

}

\section{Introduction}

\label{sec:Intro}

Coalescing comparable-mass black hole binaries are among the most likely and useful sources of gravitational waves for existing and
planned gravitational wave detectors like LIGO \cite{gw-detectors-LIGO-original},  Virgo,
\cite{gw-detectors-VIRGO-original}, the Einstein telescope \cite{2010CQGra..27s4002P}, and LISA \cite{2009astro2010S.238P,gwastro-eLISA-proposal}.
For sources in a suitable mass range, the signal these detectors receive contains significant
features from the late-stage, strong-field  dynamics of the black hole merger.
Only full numerical simulations of Einstein's equations can provide first-principle models for this epoch, including
all dynamics and emission  \cite{2010RvMP...82.3069C,gr-nr-DualJet-McWilliams-2010}.
Given the large computational cost per simulation, relatively few  well-determined models have been
produced.   Most models  thoroughly explored the physics and waveform from nonprecessing binary systems
\cite{gw-astro-EOBspin-Tarrachini2012,gwastro-Ajith-AlignedSpinWaveforms,gw-astro-nr-nina2-catalog-2012}. 
By contrast, relatively few simulations of binaries with more generic spins have been published\footnote{Though many simulations
  have been performed in the analysis of merger recoil kicks, in many cases their gravitational wave signal has not been
  described in detail \cite{2010CQGra..27k4006L,2007ApJ...659L...5C,2011PhRvL.107w1102L,2010RvMP...82.3069C}.    
}
and even fewer have provided their gravitational wave signal
\cite{2009PhRvD..79h4010C,gwastro-mergers-nr-Alignment-ROS-Polarization,PhysRevD.83.104034,Sturani:2010yv}.

Generic precessing black-hole binaries produce a rich multimodal gravitational wave signal during their inspiral and
merger.
To simplify the interpretation of these signals, several authors have proposed transforming the results of numerical
simulations, computed in the simulation frame,  into a \emph{corotating frame}
\cite{gwastro-mergers-nr-ComovingFrameExpansionSchmidt2010,gwastro-mergers-nr-Alignment-ROS-Methods,gwastro-mergers-nr-Alignment-BoyleHarald-2011,gwastro-mergers-nr-ComovingFrameExpansion-TransitionalHybrid-Schmidt2012,gwastro-mergers-nr-Alignment-ROS-Polarization}.
Empirically, a corotating frame demonstrably simplifies (and dramatically reduces in number) the  modes needed to describe the gravitational wave signal,
even during and after merger.  
On the other hand, post-Newtonian expressions for the strain also naturally decompose into two factors: an
``instantaneous'' factor describing corotating emission,  transformed by a rotation; see, e.g.,
\cite{WillWiseman:1996,gw-astro-mergers-approximations-SpinningPNHigherHarmonics}.  
A corotating frame therefore simplifies the comparison of different simulations to one another; to post-Newtonian
expressions; and to phenomenological or analytic models.  
Additionally, studies of how precession-induced modulations impact low-mass  \cite{gw-astro-SpinAlignedLundgren-FragmentA-Theory}
 and high-mass \cite{gwastro-mergers-nr-Alignment-ROS-Polarization} detection strategies also naturally express their
 results in terms of corotating frame modes and trajectories.    The nonprecessing search strategies currently being
 used miss precessing signals, roughly in proportion to two factors: (a) how much the signal precesses (geometrically)
 \cite{gwastro-mergers-nr-HigherModes-Larne2012}
 and (b) how much the corotating-frame precessing waveform resembles some nonprecessing signal model in the search
\cite{gw-astro-SpinAlignedLundgren-FragmentA-Theory,gwastro-mergers-nr-Alignment-ROS-Polarization,gwastro-mergers-nr-ComovingFrameExpansion-TransitionalHybrid-Schmidt2012}.  %

In this paper we compare the gravitational wave signal from the
merger of a generic precessing  black hole binary, 
expressed in a corotating frame, to nonprecessing merger waveforms.  
In this special frame,  a precessing merging binary emits radiation that closely resembles the signal from a
nonprecessing analog, for sufficiently short epochs.    
We determine the nonprecessing
configuration that best fits the corotating-frame's $(2,2)$ mode, as a function of  reference mass.   
Though tuned to one mode, selected examples suggest that the best-fit simulation usually reproduces \emph{multiple} modes for a comparable epoch.
This correspondence  suggests  a simple kludge to reproduce precessing merger signals, proposed directly or implicitly by authors who proposed the
construction of a corotating frame
\cite{gwastro-mergers-nr-ComovingFrameExpansionSchmidt2010,gwastro-mergers-nr-Alignment-ROS-Methods,gwastro-mergers-nr-Alignment-BoyleHarald-2011,gwastro-mergers-nr-ComovingFrameExpansion-TransitionalHybrid-Schmidt2012,gwastro-mergers-nr-Alignment-ROS-Polarization}.
If an appropriate time-dependent rotation is known, then a suitable \emph{nonprecessing} source combined with a time
dependent rotation generates an approximate synthetic precessing waveform.   
We assess this procedure constructively, comparing the
line-of-sight waveforms generated by a real precessing system and a synthetic analog.   
This procedure can synthesize good
approximations  to short waveforms from precessing black
hole binaries, using  physically-motivated choices for the precession rate and inspiral-merger-ringdown signal.  
In the most directly comparable study, \citet{gwastro-mergers-nr-ComovingFrameExpansion-TransitionalHybrid-Schmidt2012}
applied  a similar procedure to a small sample of precessing merger waveforms, derived via a post-Newtonian approximation.   As expected from the
functional form of the post-Newtonian inspiral signal \cite{gw-astro-mergers-approximations-SpinningPNHigherHarmonics},
they found that the (post-Newtonian) corotating-frame signal nearly
matched\footnote{\citet{gwastro-mergers-nr-ComovingFrameExpansion-TransitionalHybrid-Schmidt2012} adopt a completely
  different definition of ``match'', based on a single polarization and a white power spectrum.  Their quantitative
  statements  cannot be directly compared with our own.} emission from nonprecessing binaries with nearly-identical
physical parameters.   In short, this study provided concrete examples to suggest both that the inspiral signal could be
efficiently represented in the corotating frame using a nonprecessing signal with nearly identical physical parameters;
by implication, suitable hybrid precessing inspiral-merger-ringdown signals follow by adjoining precessing
post-Newtonian and numerical relativity signals in the corotating frame.  
By contrast, our study assesses the similarity  between our simulations' corotating-frame signal and well-studied models for the inspiral and merger of nonprecessing
binaries.  
Specifically, we employ  many (\nSimulationsTotal) generic merger signals;  adopt a  physically-motivated diagnostic to
assess waveform similarity;  demonstrate that even the merger phase of generic precessing signals resembles
nonprecessing binaries, in a corotating frame; and, critically, identify and  systematic limits to the accuracy of a corotating-frame approximation.

While a good zeroth approximation,  this procedure does omit  physics.  
Nonprecessing waveforms simply cannot self-consistently reproduce features tied to the system's \emph{kinematics}:
the orbital phase versus time;  the ringdown mode frequencies, set by the final black hole's mass and spin; the ringdown
mode amplitudes, which can reflect spin-orbit misalignment; et cetera. 
As a familiar example, in  post-Newtonian calculations,  time-dependent spin-orbit  and spin-spin terms must be included
in the orbital phase and calculated from suitable spin precession equations.   
In general, we find the corotating-frame waveform carries additional information (e.g., about the transverse spins) that
cannot be encoded into a nonprecessing waveform.    
The presence of these extra degrees of freedom can explain the observationally-relevant differences between
corotating-frame and simulation-frame results. 
Over sufficiently long time- and frequency-scales, the differences between the corotating frame and simulation frame
become startlingly apparent.    In particular,  early- and late-time waveforms generally resemble \emph{different} nonprecessing systems.  
As a result, while corotating waveforms  \emph{approximately} resemble nonprecessing modes, they do so only for short
periods in time and frequency.  
In practice, however, real gravitational wave data analyses are also limited to a narrow frequency interval and one line
of sight.  Using observationally-motivated diagnostics to characterize differences between signals, we find
nonprecessing systems are a surprisingly effective analog  of generic precessing sources.

We provide an executive summary and detailed outline in Section \ref{sec:ExecutiveSummary}.  
In Section \ref{sec:Compare} we review why and how we compare  simulations using only their corotating-frame $(2,2)$
modes; explain our notation;  introduce the simulations used;  and describe the   \texttt{IMRPhenomB} model \cite{gwastro-Ajith-AlignedSpinWaveforms} we use in our studies.
 In Section \ref{sec:UnderstandBestFitIfPossible} we present our results of fitting the precessing system.  We present our synthetic signal and include a discussion of the physics we miss in such a signal in Section \ref {sec:AsAKludge}.  Finally, our conclusions are presented in Section \ref{sec:Conclude}. For interested readers, we include two appendices.  Appendix
\ref{sec:MathMethods} describes the previously-developed tools and notation used in this paper  to compare detected gravitational waves and to extract a preferred direction from a precessing
binary; and Appendix \ref{sec:Simulations}, which  describes the simulations performed and numerical tests we adopted to build
quantitative confidence in our results.

\section{Executive summary and detailed outline}
\label{sec:ExecutiveSummary}

\begin{figure}
\includegraphics[width=\columnwidth]{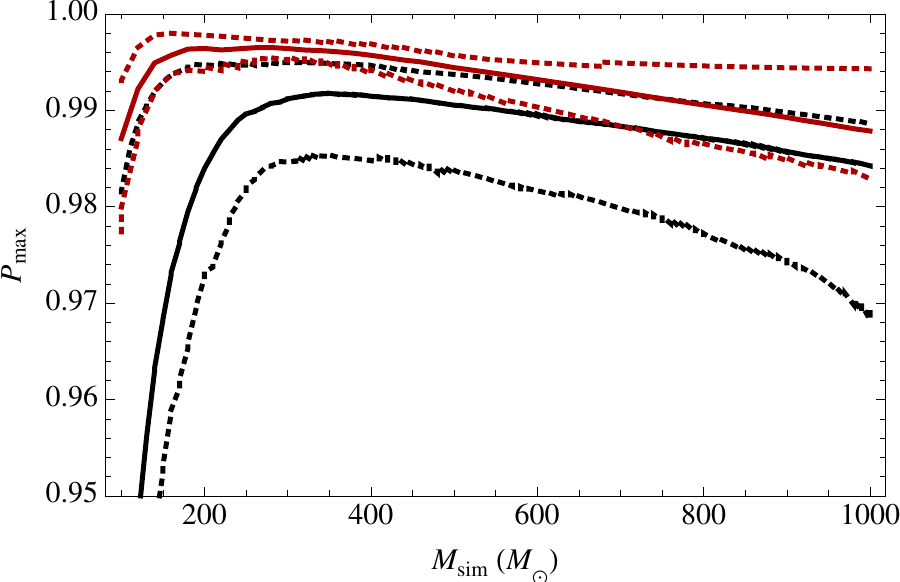}
\caption{\label{fig:ProofFitWorks} \textbf{Precessing binaries resemble nonprecessing binaries}:  Distribution of best fit complex
  overlap $P_{\rm max,corot}$ [Eq. (\ref{eq:PmaxCorot})] between the  \texttt{IMRPhenomB} and our simulated signals of mass $M$.  The solid  black curve shows
  the median-probability overlap $P_{max}$ at each mass: half of our precessing simulations have a fitting factor above the
  black solid line at each mass; the dotted black curves show the 90\% confidence interval, estimated from our set of
  \textbf{\nSimulationsTotal} simulations.
  For comparison, the red
  solid and dotted curves show the median fitting factor and 90\% confidence interval estimated from \textbf{\nSimulationsNonprecessing}
  nonprecessing simulations; these simulations are  not included in the previous list.   All calculations are performed by comparing the two models'  $(2,2)$ modes, each in a
  suitable corotating frame.
  Our simulations have significantly different initial conditions and hence durations: many shorter  simulations do not
  have enough data to reliably estimate the waveform and hence $P$ for $M\lesssim 250 M_\odot$.  Our longer simulations,
  including the nonprecessing simulations, are relatively well-fit by the \texttt{IMRPhenomB} model down to $\simeq \ReferenceMassLow{} M_\odot$.
}
\end{figure}

As shown by Figure \ref{fig:ProofFitWorks}, the pre- and post-merger leading-order emission from each simulation is
well-fit by \emph{some} nonprecessing inspiral-merger-ringdown model, in the corotating frame.    
The procedure we employ to compare two signal models is described in Section \ref{sec:Compare}.  
Specifically, for each of the simulations in our sample, described at length in Appendix \ref{sec:Simulations} and enumerated in Table \ref{tab:Simulations},  we have
transformed to a corotating frame using a suitable time-dependent rotation $R(t)$, then extracted the (angular)
$(l,m)=(2,2)$ mode time series $[R^{-1}\WeylScalar]_{2,2}(t)\equiv \int d\Omega {R^{-1}\WeylScalar} \Y{-2}_{2,2}(\hat{n})^*$ of the
Weyl scalar, as projected onto spin-weight $-2$ harmonics.
For each possible mass $M\in \ReferenceMassLow{}-\ReferenceMassHigh
M_\odot$,\footnote{As noted in the introduction and quantified in the Appendix,
  the duration of our numerical simulations was used to select a mass interval, such that only our numerical simulations
  would significantly influence the comparisons we performed.  Physically, at these masses only the late stages of
  inspiral and merger would be accessible to the advanced LIGO detector, at plausible signal 
amplitudes.}
we compare
this time series with all possible  \texttt{IMRPhenomB} phenomenological inspiral-merger-ringdown waveforms
$\WeylScalar_{PB}(\lambda)$ for nonprecessing binaries \cite{gwastro-Ajith-AlignedSpinWaveforms}.   The mass range is
chosen at the low end to avoid the use of hybrid waveforms in this study, and at the high end to ensure the inclusion of
ringdown.  Motivated by data analysis, we compare signals with a  complex ``overlap'', maximized over time and polarization
\cite{gwastro-mergers-nr-Alignment-ROS-IsJEnough,gwastro-mergers-HeeSuk-FisherMatrixWithAmplitudeCorrections}; see
Appendix \ref{sec:MathMethods} for a brief review.  We adopt a fiducial advanced LIGO noise curve (zero detuned high
power; see  \cite{LIGO-aLIGODesign-Sensitivity}) and perform integrals over $5-2000\unit{Hz}$.\footnote{For binary masses $M<300 M_\odot  (40\unit{Hz}/f_{low})$, our
  results are not strongly  sensitive to the lower frequency limit adopted for the integrand.  We adopt
  $f_{low}=5\unit{Hz}$, to insure only the detector noise power spectrum, and not arbitrary choices for $f_{low}$,
  determines our results at high mass. }   Using differential
evolution~\cite{Storn10.1023}, a hill-climbing algorithm, we select the best-fitting nonprecessing
(\texttt{IMPhenomB}) mass, spin, and mass ratio for each precessing binary merger simulation and mass.  The best-fit
parameters are denoted by $\lambda_{PB}$ and the best-fit match is denoted $P_{\rm max,corot}$:
\begin{align}
\label{eq:PmaxCorot}
P_{\rm max, corot} \equiv \max_t \frac{|\qmstateproduct{[R^{-1}\WeylScalar_{NR}]_{2,2}}{[\WeylScalar_{PB}(t,\lambda_{PB})]_{2,2}}|}{
|R^{-1}\WeylScalar_{NR}||\WeylScalar_{PB}(\lambda_{PB})|} \, .
\end{align}
Figure \ref{fig:ProofFitWorks} shows the fraction of simulations with match $P_{\rm max,corot}$ [Eqs.  (\ref{eq:def:Overlap}, \ref{eq:def:Overlap:Max})] greater than a specified threshold.
Most have a match within a few percent of unity; and, therefore, a precessing model and its nonprecessing analog would appear
very similar to a gravitational wave detector network with plausible signal amplitude
\cite{gwastro-mergers-HeeSuk-FisherMatrixWithAmplitudeCorrections}. 

\begin{figure}
\includegraphics[width=\columnwidth]{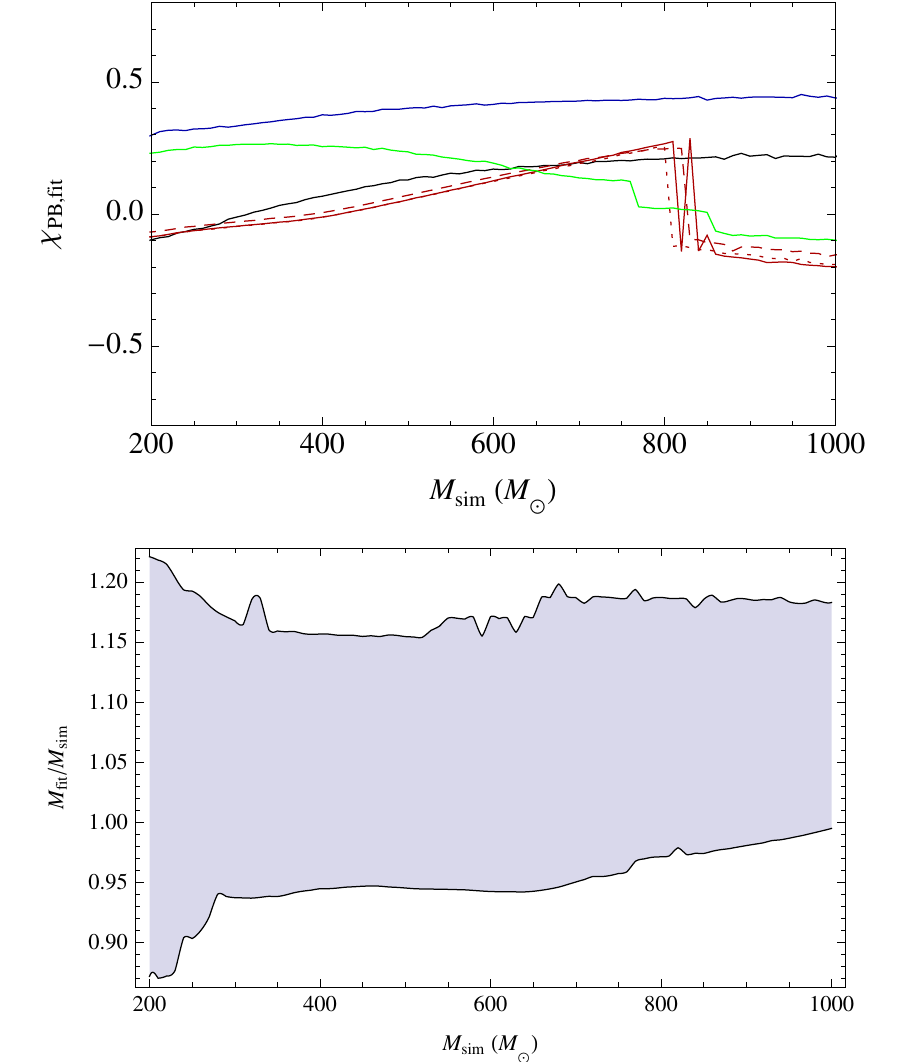}
\caption{\label{fig:BestFitCurves:Chi}\textbf{Best-fit nonprecessing parameters depend on precessing parameters
    and mass, and can differ from the simulation parameters}:  \emph{Top panel}: For selected simulations studied in this work, a plot of the best-fitting effective spin $\chi_{\rm PB}$ versus
  the simulated binary's mass.   The best-fit effective spin changes  as a function of mass: while the
  initial and final state separately resemble a nonprecessing system,  no single nonprecessing system fits the whole
  time-dependent corotating-frame mode.  Instead, the best-fitting parameters interpolate (sometimes discontinuously)
  between the initial and final state.
Colors indicate Tq(1.5, 0.4,60,10) (blue);  S(1, 0.2, 180,6.2) (black);    Sq(4, 0.6,270,6.2) (green); and several
instances of  Sq(4, 0.6,270,9) using h=M/180, M/160, and M/140 resolutions; see Appendix \ref{sec:Simulations} for details.
\emph{Bottom panel}: For all simulations used in this work, the ratio $M_{\rm fit}/M_{\rm sim}$ between the fitted and
simulated mass.  As indicated by the shaded region, the best-fitting mass can differ by up to $\simeq 10\%$ from the
simulated mass. 
}
\end{figure}

As described in Section \ref{sec:UnderstandBestFitIfPossible}, the best-fitting nonprecessing parameters  depend on precisely which
 time and frequency interval we adopt for comparison.   For the data-analysis-driven diagnostic used here, the
best-fitting parameters depend on the \emph{mass} adopted for the precessing binary.  %
Roughly speaking, at low mass the best-fitting nonprecessing model has similar mass ratio and  ``effective spin''
\cite{1995PhRvD..52..848P,2001PhRvD..64l4013D,gwastro-Ajith-AlignedSpinWaveforms,2011PhRvD..84h4037A}.  
When fitting the \texttt{IMRPhenomB} model to our data, our results suggest the best-fitting binary has the same value
of the  \texttt{IMRPhenomB} ``effective spin'' parameter $\chi_{\text{PB}}$; see Section \ref{sec:UnderstandBestFitIfPossible}.  
For illustration, Figure \ref{fig:BestFitCurves:Chi} shows how  $\chi_{\text{PB}}$ depends on simulation mass, for a one-parameter
family of equal mass binary mergers.  

Conversely, at high mass the signal produced by  the simulation and any nonprecessing model is dominated by quadrupole
radiation from the final black hole.  Inevitably, the best-fitting nonprecessing models must predict comparable final states.
Despite adopting a phenomenologically-motivated comparison, our method identifies nontrivial, physically significant
relationships between precessing and nonprecessing signals.  As demonstrated in  Section \ref{sec:UnderstandBestFitIfPossible}, the best-fit
nonprecessing model to the early time (here, low mass) signal has the physically-anticipated parameters needed to reproduce a
common orbital phase evolution: similar mass ratio and ``effective spin''.    
For this reason, we anticipate the corotating-frame modes can be naturally extended to arbitrarily early  times via
``hybridization'' with conventional post-Newtonian or effective-one-body models for the early inspiral.  
Even though we only fit the $(2,2)$ mode, detailed followup of a handful of similar systems suggests the best-fitting
parameters  reproduce several modes simultaneously.   Indeed, the striking similarity illustrated by the top panel of  Figure
\ref{fig:Prototype:NonprecessingAnalog:Extreme} and 
first noted in Figure 12 of 
\citet{gwastro-mergers-nr-ComovingFrameExpansionSchmidt2010} was used to motivate detailed analysis of the corotating
frame.    Unfortunately, we only possess  continuously-parameterized models for one  mode from a generic nonprecessing
binary; we defer  a detailed quantitative analysis of multiple modes  to a future study.

\begin{figure}
\includegraphics[width=\columnwidth]{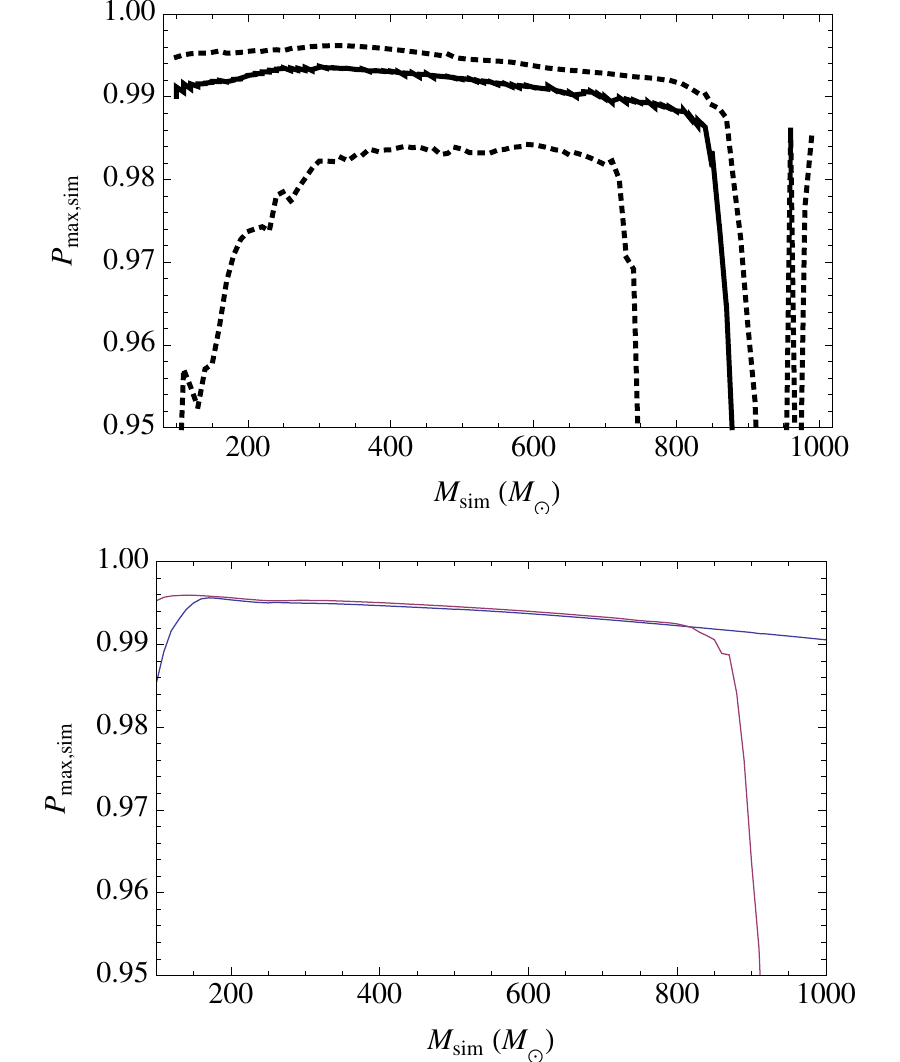}
\caption{\label{fig:ProofFitWorks:Synthetic} \textbf{Synthetic precessing waveforms resemble precessing waveforms}:  
\emph{Top panel}:  The distribution of $P_{\rm max,sim}$ [Eq. (\ref{eq:PmaxSim})], the match between
$\WeylScalar_{NR2,2}$ and $R\WeylScalar_{PB}$,  a synthetic precessing signal derived by applying a known time-dependent
rotation to a best-fitting \texttt{IMRPhenomB} waveform.  This distribution is effectively identical to  that shown in
Figure \ref{fig:ProofFitWorks}. 
\emph{Bottom panel}:  For the Sq(1.5, 0.6,45) simulation, a plot of the match  $P_{\rm max,sim}$ and $P_{\rm max,corot}$
     [Eq. (\ref{eq:PmaxCorot})].  As this example illustrates, except for very high masses $P_{\rm max,corot}\simeq
     P_{\rm max,sim}$.
 To avoid over-weighting the single worst  case, which has $P_{\rm max,corot}\simeq
0.95$ for all masses, we have explicitly eliminated
 all five resolutions of the strongly-precessing Sq(4,0.6,270,9) and the equivalent Tq(4,0.6,90,9) from this
comparison.  By contrast, all resolutions and iterations are included in Figure \ref{fig:ProofFitWorks}. 
}
\end{figure}

As noted in Section \ref{sec:AsAKludge}, this formal  similarity in a noninertial frame is directly relevant to physical observers in the inertial frame, each
limited to a single fixed line of sight that does not corotate with the
binary.   A nonprecessing signal, combined with
a suitable rotation, will generate a reasonable facsimile of a precessing signal, in the \emph{observer's frame}.   
As a result, even though detailed simulations of precessing binaries are computationally expensive, a computationally
trivial procedure can generate plausible precessing inspiral-merger-ringdown signals.  To demonstrate this agreement,
Figure \ref{fig:ProofFitWorks:Synthetic} uses the \emph{best-fit} corotating-frame \texttt{IMRPhenomB} parameters $\lambda_{PB}$ derived in
the corotating frame; generates a time-domain corotating-frame signal consisting of only the \texttt{IMRPhenomB} $(2,\pm
2)$; transforms to the corotating frame to create a synthetic multimodal signal $R(t)\WeylScalar_{PB}(t,\hat{n})$; and calculates
the complex match $P_{max}$ between  the simulation-frame $(2,2)$ modes of the synthetic signal and the original
precessing NR simulation to which the corotating-frame mode was fit\footnote{In our investigations, we  constructed
a family of synthetic precessing signals, unique up to overall mass scale $M$.  The figures provided in the text
describe the performance of the best-fitting member of that
one-parameter family.  We obtain almost identical  results 
if the procedure described in the text is followed verbatim.%
}:
\begin{eqnarray}
\label{eq:PmaxSim}
P_{\rm max,sim} = \text{max}_t  \frac{|\qmstateproduct{\WeylScalar_{NR,2,2}}{[R\WeylScalar_{PB}(t,\lambda_{PB})]_{2,2}}|}{
|\WeylScalar_{NR,2,2}| |[R\WeylScalar_{PB}(\lambda_{PB})]_{2,2}|}
\end{eqnarray}
At low mass $M\lesssim 1700 M_\odot$, these two quantities agree: $P_{\rm max,corot}\simeq P_{\rm max, sim}$.
As expected, at high mass these two quantities increasingly disagree.  This disagreement does not reflect intrinsic
dissimilarity between nonprecessing and precessing signals.  Instead, it reflects our strict application of the best-fit
parameters derived from the corotating frame to a slightly different problem: reproducing $\WeylScalar_{NR}$ in the
simulation frame.     At high mass, slightly different choices for  $\WeylScalar_{PB}$  are needed to optimally
reproduce the NR signal in each frame.\footnote{ Though the Fourier transform of 
$R(t)\WeylScalar(t)$ is well-approximated by $R(t(f))\WeylScalarFourier(f)$ at early times, when the rotation operation
changes slowly, at late times a stationary-phase approximation is not sufficiently accurate for our purposes.
Equivalently, the rotation needed at late times oscillates on a timescale comparable to the quasinormal mode frequency
spacing. }
In special cases like  Figure \ref{fig:Prototype:NonprecessingAnalog:Extreme}, a single nonprecessing model can
approximate the entire  signal.

\begin{figure}
\ifpdf{
\includegraphics[width=\columnwidth]{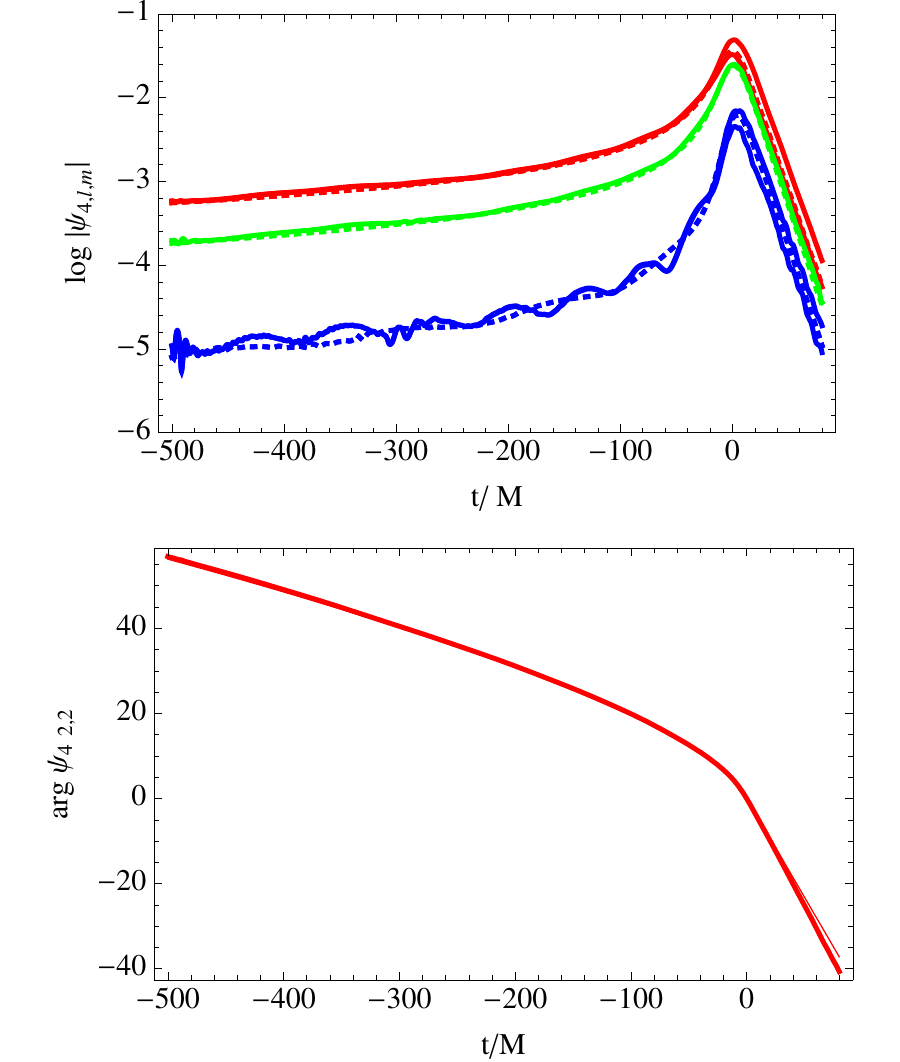}
}
\fi
\caption{\label{fig:Prototype:NonprecessingAnalog:Extreme}\textbf{Corotating signal resembles nonprecessing signal }:  
 Comparison of
  $|r \WeylScalarCorot_{l,m}|$  (top panel) and $\text{arg}\WeylScalarCorot_{2,2}$ (bottom panel) for a \emph{nonspinning} $q=4$ simulation (dotted) with the corotating
  waveform from the Sq(4,0.6,270,9) simulation (thick).   The  colored curves correspond to the ($2,\pm 2$) modes
  (red, solid); the $(2,\pm 1)$ modes (blue); and the $(3,3)$ mode (green).   A timeshift has been applied to align the two
  simulations. 
This comparison shows several modes from the same nonprecessing simulation reproduce the corotating frame modes from
another \cite{gwastro-mergers-nr-ComovingFrameExpansionSchmidt2010}.
}
\end{figure}

Our calculations demonstrate these inexpensive models provide a fast, surprisingly accurate model for generic precessing
signals.    Moreover, though not described in detail here, the rotation operation $R(t)$ can be easily modeled and fit,
particularly in a frame aligned with the total angular momentum.  In principle,  comparisons like our own allow  fits
to functions $\lambda_{PB}(\lambda)$  and $R(t|\lambda)$, hence producing a ``synthetic waveform''
$R(\lambda)\WeylScalar_{PB}(\lambda_{PB})$.   Our analysis suggests these ``synthetic waveforms'' can be used as a simple method with which to prototype data analysis strategies for precessing,
multimodal sources.  
However, because nonprecessing models omit critical physics, particularly in the high-mass regime emphasized in this
paper, we do not recommend high-precision  calibration of  these synthetic waveforms, particularly if that calibration
is limited to a study of only the leading-order $(2,2)$ mode.

\begin{figure}
\includegraphics[width=\columnwidth]{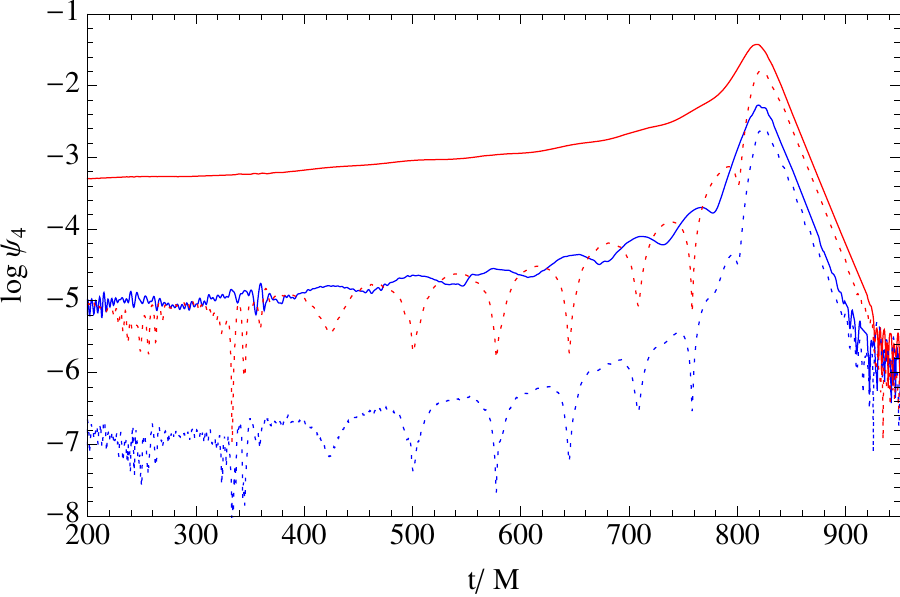}
\caption{\label{fig:Prototype:BreakReflectionSymmetry}\textbf{Precessing systems break reflection symmetry in the
    corotating frame}: For the Sq(4,0.6,90,9) simulation provided in \cite{gwastro-mergers-nr-Alignment-ROS-Polarization}, a plot of the reflection-symmetric
  (solid) and reflection-antisymmetric (dotted) parts of the corotating-frame $l=2$ modes, for $m=\pm 2$ (red) and  $\pm 1$
  (blue); see Eq. (\ref{eq:def:AB}).  
This figure suggests  that precessing binaries radiate in a way unlike any nonprecessing binary, even one viewed in a
noninertial frame; that this missing  physics has a comparable effect to higher harmonics like the ($2,\pm 1)$ modes
(e.g., dotted red vs solid blue); and that both reflection symmetry breaking and higher harmonics are required to
accurately model the merger phase of generic precessing binaries.  
}
\end{figure}

Even when viewed in a rotating frame, a nonprecessing binary retains its intrinsic symmetries: spins aligned with
the orbital plane.  Without precessing spins to break symmetry and source other multipoles, the emitted radiation has
strong symmetries: reflection symmetry through the orbital plane, insuring that at every instant the binary radiates
equal and opposite left- and right-handed radiation, in mirror-symmetric directions.   
By contrast, our previous calculations \cite{gwastro-mergers-nr-Alignment-ROS-Polarization} and long experience with
black hole superkicks demonstrate that precessing black holes radiate \emph{asymmetrically}, emitting preferentially
right-handed or left-handed radiation at any instant.    
Equivalently, even in the corotating frame, precessing binaries break reflection symmetry through the orbital plane.
In Section \ref{sec:AsAKludge} we use reflection symmetry to quantify the extent to which a nonprecessing model omits
critical physics.  To quantify the magnitude to which reflection symmetry is broken, we use conjugation symmetry to define CP-odd
($b_{l,m}(t)=-(-1)^lb_{l,-m}(t)^*$) and CP-even ($a_{l,m}(t)=(-1)^la_{l,-m}(t)^*$) parts of the corotating-frame Weyl
scalar ($\WeylScalarCorot$):
\begin{eqnarray}
\label{eq:def:AB}
[\WeylScalarCorot]_{l,m} &=& a_{l,m} + b_{l,m}  
\end{eqnarray}
Globally, this specific symmetry ${\cal C}$ corresponds to a reflection symmetry through the $z=0$ plane: 
\begin{align}
{\cal C} \psi_4 &\equiv  \WeylScalar(\pi-\theta,\phi)^* \\
 &= \sum_{l,m} [a_{l,-m}-b_{l,-m}][(-1)^l\Y{-2}_{l,m}(\pi-\theta,\phi)]^*  \nonumber \\
 &=  \sum_{l,m} [a_{l,-m}-b_{l,-m}]\Y{-2}_{l,-m}(\theta,\phi)   
\end{align}
For a nonprecessing binary, the CP-odd term $b_{l,m}=0$.  
Using a concrete precessing example, Figure \ref{fig:Prototype:BreakReflectionSymmetry} shows the CP-even (solid) and
CP-odd (dotted) parts of the leading-order
corotating-frame angular modes $\WeylScalar_{l,m}$.    Comparing the red dotted and solid lines, this figure demonstrates that after merger, the CP-odd part of the dominant mode is
significant.  In other words, this figure shows that, to accurately describe this binary's emission versus time and
angle, we \emph{must} include a CP-odd part -- a part no nonprecessing signal
could ever produce!
More broadly, because a precessing binary has more physics, its signal remains more complicated than one from a
nonprecessing binary, even in the corotating frame.

\begin{figure}
\includegraphics[width=\columnwidth]{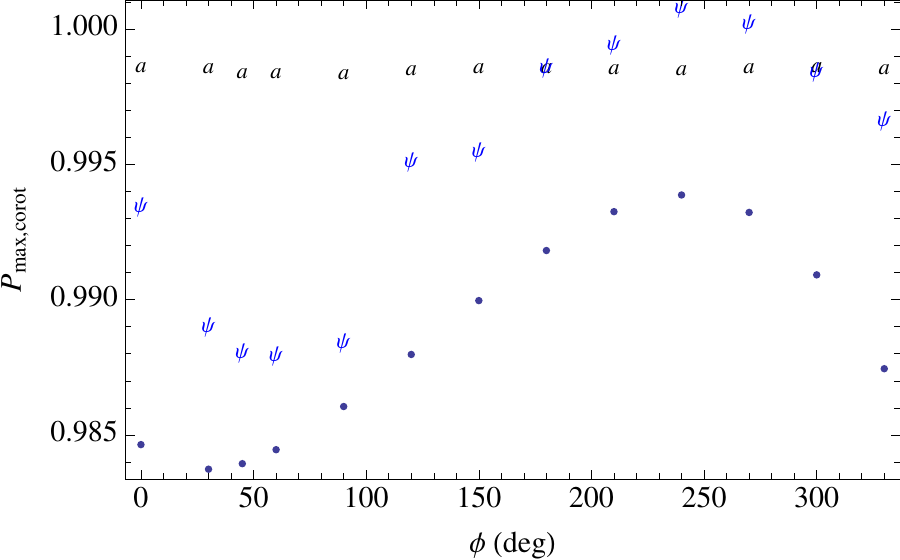}
\caption{\label{fig:ProofAsymmetryMatters}\textbf{Does it matter if we omit the transverse spin?}:   This figure
  demonstrates that transverse spins have a significant effect on the gravitational wave signal, via
  reflection-symmetry-breaking  between the $(2,\pm 2)$ modes.   
For the special-purpose  $V(a,\theta,\phi)$ family of precessing simulations described in the text and  Appendix
\ref{sec:Simulations}, points show  the match $P_{\rm max,corot}$ versus the angle $\phi$, for the fixed angle $\theta=34^o$.   For comparison, this figure
uses $\psi$ to indicate  the match $P_{\rm max,NR}$ [Eq. (\ref{eq:Pmax:NR})] evaluated between the corotating-frame $(2,2)$ mode for one member of that family
[$V(0.6,34,240)$] and all others, directly demonstrating how much these signals differ from one another.    
For another comparison, this figure uses $a$ to indicate the match $P_{\rm max,NR,a}$  [Eq. (\ref{eq:Pmax:NR:a})] between the reflection-symmetric part of the corotating-frame $(2,2)$
 ($a_{2,2}$) of that same simulation and all other angles $\phi$.  The strong variability in the first set ($\psi$) and
lack of variability in the second ($a$) suggests that reflection asymmetry is the dominant source of variation between
different simulations.  
This figure illustrates strong, generic correlations between   the precise magnitude and direction of the transverse
spins; reflection asymmetry between the $(2\pm 2)$ modes in the corotating frame; and the degree of similarity between the $(2,2)$
corotating-frame mode and conventional non-spinning approximations.  
}
\end{figure}

Figure \ref{fig:ProofAsymmetryMatters} illustrates
the practical implications of these asymmetries.   This figure shows the match $P_{\rm max, corot}$ for a one-parameter family of  high-symmetry
simulations: binaries with $m_1=m_2$,  $\vec{a}_1 = a \hat{n}(\theta,\phi)$, and $\vec{a}_2=a\hat{n}(\theta,\phi+\pi)$
as a function of $\phi$.    By construction, each member of this one-parameter family has an identical projection of the
spin along the orbital angular momentum axis ($\chi_{\rm PB}$) and more generally identical total spin ${\bf S}_1+{\bf S}_2$; evolves without precession of the orbital plane; and produce black holes whose final mass and spin does not
 change significantly with $\phi$ [Table \ref{tab:Simulations}].    
Moreover,   these binaries produce similar \emph{symmetric}
radiation $a_{2,2}$ for all orientations $\phi$.   To quantify this similarity, Figure \ref{fig:ProofAsymmetryMatters}
shows the value of 
\begin{eqnarray}
\label{eq:Pmax:NR:a}
P_{\rm max, NR,a} =  \max_t \frac{
|\qmstateproduct{a_{2,2}}{a' _{2,2}(t)}|
}{
|a_{2,2}||a'_{2,2}|
}
\end{eqnarray}
evaluated between one reference simulation and all others, at one reference mass; the high and nearly constant match demonstrates  $a_{2,2}$ is
effectively independent of $\phi$.\footnote{Similar results occur at all masses.  In the interests of brevity, we do not
  plot all of the functions  $a_{2,2}(\phi)$, even though their manifest similarity is immediately apparent to the eye. }
Nonetheless, the total gravitational wave signal $[\WeylScalarCorot]_{2,2}=a_{2,2}+b_{2,2}$ emitted by these binaries
measurably changes with $\phi$.  These differences can be easily assessed simply by noting the best-fitting parameters
and 
 $P_{\rm max, corot}$ recovered when comparing with \texttt{IMRPhenomB} vary with mass; as an example of the latter
diagnostic, see the points in Figure
\ref{fig:ProofAsymmetryMatters}.  More directly, these differences can be demonstrated by computing the overlap between
different simulations $[\WeylScalarCorot]_{2,2}$ for the values $\phi,\phi'$, via 
\begin{eqnarray}
\label{eq:Pmax:NR}
P_{\rm max NR} =  \max_t \frac{
|\qmstateproduct{[R^{-1}\WeylScalar_{NR}]_{2,2}}{[R^{' -1}\WeylScalar_{NR}']_{2,2}}|
}{
|R^{-1}\WeylScalar_{NR}| [R^{' -1}\WeylScalar_{NR}']_{2,2}
}
\end{eqnarray}
One such example is provided by the markers labeled $\psi$ in that figure.  
Figure \ref{fig:ProofAsymmetryMatters} suggests  the asymmetry ($b_{2,2}$) is principally responsible both for a
significant fraction of the mismatch $1-P_{\rm max,corot}$ and for all the fluctuation in the match versus $\phi$.    
This correlation is generic:  all other generic binaries preferentially radiate asymmetrically as the spins precess,  above and below the
instantaneous orbital plane.  To a first approximation, the dominant, symmetric part $a_{2,2}$ resembles a nonprecessing
signal; a comparison between $a_{2,2}$ and \texttt{IMRPhenomB}  approximately determines the best-fitting parameters.
Particularly during at late times when   asymmetries between $(2,\pm 2)$ become particularly significant, however, the
influence of asymmetry 
can significantly diminish (or enhance) the similarity between nonprecessing and precessing signals  -- in our context, change the
best-fitting \texttt{IMRPhenomB} parameters and match $P_{\rm max,corot}$.  
Broadly speaking,  our results  suggest larger degrees of asymmetry between $(2,\pm 2)$ correlate with a generally larger but spin-orientation-dependent mismatch between the corotating-frame (2,2) modes and
\texttt{IMRPhenomB}.

To summarize, our study suggests that nonprecessing signals, suitably rotated, \emph{resemble} but cannot reproduce with
high precision the gravitational wave signal from  generic precessing binaries.   With few
exceptions,\footnote{For special quantities constrained
by symmetry, like the opening angle of the precession cone, parameter estimation should be reliable, independent of
whether the corotating-frame model works well, due both to separation of timescales and due to the way polarization
modulations enter into the signal} synthetic precessing signals generated from nonprecessing binaries will not be
adequate for high-precision parameter estimation, unless augmented by new physics (i.e., transverse spins) and multiple harmonics.  That said, given the robust similarity of nonprecessing and precessing merger signals over observationally relevant intervals, we
anticipate synthetic precessing signals will be extremely useful \emph{as plausible signals}.    We
strongly recommend that existing gravitational wave search strategies for high-mass merger signals test how reliably they can recover the
complex multimodal from precessing binaries with $M>\ReferenceMassLow{} M_\odot$ using ``synthetic'' multimodal inspiral-merger-ringdown
signals, generated simply by viewing a nonprecessing binary in a
precessing frame.

Finally, motivated by previous studies and our own results [e.g., Figure \ref{fig:Prototype:BreakReflectionSymmetry}],  we anticipate that detailed parameter
estimation of high-mass ($M>\ReferenceMassLow{} M_\odot$) binary mergers will require detailed modeling of multiple modes of
generically-precessing binaries.    A corotating frame and a nearly-
nonprecessing CP-even
approximation can reduce the number of functions to fit; in this mass region, however, higher harmonics cannot be neglected.

\section{Comparing the corotating and simulation frame}
\label{sec:Compare}
\subsection{Why compare nonprecessing and precessing binaries in the corotating frame?}
\label{sec:RationaleAndReview}

\begin{figure}
\includegraphics[width=\columnwidth]{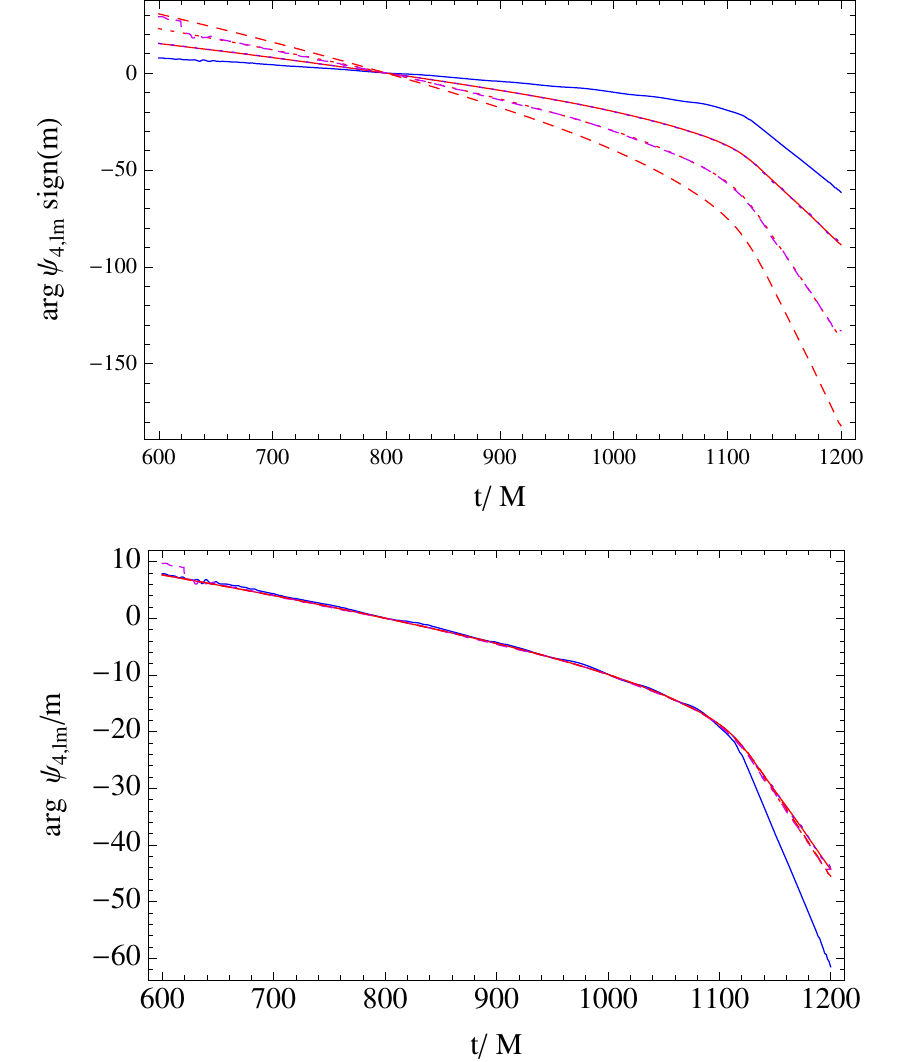}
\caption{\label{fig:SymmetryExample:Corotating}\textbf{Corotating modes evolve nearly in phase prior to merger}: Phases of corotating modes in an absolute scale (top panel) and
  scaled in proportion to their angular order
  $\phi_{l,m}/|m|$ (bottom panel), for the  Tq(2,0.6,90) simulation.  To a good approximation all evolve in phase prior
  to merger.  After merger, all angular harmonics  shown continue to evolve nearly in phase, except  the $(2,1)$ mode.  
Colors indicate the $(l,l)$ modes (red, with solid as $l=2$, dotted as $l=3$, and dashed as $l=4$); the $(2,1)$ mode
(blue); the (4,3) mode (dashed purple); and the (3,2) mode (dotted purple).
}
\end{figure}

As we will discuss at greater length in Section \ref{sec:AsAKludge}, with more degrees of freedom in their underlying kinematics, precessing binary mergers never look exactly like a
nonprecessing merger, even in the corotating frame.  In the corotating frame, the gravitational waves from precessing binaries  break symmetries
and carry more information than the signal from the best-fitting nonprecessing analog.   Nonetheless, the additional degrees of freedom are very difficult to excite in a
quasicircular inspiral.    In surprisingly many scenarios,  a precessing binary may be well-approximated by the emission
from a nonprecessing binary, plus a slowly-changing orientation. 
Qualitatively speaking, the corotating-frame mode amplitudes  $\WeylScalarCorot_{l,m}$ extracted from a precessing binary
have similar characteristics as the corresponding modes seen in a nonprecessing binary.     
In the time domain, all of the corotating-frame modes $\WeylScalarCorot$ have smooth phase evolution and (roughly)
smooth amplitude evolution.\footnote{The corotating-frame   $|\WeylScalarCorot_{l,m}|$ are weakly modulated during the
  inspiral, in two ways.  First, the $(l,\pm m)$ modes can increase or decrease at the same time (in phase),  due to
  either to residual eccentricity or suitable spin precession (for suitable modes). The $(l,\pm m)$ modes can increase or decrease out of phase, as precessing spins source
  symmetry-breaking multipole moments.   For example, during inspiral the leading-order asymmetry occurs from current quadrupole
radiation, sourced
by the transverse spins being transported around the orbit.     The theory underlying the latter scenario is briefly reviewed in the text and
  Appendix \ref{sec:MathMethods}.} 
  This smooth and simple behavior  in the amplitude and phase  has already been noted
previously
\cite{gwastro-mergers-nr-ComovingFrameExpansionSchmidt2010,gwastro-mergers-nr-Alignment-ROS-Methods,gwastro-mergers-nr-Alignment-BoyleHarald-2011,gwastro-mergers-nr-ComovingFrameExpansion-TransitionalHybrid-Schmidt2012}.  
As a first approximation,  we also find corotating modes resemble suitable nonprecessing ones; see Figure
\ref{fig:Prototype:NonprecessingAnalog:Extreme} as an example.
Because of this correspondence, the  corotating frame waveforms roughly satisfy the same symmetries as nonprecessing binaries.  For example, during the inspiral the modes evolve in phase (i.e., $\propto
m^{-1}\text{arg}\WeylScalarCorot_{l,m}$); see Figure \ref{fig:SymmetryExample:Corotating}.  
More generally, all modes are
approximately phase conjugate (i.e., $\text{arg}\WeylScalarCorot_{l,m} \simeq - \text{arg}(-1)^l\WeylScalarCorot_{l,-m}$).   
By contrast, as has repeatedly been demonstrated in the literature, the simulation-frame modes are both significantly
different and more complicated
\cite{gwastro-mergers-nr-ComovingFrameExpansionSchmidt2010,gwastro-mergers-nr-Alignment-ROS-Methods,gwastro-mergers-nr-Alignment-BoyleHarald-2011,gwastro-mergers-nr-ComovingFrameExpansion-TransitionalHybrid-Schmidt2012,gwastro-mergers-nr-Alignment-ROS-Polarization}.  
Except for  nearly-nonprecessing binaries, the substantial differences between the corotating-frame and
simulation-frame waveforms are much more significant than the relatively small numerical uncertainties in this mass
regime. %

In short, the corotating frame provides a significantly different, simpler waveform with many approximate symmetries.
Previous experience suggests these symmetries facilitate interpretation and quantitative calculations for precessing
merging binaries \cite{gwastro-mergers-nr-Alignment-ROS-Polarization,gw-astro-SpinAlignedLundgren-FragmentA-Theory}.
Moreover, even though observers cannot co-rotate with the binary, analytic calculations suggest the corotating frame
signal has direct observational relevance, based on separation of timescales.  
For example, as we highlighted in our executive summary, comparisons between the (unobservable) corotating-frame signals are
directly comparable to comparisons between (observable) simulation-frame quantities: $P_{\rm max,corot} \simeq P_{\rm
  max,sim}$ [Eqs. (\ref{eq:PmaxCorot},\ref{eq:PmaxSim}); see Figure \ref{fig:ProofFitWorks:Synthetic}].

\subsection{Comparing a corotating mode with a phenomenological model using a complex overlap}
\label{sec:Quantitative22Comparison}

Following previous studies
\cite{gwastro-mergers-nr-Alignment-ROS-IsJEnough,gwastro-mergers-nr-Alignment-ROS-Polarization,gwastro-mergers-HeeSuk-FisherMatrixWithAmplitudeCorrections},
we use a complex overlap to assess the observationally-relevant differences between two signals $a,b$ 
\begin{align}
\qmstateproduct{a}{b} &\equiv 2 \int_{-\infty}^{\infty} \frac{a^*(f)b(f)}{S_h    (2\pi f)^4} 
\end{align}
where each of $a,b$ represent some value for the outgoing Weyl scalar $\WeylScalar_{2,2} \equiv \qmstateproduct{2,2}{\WeylScalar}= \int [\Y{-2}_{l,m}]^* \WeylScalar d\Omega$.   
see, e.g., Eq. (24) in  \cite{gwastro-mergers-nr-Alignment-ROS-Polarization} and the detailed review in Appendix
\ref{sec:MathMethods}.  
In this work, we apply this diagnostic to pairs of signals in the corotating frame [Eq. (\ref{eq:PmaxCorot})] and in the
simulation frame [Eq. (\ref{eq:PmaxSim})].

\subsection{The \texttt{IMRPhenomB} family as a continuous reference model}
\label{sec:IMRPhenom}

Though numerous, our numerical simulations of nonprecessing and precessing binaries only discretely sample the model
space.   For this reason, rather than directly compare pairs of simulations,  we compare our simulations against
\texttt{IMRPhenomB} \cite{gwastro-Ajith-AlignedSpinWaveforms}, a model for $h_{2,2}$ as a function of total binary mass $M$, mass ratio $\eta=m_1
m_2/(m_1+m_2)^2$, and a single effective spin parameter $\chi_{PB}$:
\begin{subequations}
\label{eq:def:AlignedSpinOptions}
\begin{eqnarray}
\mathbf{\chi}_{\pm} &\equiv& \hat{L}\cdot ( {\bf S}_1/m_1^2 + {\bf S}_2/m_2^2)/2  \\
\chi_{\text{PB}} &=&  \hat{L}\cdot (m_1 {\bf S}_1/m_1^2 +m_2 {\bf S}_2/m_2^2)/M \nonumber \\
&=& [(1+\delta) ({\bf S}_1/m_1^2) + (1-\delta){\bf S}_2/m_2^2]/2 \nonumber \\
&=& \chi_++ \delta \chi_- 
\end{eqnarray}  
\end{subequations} 
where $\delta = (m_1-m_2)/M$.   
This model employs a physically-motived piecewise-continuous expression for $\tilde{h}_{22}(f)$, expressed as an
amplitude $|h_{22}|$ and phase $\text{arg}h_{22}$.  At low
frequencies their expressions reproduce conventional stationary-phase  approximations derived from post-Newtonian
theory; conversely,  at high frequencies, their amplitude model $|h_{22}(f)|$ has the Lorentzian form expected from a
quasinormal mode ringdown.  Parameters of this hybrid, phenomenological model were set by comparing to numerical
simulations and to an extreme-mass-ratio limit. 
This model has been calibrated against numerical simulations;  includes the effect of aligned spins;  and has been made publicly-available
via the \texttt{lalsimulation} toolkit.    
As this code provides a model for the strain ($h_{PB}$) rather than the Weyl scalar, we explicitly convert between
representations using the Fourier transform
\begin{align}
\WeylScalarFourier_{PB}(f|&M,\eta,\chi_{PB},t_{mgr}) = -  \frac{1}{(2\pi f)^2} \tilde{h}_{PB}(f) 
\end{align}

Because the inspiral and final black hole depend on the component masses and spins in distinctly
different ways, no one \texttt{IMRPhenomB} model can reproduce a generic corotating-frame $(2,2)$ mode for all
time.  For this reason, we distinguish between the simulated binary's  parameters ($M_{\rm sim},\eta_{\rm sim},
\chi_{PB,\rm sim}$) and the parameters of some best-fitting \texttt{IMRPhenomB} model  ($M_{\rm
  fit},\eta_{\rm fit},\chi_{\rm PB,fit})$.

By comparing a model against our simulations rather than comparable simulations against one another, we introduce
systematic error.  To quantify the degree of similarity between the \texttt{IMRPhenomB} model  and our nonprecessing
simulations, we have evaluated $P_{\rm max,sim}=P_{\rm max,corot}$ between this model and an array of 34 nonprecessing
simulations.  As indicated by the red curves in Figure \ref{fig:ProofFitWorks}, for most simulations and a wide range of
masses, some member of the  \texttt{IMRPhenomB} model matches each of our simulations to better than $\simeq 1-2\%$.
The best-fitting parameters are always close to our simulation parameters. %

Since \texttt{IMRPhenomB} was first published,  other models have been developed that may even more accurately reproduce the merger of precessing binaries over a
wide range of masses and spins \cite{gwastro-nr-Phenom-Lucia2010,gw-astro-EOBspin-Tarrachini2012}.  
While we have compared these models against some of our numerical simulations, in this work we limit our comparisons to the
sufficiently accurate and easy-to-evaluate \texttt{IMRPhenomB} model.

\subsection{In the corotating frame, nonprecessing and precessing binaries resemble one another }
\label{sec:sub:Resemblance}

 Figure \ref{fig:ProofFitWorks} summarizes the results of our comparison between \texttt{IMRPhenomB} and each
 numerical simulation, in the corotating frame.  In short, the two nearly agree: for every simulation, some
 nonprecessing \texttt{IMRPhenomB} model exists which nearly reproduces that simulation's corotating $(2,2)$ mode
 ($[\WeylScalarCorot]_{2,2}$) over an observationally relevant interval.  

A closer investigation of Figure \ref{fig:ProofFitWorks}, however, suggests that systematic differences exist between
 corotating-frame waveforms and nonprecessing signals.  In that figure, the red curves show that truly nonprecessing
 simulations are better fit by \texttt{IMRPhenomB} (red curves) than the corotating-frame signal from a precessing
 binary (black curves), with matches of  order $1-3\%$ better.   
This level of disagreement is large enough to be observationally accessible: typically, parameter estimation strategies
can resolve differences between models when their matches differ by $1/\rho^2$, expected to be of order $ \gtrsim
1/10^2$ for the first detections at signal amplitude $\rho \simeq 10$.  
As highlighted in the executive summary  and as we will describe in detail later, these differences between the
corotating-frame and simulation-frame signal are expected on physical grounds: the real binary has more degrees of
freedom, reflected in the orbit and critically current quadrupole radiation.     These differences fundamentally limit
the accuracy of synthetic signals that hope  to reproduce precessing signals using suitably-rotated nonprecessing
binaries.   Equivalently,  the difference between the red and black curves in Figure \ref{fig:ProofFitWorks} suggests an
accuracy threshold for proposals that hope to carefully calibrate such models against  numerical merger signals.

For comparison, we have repeated the fitting process in the simulation frame, computing $P_{\rm max}$ by directly comparing
$\WeylScalar_{PB,2,2}$ to $\WeylScalar_{NR,2,2}$.  Most of our simulations have
approximately aligned  initial orbital ($\hat{L}$) and total ($\hat{J}$) angular momentum
directions.   As a result, the fitting procedure generally finds a similar best fit $P_{\rm max}$ at a similar
parameter location.  The small offset between the best-fit simulation frame and corotating frame parameters can be
qualitatively understood: rapid in-band precession shifts the best-fit parameters by an amount proportional to the
post-merger precession frequency.   
The question  of directly comparing precessing simulations to nonprecessing models has considerable practical interest,
particularly for efforts to estimate binary parameters using simple models for the merger signal.  That said, the
results of this comparison are beyond the scope of our current study and not provided here.

\section{Understanding the best-fit parameters}
\label{sec:UnderstandBestFitIfPossible}

As described above, for each simulation and mass, we have found the nonprecessing model parameters ($M_{\rm
  sim},\eta_{\rm sim},\chi_{\rm PB,sim}$) %
such
that the corotating-frame $(2,2)$ mode and the \texttt{IMRPhenomB} mode are most similar.   
As illustrated by Figure
\ref{fig:BestFitCurves:Chi}, 
the best-fit nonprecessing
parameters are neither constant nor trivially related to the precessing binary being simulated.   In fact, on physical
grounds we expect and our calculations show that the 
best-fit nonprecessing model to  evolve from a \emph{low-mass} limit that reproduces qualitative features of the
corotating-frame inspiral to a \emph{high-mass} limit that reproduces qualitative features of the corotating-frame final
black hole's ringdown.

\subsection{Low-mass limit }

\begin{figure}
\includegraphics[width=\columnwidth]{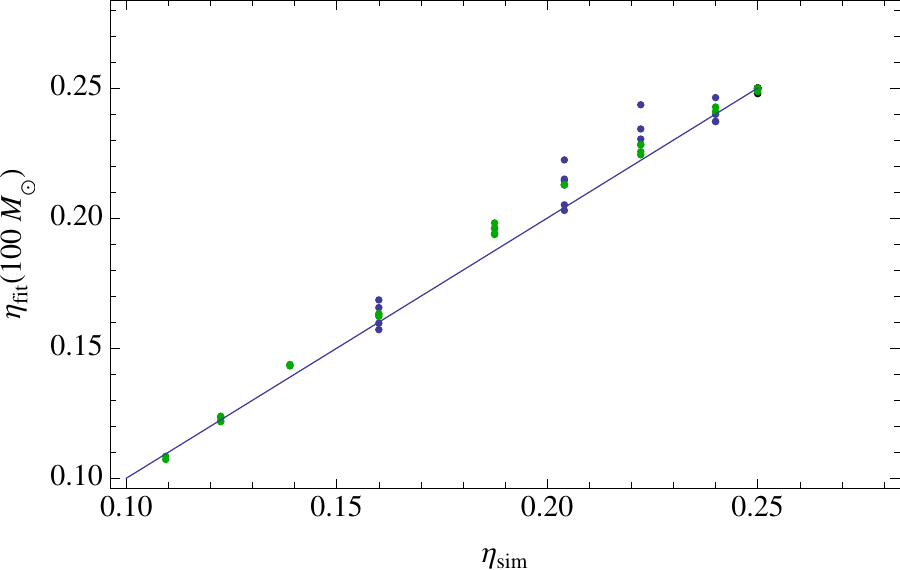}
\caption{\label{fig:ParameterRecovery:LowMass:MassRatio}\textbf{Low mass limit: Recovering the mass ratio}: 
At low masses and therefore early times, the best-fit
  corotating-frame model reflects the initial binary.  These panels show  comparisons between the best-fit \texttt{IMRPhenomB}
  parameter $\eta_{fit}$ and the physical binary mass ratio $\eta$, using best-fit values at $M=\ReferenceMassLow{} M_\odot$;  compare to Figure
  \ref{fig:ParameterRecovery:LowMass:Spins}.  
  This figure explicitly excludes all of our
  shorter simulations ($d<7M$).    Green points are nonprecessing
  spin-aligned systems; black points are long-duration simulations from the Sq series; and blue points are long-duration
  simulations from the Tq series.   A solid blue line at $\eta_{\rm sim}=\eta_{\rm fit}(\ReferenceMassLow{} M_\odot)$ is shown to guide
  the eye.  [An identical color scheme is adopted in the subsequent Figure
    \ref{fig:ParameterRecovery:LowMass:Spins}.]
}
\end{figure}

In the advanced LIGO sensitive band, the gravitational wave signal  from a low mass binary $M<300 M_\odot$ is produced
principally \emph{prior} to merger, when the two objects can be distinguished as separate objects.    
At low mass the best-fitting nonprecessing model should resemble the initial binary, reproducing its
orbital evolution.  In other words, not only will the \texttt{IMRPhenomB} model fit, but the best-fitting model
parameters are actually physical properties of the binary.  Specifically, the best-fitting nonprecessing model $\WeylScalar_{PB}(\lambda_{PB})$ should have similar  mass ratio and  ``effective spin''
\cite{1995PhRvD..52..848P,2001PhRvD..64l4013D,gwastro-Ajith-AlignedSpinWaveforms,2011PhRvD..84h4037A}.  
Figure \ref{fig:ParameterRecovery:LowMass:MassRatio} compares the simulation's mass ratio $\eta_{\rm sim}$ to the best-fitting \texttt{IMRPhenomB} mass ratio
$\eta_{\rm fit}$ for all of our long-duration simulations (i.e., an initial separation $d>7$).  To emphasize the inspiral,
this figure adopts the best-fit parameters at $M=\ReferenceMassLow{}M_\odot$.     Despite systematic differences between the
\texttt{IMRPhenomB} model and our nonprecessing simulations (green points), this figure and Figure
\ref{fig:ProofFitWorks} together suggest that both nonprecessing
and precessing simulations  and the \texttt{IMRPhenomB} model all produce nearly-indistinguishable estimates for
the inspiral phase of $\WeylScalarCorot_{2,2}$ at the same physical parameters.    

In practice, the short duration of most of our precessing  simulations limits our ability to recover the inspiral signal
in isolation.   
Equivalently, as illustrated in  Figure \ref{fig:ProofFitWorks}, the best fit between \texttt{IMRPhenomB} and our
precessing signals is relatively poor below $250 M_\odot$.   
That said, because \texttt{IMRPhenomB} includes both inspiral and merger, we anticipate the best-fit parameters at
$M=\ReferenceMassMiddle{} M_\odot$ will still correspond to binary parameters that approximate the simulated system just prior to merger,
in a corotating frame; see the Appendix for further discussion.  
  Figure \ref{fig:ParameterRecovery:LowMass:Spins}   shows an example of the best-fitting \texttt{IMRPhenomB}
spin parameter $\chi_{\text{PB,fit}}$ for a one-parameter family of simulations, evaluated at $M=\ReferenceMassMiddle{} M_\odot$.
 Given the fundamental physical differences between the \texttt{IMRPhenomB}
model and simulations and given the mass used for comparison, we are not surprised that the best fit parameters do not
lie precisely on any theoretically anticipated correlation.

\begin{figure}
\includegraphics[width=\columnwidth]{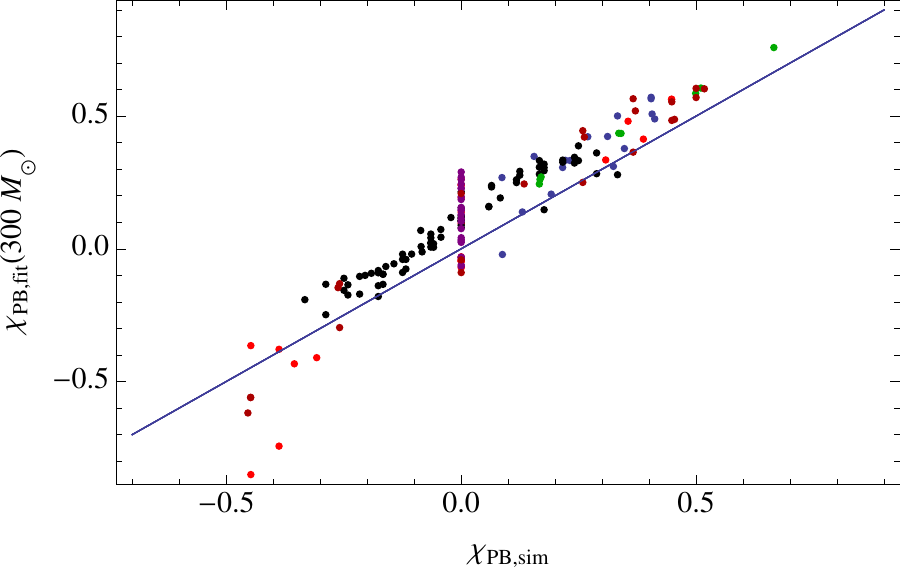}
\caption{\label{fig:ParameterRecovery:LowMass:Spins}\textbf{Low mass limit: Recovering the effective spin}: At low masses and therefore early times, the best-fit
  corotating-frame model reflects the initial binary.  These panels show  comparisons between the best-fit \texttt{IMRPhenomB}
  parameter $\chi_{eff}$ and $\chi_{\rm PB}$ %
  [Eq. (\ref{eq:def:AlignedSpinOptions})], using best-fit values at $M=\ReferenceMassMiddle{} M_\odot$.  Colors indicate the Sq series
  (black); the Tq series (blue); and the T series (red).   
Colors indicate the Sq series (black); precessing binaries in the Tq series (blue); aligned spin binaries from the Tq
series (green);  and the short simulations of the T series (red), Eq series (purple), and Lq series (dark red).     
 A solid blue line at $\chi_{\rm PB,sim}=\eta_{\rm PB,fit}(\ReferenceMassLow{} M_\odot)$ is shown to guide
  the eye.
Though provided at high mass ($M=\ReferenceMassMiddle{}M_\odot$) to
insure short simulations are well-resolved, a similar distribution is recovered when using earlier epochs of longer
signals (i.e., $M\simeq \ReferenceMassLow{} M_\odot$).  
}
\end{figure}

Even in the best of circumstances, we do not anticipate being able to accurately reproduce all parameters perfectly with
our fitting process.   At any mass, mass ratio, and spin,  gravitational wave signals like $\WeylScalar_{PB}(\lambda)$ with neighboring
parameters $\lambda$ closely resemble one another.   Depending on the particular signal model and parameters $\lambda$,
some parameter combinations are easier to identify than others via a fitting process.   The easy-to-measure and
hard-to-measure parameters can be identified via a Fisher matrix \cite{CutlerFlanagan:1994}; as an example, in the
interpretation of low-mass nonprecessing binaries,  a particular correlated combination of mass,
mass ratio, and spin is exceptionally difficult to constrain observationally \cite{CutlerFlanagan:1994}.  
The accuracy of our fit  is also limited by   by systematic differences between \texttt{IMRPhenomB} and our signal.   
In this proof-of-concept study we will not provide a detailed analysis of the systematic uncertainties associated with
this fitting procedure. 

\subsection{High-mass limit}

The late-time gravitational wave signal from a merging binary should be dominated by its leading-order quadrupole
quasinormal mode, both in the simulation and corotating frame.  On physical grounds, we therefore expect the
best-fitting nonprecessing model to predict a similar final resonant frequency and decay timescale -- or, equivalently,
a similar final black hole mass $M_{\rm f}$ and spin $a_{\rm f}$.   
As before, we will distinguish between the final black hole mass and spins derived from our simulations ($M_{\rm
  f,sim},a_{\rm f,sim}$) and derived from the best-fitting \texttt{IMRPhenomB} parameters ($M_{\rm f,fit},a_{\rm
  f,fit}$).  

The  \texttt{IMRPhenomB} parameters $\chi_{\rm sim}, \eta_{\rm sim}$ are not transparently related to the final
quasinormal mode frequencies and hence to the assumed final black hole mass and spin.   To transform the best-fitting
parameters $\eta_{\rm fit},\chi_{\rm PB,fit}$ into a specific final black hole mass $M_{\rm f,fit}$ and spin $a_{\rm
  f,fit}$,    we adopt relationships between the
``initial'' and post-merger state that reproduce the correlations seen in our simulations.

\begin{figure}
\includegraphics[width=\columnwidth]{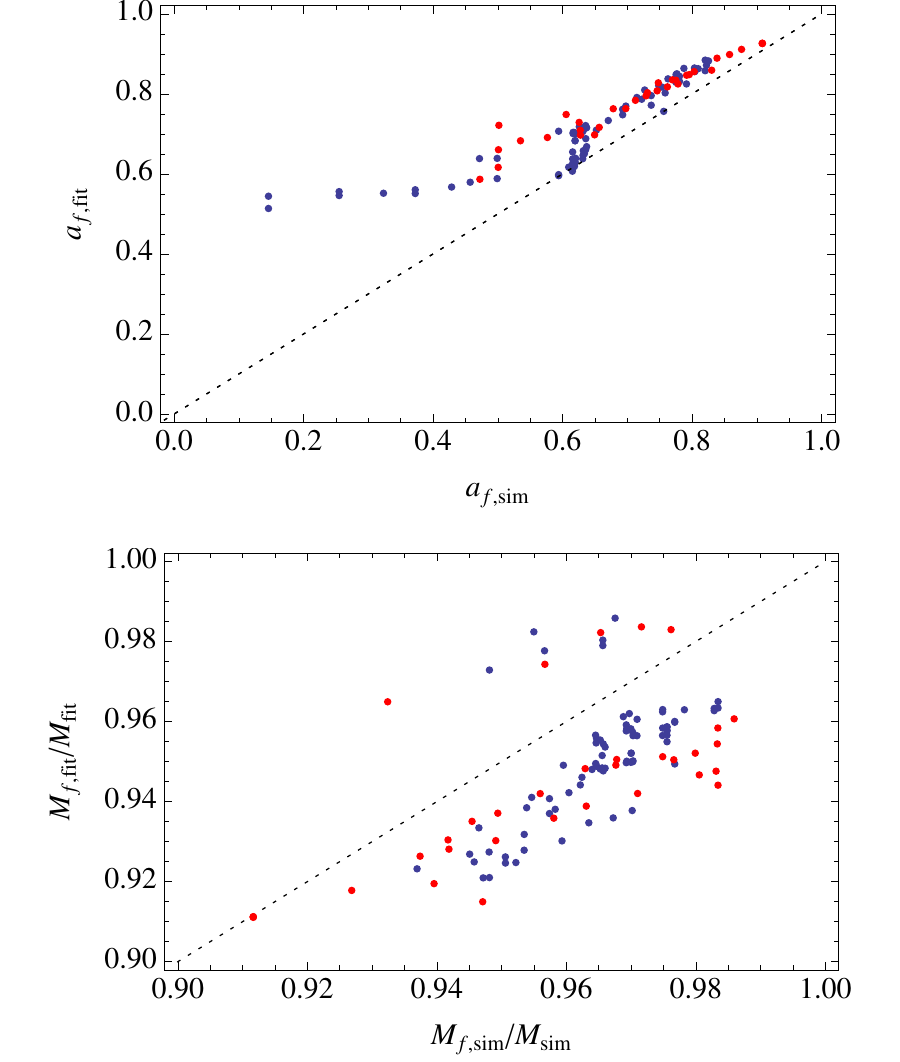}
\caption{\label{fig:UnderstandRecovered:HighMassLimit}\textbf{High mass: Best fit approximately  recovers the final state}:  At
  $M\simeq \ReferenceMassHigh{}
  M_\odot$, a scatter plot of the true ($M_{\rm sim}, a_{sim}$) and best-fit ($M_{\rm fit},a_{\rm fit}$) final black hole
  masses (bottom  panel) and spins (top  panel), derived from our precessing simulations (blue) and nonprecessing
  simulations (red).    The best-fit final black hole properties are derived from the
  \texttt{IMRPhenomB} best-fit parameters using Eqs. (\ref{eq:FinalState:Mass},\ref{eq:FinalState:Spin}).   
In both figures, a dotted black line is provided to guide the eye onto $a_{\rm f,sim}=a_{\rm f, fint}$ and similarly.
Because  the \texttt{IMRPhenomB} model claims to reproduce nonprecessing binaries,  the distribution of the red points
can be used to estimate systematic uncertainties in this method.  For example, in the top panel, the systematic offset
between the red points and the line $a_{\rm f,sim}=a_{\rm f,fit}$ suggests significant systematic differences between
our simulations' ringdown frequencies and the model of \texttt{IMRPhenomB}.  In the bottom panel, the significant
scatter in both red and blue points reflects our inability to measure total masses to better than a few percent by a
fitting procedure; see also the bottom panel in Figure \ref{fig:BestFitCurves:Chi}.  
This figure includes a separate point for each individual simulation, including some simulations with physically
identical parameters but performed with different resolutions.   
To better resolve the final black hole quasinormal modes, we have explicitly excluded all simulations performed with
$h=M/77$.  
}
\end{figure}

The transformation from nonprecessing binaries to final black holes is degenerate: the initial state is specified by
three parameters (mass, mass ratio, and at least one ``typical''
spin\footnote{Physically, a nonprecessing binary has \emph{four} parameters: mass, mass ratio, and \emph{two} spins.
  Due to symmetry, a nonprecessing binary's  $\WeylScalar_{2,2}$ must be  reflection- and exchange-symmetric, implying it
  cannot trivially (e.g., linearly) depend  on an antisymmetric combination of the binary's spins.  To an excellent
  approximation, the leading order quadrupole ($\WeylScalar_{2,2}$) produced by a nonprecessing binary depends on only
  \emph{three} parameters: mass, mass ratio, and an effective spin. }) while the final black
hole has only two (mass and spin).  
Expressions exist which relate the pre- and post-merger parameters
\cite{gr-nr-io-fitting-Boyle2007,2008PhRvD..78b4017B,gr-nr-io-review-Rezzolla2008,2012ApJ...758...63B,2010PhRvD..81h4023L,2010CQGra..27k4006L,2012ApJ...758...63B}.
For generic binaries, because spins precess between formation and merger, the precise relationship between the pre- and post-merger state
depends on precisely when the initial binary's parameters are specified
\cite{2010PhRvD..81h4054K,2010PhRvD..81h4023L,2010ApJ...715.1006K}.  
In this work, however, these approximate analytic relationships will only be applied to nonprecessing initial
parameters.\footnote{The expressions provided in the text have been extremely accurately tuned for aligned-spin
  binaries.  Though significantly less accurate for precessing binaries, these expressions are a qualitatively consistent
relationship between our simulations'  initial and final states when the binary has two precessing spins.  In
particular, these expressions usually predict the final black hole mass $M_{\rm f, sim}/M_{\rm sim}$ for generic
precessing binaries to well within  $1\%$ of their true value.  This scale can be helpfully compared with the
significant scatter visible in the top panel of in Figure \ref{fig:UnderstandRecovered:HighMassLimit}.
 }
In that case, the final black hole mass can be estimated using the following combination the pre-merger mass, mass ratio, and spins  using  \cite{2012ApJ...758...63B}
\begin{subequations}
\label{eq:FinalState:Mass}
\begin{align}
\frac{E_{rad}}{M} &= 1-M_f/M =  [1 - E_{ISCO}(\bar{a})] \nu \\
& + 4 \nu^2 [4 p_o + 16 p_1 \bar{a}(\bar{a}+1) + E_{isco}(\bar{a})-1] \\
E_{ISCO} &=\sqrt{1- 2/3 r_{ISCO}(a)} \\
\bar{a} &\equiv \frac{\hat{L} \cdot (S_1+S_2)}{M^2} \\
(p_o,p_1) &\simeq (0.04827, 0.01707)
\end{align}
\end{subequations}
where $r_{ISCO}(a)$ is the (dimensionless) radius of the Kerr horizon in Boyer-Lindquist coordinates.  
This approximation assumes the final black hole mass can depend on the spins only through $\bar{a}$ -- specifically,
only through the total spin, projected along the orbital angular momentum; other approximations allow for more degrees
of freedom and physics  \cite{2010CQGra..27k4006L}.  
Similarly, the final black hole \emph{dimensionless spin} can be computed from the pre-merger mass, mass ratio, and spins  using either
Eqs. (6-7) from \citet{2010CQGra..27k4006L}\footnote{ To implement this expression exactly  requires knowledge of the spins just
prior to plunge and merger.  We only apply this expression to nonprecessing systems -- in fact, only to the
\texttt{IMRPhenomB} best-fit parameters. } or the simpler Eq. (6) from \cite{2009ApJ...704L..40B}:
\begin{subequations}
\label{eq:FinalState:Spin}
\begin{align}
a_{\rm f} \equiv J_{\rm f}/M_{\rm f}^2 &= \frac{|\vec{a}_1+\vec{a}_2 q^2 + \hat{L}|\ell|q|}{(1+q)^2}\\
|\ell| &= 2\sqrt{3} + t_2 \eta + t_3 \eta + \frac{s_4|\vec{a}_1+\vec{a}_2 q^2|^2}{(1+q^2)^2} \nonumber \\
 &+  \frac{s_5 \eta + t_o + 2}{1+q^2} \hat{L}\cdot(\vec{a}_1+q^2 a_2) \\
(t_0,t_3,t_2) &= (-2.89, 2.57, -3.52) \\
(s_4,s_t) &= (-0.1229, 0.45)
\end{align}
\end{subequations}
These expressions accurately reproduce the results of our nonprecessing simulations. 
For example, given the initial spins, these
expressions reliably reproduce the final black hole mass we derive from the final horizon to
significantly better  than $1\%$ in all well-resolved simulations.

In terms of these relationships, we derive the final black hole  mass and spin implied by the best-fitting
\texttt{IMRPhenomB} parameters  $M_{\rm fit},\chi_{\rm
  PB,fit},\eta_{\rm fit}$ by direct substitution.  For example, we evaluate the final black hole mass $M_{\rm f,fit}$ by
assuming $S_1=\chi_{\rm PB,fit}m_1^2
\hat{L}$ and ${\bf S}_2 = \chi_{\rm PB,fit}m_2^2 \hat{L}$, where $m_{1,2}$ are derived from $\eta_{\rm fit}$, in the
expression above:
\begin{eqnarray}
M_{\rm f,fit}/M_{\rm fit} = 1- \eta_{\rm fit}(1-E_{ISCO})(\bar{a}) 
\end{eqnarray}

As shown by example in  Figure \ref{fig:UnderstandRecovered:HighMassLimit}, the best-fit parameters $M_{\rm f,
  fit},a_{\rm f,fit}$ should be
close to the final black hole's state.   This figure compares the final black hole state identified in each simulation
with the final state implied by the best-fitting \texttt{IMRPhenomB} parameters. %
At the late times and high masses studied here, both the signal model and \texttt{IMRPhenomB} will be dominated by
late-time quasinormal mode decay, with some characteristic frequency $\omega_{2,2}(M_{\rm f},a_{\rm f})$.
Qualitatively speaking, we expect the fitting process enforces near-equality between $\omega_{2,2}$ derived from
$\WeylScalar_{PB}$ and the late time ringdown frequency in the corotating frame.    The functional dependence of
$\omega_{2,2}$ on $M_f,a_f$ determines the key features seen in Figure
\ref{fig:UnderstandRecovered:HighMassLimit}.  First, the  mass ratio
$\text{Im}\omega_{2,2}/\text{Re}\omega_{2,2}$ depends strongly on the final black hole spin $a_f$; and, therefore, this fitting procedure
should reliably determine the final black hole spin parameter; see, for example, the bottom panel in Figure \ref{fig:UnderstandRecovered:HighMassLimit}.   Second, because of the extremely limited number of
cycles available to constrain the final black hole's properties, the total black hole mass cannot be reliably
measured.\footnote{A more detailed treatment of parameter estimation uncertainties in the ringdown signal is beyond the
  scope of this paper.}  In particular, because relatively little mass is loss to infinity
($M_f/M \simeq 1$), the fraction of mass lost to infinity cannot be reliably determined; see the top   panel of Figure \ref{fig:UnderstandRecovered:HighMassLimit}.

Above and beyond the fundamental limitations set by the functional form of $\omega_{2,2}$,  the best-fit \texttt{IMRPhenomB} final spin differs
systematically from the physical spin parameter for three reasons.  
Systematic differences exist between the \texttt{IMRPhenomB} model and our simulations, even for
aligned spins; see the red curves in Figure \ref{fig:ProofFitWorks}.  These differences arise due principally  due to physical
differences between the \texttt{IMRPhenomB} model and our merger simulations.\footnote{We have simulated identical nonprecessing systems at different resolutions and compared them with
  \texttt{IMRPhenomB}; all results for these simulations are included in Figure \ref{fig:ProofFitWorks}. The best-fit match and recovered parameters depends
  only weakly on resolution; by contrast, significant differences exist between  \texttt{IMRPhenomB} and any of our models.}
In addition,  the \texttt{IMRPhenomB}
approximation uses its own implicit relationship between initial binary and final black hole, encoded in the final
ringdown frequency \cite{gwastro-Ajith-AlignedSpinWaveforms}.    As indicated in  Figure
\ref{fig:UnderstandRecovered:HighMassLimit} using red points, comparable differences between the physical final state and a
final state derived from the best-fit  \texttt{IMRPhenomB} model appear when this procedure is applied both to
precessing and nonprecessing binaries.     
We have eliminated physical degrees of freedom and introduced systematic error by requiring the final black hole
mass and spin depend only on the black hole spins through $\chi_{\rm  PB}$.  By comparison, neither expressions we use
to relate the initial and final state [Eqs. (\ref{eq:FinalState:Mass}, \ref{eq:FinalState:Spin})] depend on spins through
$\chi_{\rm PB}$!  Instead, both expressions used here depend on the total spin ${\bf S}_1+{\bf S}_2$.    Alternatively,
Table \ref{tab:Simulations}  has several examples of simulations with identical mass ratio and $\chi_{\rm PB}$ but producing
significantly different $a_{\rm f}$; see, for example, Sq(4,0.6,90) and Sq(4,0.6,270).  
Finally, the corotating-frame transformation involves a rotation at an appreciable proportion of the final black hole's
quasinormal mode frequency \cite{gwastro-mergers-nr-Alignment-ROS-Polarization}.    As a result, we fully expect the
best-fit parameters to be  offset from the physical parameters of the final black hole.

\section{Synthetic signals and Missing physics}
\label{sec:AsAKludge}

\subsection{Synthetic signals}
As described in previous work on corotating frames \cite{gwastro-mergers-nr-ComovingFrameExpansionSchmidt2010,gwastro-mergers-nr-Alignment-ROS-Methods,gwastro-mergers-nr-Alignment-BoyleHarald-2011,gwastro-mergers-nr-ComovingFrameExpansion-TransitionalHybrid-Schmidt2012}, the similarity between corotating-frame modes and nonprecessing systems suggests a simple strategy for synthetic
waveform generation.  
In this procedure, we join some nonprecessing set of modes $\WeylScalar_{l,m}^{ROT}$ to a time-dependent rotation $\hat{R}$, constructively generating a synthetic
waveform for each line of sight.   
For example, each of the  corotating-frame modes could be interpolated phenomenologically, starting with our collection
of corotating-frame modes.  
Likewise, prior to merger the rotation operation could track expected  trajectory based on adiabatic quasicircular
spin-orbit evolution; after merger,  the rotation operation could be measured and calibrated to mimic suitable
precession.   
This very aggressive strategy requires careful, complicated interpolations: each mode and spins depends on several
parameters (two masses and two spin vectors).  
Less aggressive strategies could adopt simpler models for precession or the inspiral signal.  For example, a
well-motivated, parameterized  inspiral-merger-ringdown model  like the effective-one-body model could track the orbital phase.  

To assess how well this strategy could perform in optimal circumstances, we assume  the physical rotation and
best-fitting \texttt{IMRPhenomB} can be determined for each candidate physical system.  In this most optimistic case,
the match $P_{\rm max,sim}$ [Eq. (\ref{eq:PmaxCorot})] quantifies the degree of similarity between the simulation-frame
 $(2,2)$ mode $\WeylScalar_{NR,2,2}$ and the $(2,2)$ mode generated by this synthetic procedure:
 $[\WeylScalarCorot_{PB}(\lambda_{PB})]_{2,2}$.  
The top panel of Figure \ref{fig:ProofFitWorks:Synthetic} shows the distribution of $P_{\rm max,corot}$.  In short, this
strategy will be strikingly successful, fitting the IMR signal from our generic mergers to better than a few percent.  

More striking still, our simulations generally show $P_{\rm max,corot}\simeq P_{\rm max,sim}$.  In other words,  we only needed to assess how similar the corotating-frame signal is to a
nonprecessing signal model, to determine how well a
synthetic precessing search will perform.

Serious challenges exist before this synthetic waveform can be implemented in practice.  For example, the relationship
between the physical parameters of the binary, the physical rotation versus time, and the best-fitting nonprecessing
model parameters must be carefully tabulated before  these
synthetic signals could be applied to  parameter estimation.  
Also, this model omits physical degrees of freedom and cannot reproduce the signal perfectly, even with the
best-possible nonprecessing signal model.  As a practical matter, the latter limit sets the accuracy to which this
program should be pursued in detail.  

Our study also allows us to better understand the essential features a more complete interpolation strategy must have.  
Because the best-fitting nonprecessing systems are different at low and high mass, a robust strategy
must produce waveforms that resemble different nonprecessing systems early and late in the signal.  This procedure could
work physically, with a well-motivated IMR model; somewhat phenomenologically, using two basis epochs and  interpolating the junction conditions versus
spins; or completely phenomenologically, interpolating the whole waveform.   

\subsection{Limitations of nonprecessing analogs and synthetic precessing waveforms  }
\label{sec:Limits}

Precessing binaries break a symmetry: reflection through the orbital plane.  As a result, precessing binaries have
waveforms with fundamentally more complexity, \emph{even in the corotating frame}.   No nonprecessing waveform, no
matter how rotated, can reproduce that symmetry breaking; see Appendices  \ref{sec:sub:OddVsEvenExpand} and \ref{sec:sub:ResidualFreeParameters}.
As a result, the straightforward approach described above has limited accuracy, applied to generic precessing systems;
see the executive summary and Figure \ref{fig:ProofFitWorks:Synthetic}.  
In this section, we describe the physics missed when one assumes the corotating-frame waveform has an instantaneous
reflection symmetry.  

The straightforward approach  approach described above  has not included the \emph{direct} feedback of precessing spins on the waveform.  For
example, during the inspiral the orbiting spins source  current quadrupole radiation.  %
By breaking  reflection symmetry about the instantaneous orbital plane, the current quadrupole and similar
precession-sourced terms introduce behavior which cannot be reproduced by nonprecessing waveforms, no matter how
rotated.   
As described in detail in the executive summary, 
 Figure
\ref{fig:Prototype:BreakReflectionSymmetry} shows the reflection-symmetric and reflection-asymmetric parts of the $l=2$
corotating-frame waveform, illustrating the small but increasingly important impact of symmetry-breaking terms.  
Prior to merger, the symmetry-breaking terms are small, typically several PN orders smaller than the dominant term for
each mode; see Appendix \ref{sec:sub:OddVsEvenExpand} for a symmetry decomposition of the PN expansion.  
During and after merger, however, these symmetry-breaking terms become a significant fraction of the overall amplitude
$|\WeylScalar|$.  
These symmetry-breaking terms reflect an instantaneous bias towards preferentially left- or right-handed emission
\cite{gwastro-mergers-nr-Alignment-ROS-Polarization}.   
In Figure \ref{fig:Prototype:BreakReflectionSymmetry} as in most cases, the dominant asymmetry occurs between the
$(2,\pm 2)$ modes.

The effect of precessing spins and bias towards one helicity or another is far from academic, particularly at high mass.
Using a data-analysis-motivated comparison in the \emph{simulation frame},
\citet{gwastro-mergers-nr-Alignment-ROS-Polarization} have already showed that this oscillating bias towards one
handedness or the other leads to a preferred handedness in the detected signal, changing slowly as a function of mass
and line of sight.  
These modulations distort the $2,2$ mode and generally cannot be perfectly reproduced by a nonprecessing waveform.
Figure \ref{fig:Prototype:BreakReflectionSymmetry} and comparable calculations for generic sources suggests that
strong (tens of percent) symmetry breaking occurs ubiquitously in precessing mergers.

The synthetic procedure described above also neglects higher harmonics.  Higher harmonics are well-known to produce
observationally-accessible modulations of the gravitational wave signal along generic lines of sight for generic
high-mass mergers \cite{gwastro-mergers-nr-Alignment-ROS-IsJEnough,gwastro-mergers-nr-HigherModes-Larne2012}.   As
suggested by Figure 
\ref{fig:Prototype:BreakReflectionSymmetry}, higher harmonics are \emph{at least} as important as (and difficult to
disentangle from) spin precession effects.   Our calculations support previous results suggesting that the observationally-accessible information requires
detailed models for several harmonics, beyond the leading order  \cite{gwastro-mergers-nr-Alignment-ROS-IsJEnough,gwastro-mergers-nr-HigherModes-Larne2012}.

Our results, along with the previously mentioned helicity bias described in
\cite{gwastro-mergers-nr-Alignment-ROS-Polarization}, strongly suggest that high-mass precessing binaries cannot be
well-understood without breaking reflection symmetry through the orbital plane.  In turn, these symmetry-breaking terms
cannot be understood without modeling  transverse spin dependence in detail.  
For strongly precessing asymmetric binaries, we do not believe the waveform generation problem cannot be completely
decoupled from kinematics, at an observationally relevant level.
Finally, as in previous studies, we anticipate higher harmonics will be required to interpret the merger signal from
generic precessing sources.

\section{Conclusions }
\label{sec:Conclude}

In this paper we explore a simple  ``synthetic'' model for the leading-order gravitational wave signal from precessing,
merging binaries: a suitable nonprecessing binary, viewed in a suitable noninertial frame.  
Using a data-analysis-driven diagnostic, we compare the $(2,2)$ modes extracted from a large collection of binary black
hole merger simulations to such a synthetic model, both in an inertial frame and a ``corotating'' frame which tracks the
evolution of the binary.    In all cases explored here, we find that the late-time inspiral and merger signal from generic black hole mergers can
 be reasonably approximated by a nonprecessing binary seen in a corotating frame.   
Moreover, as expected on physical grounds, at early times the corotating-frame signal resembles   emission from a
nonprecessing binary with physically similar  parameters: similar mass,
mass ratio, and ``effective spin.''  
Our study restricted attention to our numerical simulations, without  analytic extrapolations 
at early (``hybridization'') or late times; that said, our results suggest that  followup investigations with suitable
semi-analytic hybrids can reasonably approximate generic precessing merger signals for all time.

Though this approximation neglects significant physics and has correspondingly limited accuracy, this ``synthetic'' approach allows the efficient
generation of complicated, multimodal signals from generic merging sources.   Because this procedure bootstraps
experience gained from nonprecessing sources, we  strongly recommend more effort be devoted to modeling, approximating,
or hybridizing the observationally-relevant higher harmonics from generic two-spin nonprecessing binaries; to
modeling precession before, during, and after merger; and hence to generating qualitatively realistic synthetic precessing merger signals.  

While a qualitatively adequate zeroth approximation,  this procedure does not easily generalize to a high-precision
quantitative approximation, with controlled error estimates.  
Nonprecessing waveforms simply cannot self-consistently reproduce features tied to the system's \emph{kinematics}:
the orbital phase versus time;  the ringdown mode frequencies, set by the final black hole's mass and spin; the ringdown
mode amplitudes, which can reflect spin-orbit misalignment; et cetera.
As a familiar example, in  post-Newtonian calculations,  time-dependent spin-orbit  and spin-spin terms must be included
in the orbital phase and calculated from suitable spin precession equations.   
More significantly, nonprecessing waveforms have a \emph{reflection symmetry} and thus cannot reproduce current quadrupole
or similar asymmetric radiation modes.    
Our study emphasized reproducing the principal $l=2$ emission from merging binaries; generic asymmetric precessing
systems possess several strong higher harmonics, many of which must be included to accurately reproduce even a
nonprecessing source \cite{gwastro-mergers-nr-HigherModes-Larne2012}.  
Our calculations support previous results suggesting that the observationally-accessible information requires
detailed models for several harmonics, beyond the leading order  \cite{gwastro-mergers-nr-Alignment-ROS-IsJEnough,gwastro-mergers-nr-HigherModes-Larne2012}.  
We anticipate detailed parameter estimation of high-mass ($M>\ReferenceMassLow{} M_\odot$) binary mergers will require detailed modeling
of multiple modes of generically precessing binaries, an effort in support of which a considerably larger sample of
generic binary mergers are required. 
Moreover, because of the significant role the merger plays in comparable-mass binary black hole mergers  with $M\gtrsim 20 M_\odot$, we
suspect that high-precision parameter estimation of generic precessing low-mass systems will  also require a 
model for precession during merger that has been  carefully calibrated against numerical simulations.

To the best of our knowledge, gravitational wave detection strategies have never been tested against  generic merger signals that self-consistently
include precession and ringdown.   Based on the quantitative similarities characterized in this work,
we strongly recommend that gravitational wave data analysis strategies for high-mass binaries  ($M>50 M_\odot$) be tested against 
simple synthetic precessing inspiral-merger-ringdown waveforms, generated by combining plausible rotations and
nonprecessing binary merger signals. 

\appendix

\section{Mathematical methods}

\label{sec:MathMethods}

\subsection{Extracting the corotating waveform}

Particularly early in the inspiral, the gravitational wave signal from merging binaries can be approximated by
the emission from instantaneously nonprecessing binaries, slowly rotated with time as the orbital plane precesses
\cite{gwastro-mergers-nr-ComovingFrameExpansionSchmidt2010,ACST,gw-astro-mergers-approximations-SpinningPNHigherHarmonics}.
At late times, the gravitational wave signal will reflect perturbations of a single black hole with a well-identified
spin axis.
In both cases and in between, a well-chosen instantaneous or global frame can dramatically simplify the decomposition of
$\WeylScalar(\hat{n},t)$ in terms of spin-weighted harmonic functions $\WeylScalar_{l,m}(t)$.
These simplifications make it easier to distinguish physically relevant from superfluous modulations; to model emission
and generate hybrids; and to formulate tests of general relativity itself.

In this paper, we adopt a preferred direction $\hat{V}$ aligned with the principal axes of $\avL$
\cite{gwastro-mergers-nr-Alignment-ROS-Methods}.  The tensor $\avL$ is defined by the following angular integral, acting
on a symmetric tensor constructed from the rotation group generators ${\cal L}_a$ acting on the asymptotic Weyl scalar:
\begin{eqnarray}
\label{eq:def:avL}
\avL &\equiv& 
 \frac{\int d\Omega \WeylScalar^*(t) {\cal L}_{(a}{\cal L}_{b)} \WeylScalar(t)
  }{
   \int d\Omega |\WeylScalar|^2
}
 \\
 &=& \frac{\sum_{lmm'} \WeylScalar_{lm'}^*  \WeylScalar_{lm}\qmoperatorelement{lm'}{{\cal L}_{(a}{\cal L}_{b)}}{lm} }{\int d\Omega |\WeylScalar|^2}  \nonumber
\end{eqnarray}
where in the second line we expand $\WeylScalar= \sum_{lm} \WeylScalar_{lm}(t)\Y{-2}_{lm}(\theta,\phi)$ and perform the
angular integral.   The action of the rotation group generators ${\cal L}_a$  on basis states $\qmstate{lm}$  is
well-understood, allowing us to re-express the tensor  $\avL$  as \cite{gwastro-mergers-nr-Alignment-ROS-PN}:
\begin{subequations}
\label{eq:OrientationReference}
\begin{eqnarray}
I_2&\equiv&  \frac{1}{2}\,(\psi,L_+L_+\psi) \nonumber \\
 &=& \frac{1}{2}\,\sum_{l,m} c_{l,m}c_{l,m+1} \psi_{l,m+2}^*\psi_{l,m} \\
I_1 & \equiv &(\psi,L_+(L_z+1/2)\psi)  \nonumber \\
 &=& \sum_{l,m} c_{l,m}(m+1/2) \psi_{l,m+1}^*\psi_{l,m} \\
I_0 &\equiv& \frac{1}{2}\left(\psi| L^2 - L_z^2 |\psi\right) \nonumber\\
 &=& \frac{1}{2}\sum_{l,m} [l(l+1)-m^2]|\psi_{l,m}|^2  \\
I_{zz} &\equiv& (\psi,L_z L_z \psi) = \sum_{l,m} m^2 |\psi_{l,m}|^2 
\end{eqnarray}
where $c_{l,m} = \sqrt{l(l+1)-m(m+1)}$.
In terms of these expressions, the orientation-averaged tensor is
\begin{align}
\avL &=& \frac{1}{\sum_{l,m}|\psi_{l,m}|^2}
\begin{bmatrix}
 I_0 + \text{Re}(I_2) & \text{Im} I_2  &   \text{Re} I_1 \\
  &   I_0 - \text{Re}(I_2) & \text{Im} I_1 \\
 & & I_{zz}
\end{bmatrix}
\end{align}
\end{subequations}
The dominant eigendirection $\hat{V}$ of this tensor specifies two of the three Euler angles  needed to
specify a frame: 
\begin{eqnarray}
\hat{V} = (\cos \alpha \sin \beta,\sin \alpha \sin \beta, \cos \beta) \; .
\end{eqnarray}
To determine the remaining Euler angle ($\gamma$), we self-consistently adjoin a rotation in the  plane
transverse to this orientation, to account for the gradual buildup of transverse phase due to precession
\cite{gwastro-mergers-nr-Alignment-BoyleHarald-2011}:
\begin{eqnarray}
\gamma(t)  = -\int_0^t \cos \beta \frac{d\alpha}{dt} dt 
\end{eqnarray}

Having specified the three Euler angles that define a new frame, we rotate the simulation-frame $\Y{-2}_{l,m}$
coefficients of $\WeylScalar$ to the new, time-dependent frame: 
\begin{eqnarray}
\WeylScalar_{l,m}^{ROT} &=&  \sum_{\bar{m}}D^l_{m\bar{m}}(R(\alpha,\beta,\gamma)^{-1}) \WeylScalar_{l\bar{m}} \\
 &=& \sum_{\bar{m}} e^{i\bar{m}\gamma} d_{\bar{m}m}(\beta) e^{im\alpha} \WeylScalar_{l\bar{m}}
\end{eqnarray}
where $R(\alpha,\beta,\gamma)$ carries the $\hat{z}$ axis to $\hat{V}$, plus a rotation transverse to that direction by $\gamma$.

All simulations of the same physical system (with the same tetrad normalization) will agree on
$\WeylScalar(t,\hat{n})$.  The choice of frame at future infinity  reparameterizes the same results.   While our choice
for the preferred frame continues to precess during and after merger, to the extent our simulations have so far
resolved, some future choice for the preferred frame could conceivably converge to a fixed frame, aligned with the final total angular
momentum direction $\hat{J}_f$.  
The choice of corotating frame depends on convention  \cite{gwastro-mergers-nr-Alignment-ROS-PN}.  
As a result, the corotating-frame waveforms we describe below can differ from those extracted using other conventions,
with differences increasing at late times.  
For the purposes of this paper -- comparison with nonprecessing binaries, principally of the leading-order mode -- we
anticipate these differences are small.   

\ForInternalReference{
\begin{figure}
\includegraphics[width=\columnwidth]{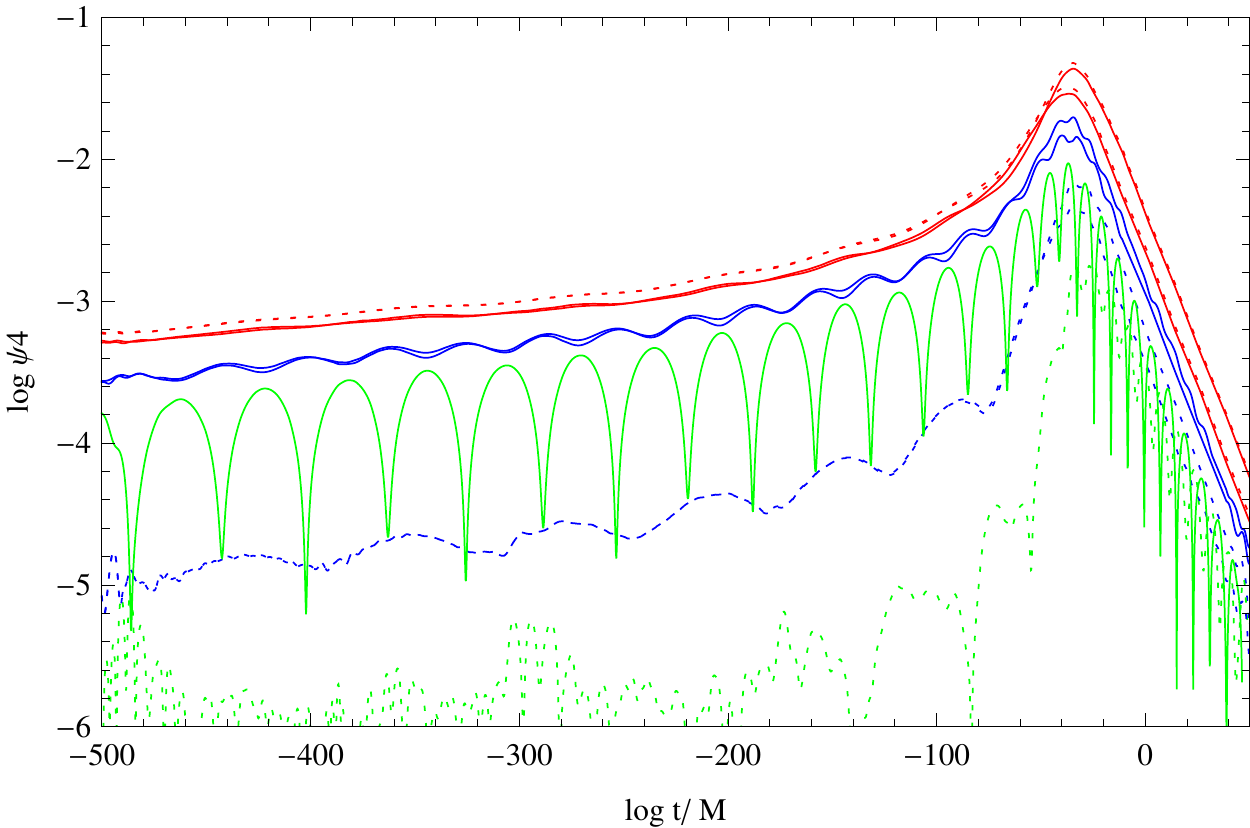}
\caption{\label{fig:FiducialWaveforms:AlignedOnJFinal}\textbf{Why not use the final BH spin frame?}: Comparison of waveforms in a frame
  aligned with the final black hole spin (solid) to the corotating frame (dotted), for the fiducial Sq(4,0.6,90,9)
  simulation.   
In this figure, red, blue, and green indicate $(2,\pm 2)$, $(2,\pm 1)$, and $(2,0)$ modes respectively.  
Prior to merger, a corotating frame captures precession-induced modulations and insures a strong mode amplitude
hierarchy.  After merger, a corotating frame produces smoother, more easily-modeled amplitudes and phases.  
}
\end{figure}
}

As shown in prior work  \cite{gwastro-mergers-nr-Alignment-ROS-Polarization} and illustrated again in the bottom panel
of Figure \ref{fig:ResolutionTest:3}, the preferred orientation $\hat{V}$ and
rotation $\gamma$ evolve smoothly, usually precessing around the total angular momentum direction.    
     In this work, we do not construct synthetic waveforms and therefore do not report on a functional approximation
     to the Euler angles as a function of time.  That said, using a frame aligned with the (initial) total angular momentum J to
     define the Euler angles, the opening angle $\beta$ is nearly constant; the precession angle $\alpha$ evolves steadily
     forward, at a precession rate set either by spin-orbit coupling (in the inspiral) or the final quasinormal mode
     frequencies (during ringdown); see Figure 8 in \cite{gwastro-mergers-nr-Alignment-ROS-Polarization}.
We therefore expect the rotation $R(t)$ can be easily and reliably fit as a function of time.

\subsection{Complex overlap and mass-weighted comparisons}
We coherently compare the (noise-free) signal expected along any pair of orientations
with a complex inner product motivated by the detector's noise power spectrum \cite{gwastro-mergers-nr-Alignment-ROS-Methods,gwastro-mergers-nr-Alignment-ROS-IsJEnough}.
For our purposes, numerical relativity simulations take as inputs binary black hole parameters and desired line of sight
(denoted by $\lambda$) and return  the Weyl scalar $\psi_4(t)$,  a complex-valued function of time evaluated along that
line of sight.
For any pair of simulations and lines of sight,  we compare $\WeylScalar$ and $\WeylScalar'$ by a complex
overlap
\begin{subequations}
\label{eq:def:Overlap}
\begin{eqnarray}
P(\lambda,\lambda')  &\equiv & 
 \frac{\left( r \WeylScalar |r \WeylScalar' \right) }{|r\WeylScalar| |r\WeylScalar'|} \\
(A,B) &\equiv&  \int_{-\infty}^{\infty} 2 \frac{df}{(2\pi f)^4 S_h} \tilde{A}(f)^* \tilde{B}(f)
\end{eqnarray}
\end{subequations}
where $S_h$ is a detector strain noise power spectrum.  
 In this and subsequent expressions we used unprimed and primed variables to distinguish between the two waveforms being
compared, involving potentially distinct parameters $\lambda'$ and lines of sight $\hat{n}'$.
For simplicity and to avoid ambiguity, in this paper, we adopt
a semi-analytic model for the initial LIGO 
sensitivity \cite{gw-astro-mergers-NRParameterEstimation-Nonspinning-Ajith}.
As with the single-detector overlap, the complex overlap can be  maximized over the event time and \emph{polarization}  ($t_c,\psi_c)$ and  by a simple Fourier transform:
\begin{eqnarray}
\label{eq:def:Overlap:Max}
P_{max}&\equiv& \text{max}_{t_c,\psi_c} |P| \\
& =&\frac{1}{|\WeylScalar| |\WeylScalar'|} 
\left| \int_{-\infty}^{\infty} 2 \frac{df}{(2\pi f)^4 S_h} \FourierWeylScalar(f)^* \FourierWeylScalar(f) e^{i(2\pi f t_c + \psi_c)} \right|  \nonumber
\end{eqnarray}

The overlap $P_{max}$ is unity for identical simulations and lines of sight.   How different must $P_{max}$ be from 1 to
be significant?    Roughly speaking, mismatch leads to detectable effects when  $1-P>1/\rho^2$
\cite{gwastro-mergers-HeeSuk-FisherMatrixWithAmplitudeCorrections}, for $\rho$ the signal amplitude.    
Given the expected signal amplitudes for the first few gravitational wave events, a  nonprecessing analog is for
practical purposes indistinguishable from a corotating waveform with  $P_{max}>99\%$.

Unlike other authors, in this paper we only employ waveforms extracted from the fully simulated spacetime, without
making any attempt to hybridize that signal onto a post-Newtonian precursor.    
At the same time, we adopt a data-analysis-driven comparison, driven by the unique bandpass of a plausible detector.  
This comparison requires us to adopt \emph{physical} timescales ($\propto M$) and frequency scales ($\propto 1/M$). 
Because the simulation has finite duration and dynamics with a finite frequency range, our simulations are physically
relevant only for a specific range of masses.    The limit at the low mass end is set by the simulations' initial
orbital frequency, which at very low masses ($\simeq 100 M_\odot$) can lie in a detector's sensitive band.  At the high
mass end, the exponential decay of post-merger oscillations implies that our comparisons are contaminated by numerical
noise at above $\simeq \ReferenceMassHigh{}M_\odot$.  

To determine these two limits unambiguously, we perform the following quantitative tests.  To set the upper mass limit,
we use a one-parameter family of simulations $k$  similar to Sq(4, 0.6, 270, 9), performed at different resolutions.  We
require the overlap $\qmstateproduct{\WeylScalar_{2,2,k}}{\WeylScalar_{2,2,k'}}/|\WeylScalar_{2,2,k}||\WeylScalar_k'$ be
greater than 0.97 for \emph{all} pairs of resolutions.  The largest mass for which this bound holds is $\simeq 350
M_\odot$.
To set the lower mass limit, we explored how the overlap between a signal and a truncated copy of itself changed,
depending on how much of the early inspiral was removed; see Appendix \ref{ap:Resolution} and Figure \ref{fig:ResolutionTest:3}.  
To conservatively insure the signal duration had less than a $1\%$ influence on the overlap, we extremely conservatively limited
$M>200  M_\odot$ for our typical short-duration signals ($r_{\rm start}=6.2 M$). %
Advanced detectors can nominally be sensitive to \emph{extremely} high-mass objects $M>500 M_\odot$, if frequencies
below $40\unit{Hz}$ are properly calibrated.  In the text, we have optimistically assumed all frequencies above
$5\unit{Hz}$ will be calibrated, allowing ground-based interferometers to detect and measure the properties of binaries
with total masses up to $\simeq 2500 M_\odot$.
Realistically, however, ultra-low-frequency detector noise and calibration remains a significant challenge, owing to the complicated and historically nongaussian noise in this
regime \cite{Abadie2010223}.   
Given the suspension and other realistic constraints, advanced detectors may be calibrated only for frequencies above
$10\unit{Hz}$, limiting gravitational wave detection to $\lesssim 1500 M_\odot$.

\subsection{How many degrees of freedom are eliminated in going to a corotating frame?}
\label{sec:sub:ResidualFreeParameters}
At each time step, the corotating frame expansion uses the extracted waveform data to reconstruct 3 Euler angles.  Two of
the three Euler angles, specifying the direction of $\hat{V}$, are reconstructed from the instantaneous value of $\avL$
and do not depend on the past history of the binary.\footnote{Since the orientation tensor $\avL$ cannot distinguish
  between $\pm V$,  we additionally require $\hat{V}$ be continuous and start with $\hat{V}\cdot \hat{L}>0$.}  
The third Euler depends weakly on the past history of the binary, through the minimal-rotation condition.  
By contrast, each constant-$l$ subspace has $2\times (2l+1)$ real degrees of freedom in the amplitudes and phases of its
$(2l+1)$ modes.  For example, the $l=2$ subspace alone has $10$ real degrees of freedom, while  the set of modes $l\le 4$ has
$42$.  
By simple parameter counting, a corotating frame expansion cannot eliminate as many degrees of freedom as can exist in
the waveform.

By contrast, nonprecessing simulations exhibit many extremely strong symmetries between different modes.
First, nonprecessing modes are always \emph{chiral}:  
\begin{eqnarray}
\WeylScalarFourier_{l,m}(f)=0 \quad \text{m f}< 0
\end{eqnarray}
Second, nonprecessing binaries have amplitude and phase conjugation symmetry:
$\WeylScalar_{l,m}^*=(-1)^l \WeylScalar_{l,-m}$.    
Third, during the inspiral, all modes evolve in phase, with $\text{arg}\WeylScalar_{l,m} = m \Phi_{orb}$.  
Repeatedly corroborated empirically  \cite{2008PhRvD..78d4046B}, this fact suggests that the inspiralling binary emits as if a rigid body.  
These symmetries significantly reduce the number of degrees of freedom needed to specify a nonprecessing source.   To
use the $l=2$ subspace as an example and omitting the three (constant) Euler angles needed to describe the system, prior
to merger only one phase and $3$ amplitudes are needed to describe the system, while after
merger,  $2$ phases and  three amplitudes are needed.

Our corotating expansion does not impose any of these properties.  
As a concrete example, consider a  fictitious source producing only $(2,2)$ and $(2,-2)$ modes in some constant frame
misaligned with the global reference frame.  This source can produce each mode independently, with arbitrary amplitude
and (chiral but otherwise arbitrary) phase as a function of time. 
Our corotating expansion would identify  the (constant) orientation of that frame and the (arbitrary) functional form of
the two basis sources.  
That said, a phase-conjugate ``corotating'' signal ($\WeylScalar_{l,m}=(-1)^l\WeylScalar_{l,-m}^*=a_{l,m}$) plus an arbitrary time-dependent rotation
has enough parameters to fit \emph{many but not all} sources.   
Counting parameters, a phase-conjugate source would have $l+1$ amplitudes and $l$ phases in each constant-$l$ subspace.
Including an arbitrary rotation lets a phase-conjugate source fit $2l+4$ degrees of freedom.    For the $l=2$
subspace, however, two degrees of freedom remain that this form, though relatively generic, cannot fit: the two
conjugate-\emph{antisymmetric} (``odd'') moments ($\WeylScalar_{l,m}=- (-1)^l\WeylScalar_{l-m}^*=b_{lm}$ for $m=1,2$).
More generally, a phase-conjugate corotating source will never fit the odd moments $b_{l,m}$ in each constant-$l$
subspace.
Additionally, even a phase-conjugate source can conceivably fit \emph{different orientations to each constant-$l$ subspace}.

To conclude, nonprecessing simulations exhibit many strong symmetries.  While we \emph{hope}  any ``nonprecessing
analog'' will  satisfy them, our corotating frame will not enforce them.
First, each constant-$l$ subspace has a consistent preferred direction at each time.
Second, prior to merger, the corotating modes have common phase evolution.  
Finally, nonprecessing systems cannot source current moments and must emit symmetrically when reflected through the
orbital plane.

\subsection{Symmetry expansions in the corotating frame}
\label{sec:sub:OddVsEvenExpand}

To better identify and discuss small features that spin precession introduces into the waveform, we sometimes separate
the \emph{corotating-frame} Weyl scalar $\WeylScalarCorot$ into different parts, reflecting symmetries.  
One way to split $\WeylScalar$  is the usual ($l+1$ derivatives of the) ``mass''
and ``current'' quadrupole moments ${\cal I}_{l,m}$ and ${\cal S}_{l,m}$ respectively, which we define as \cite{Thorne-STF,2008PhRvD..77d4031S} 
\begin{eqnarray}
\WeylScalar_{l,m} = - \frac{1}{\sqrt{2} r M}[ {\cal I}_{l,m}  - i {\cal S}_{l,m}] 
\end{eqnarray}
where ${\cal I}_{l,m}=(-1)^m{\cal I}_{l,-m}^*$ and ${\cal S}_{l,m}=(-1)^m{\cal S}_{l,-m}^*$ have the usual symmetry \cite{Thorne-STF}.  
This operation defines two projections $\WeylScalar =\WeylScalar^M+\WeylScalar^S$ that uniquely separate $\WeylScalar$
into mass and current contributions. 

Unfortunately,  even the leading-order emission from nonspinning binaries produces both mass and
current moments 
\cite{2008PhRvD..77d4031S,gw-astro-mergers-approximations-SpinningPNHigherHarmonics,Thorne-STF}.  The standard mass and
current decomposition does not provide a high-precision tool to  distinguish between the nonprecessing and precession-only contributions.

For our phenomenological purposes, a more productive decomposition uses classical axial and polar parity.  For
notational  convenience, we define this split on a mode-by-mode basis, using conjugation symmetry to define CP-odd
($b_{l,m}=-(-1)^lb_{l,-m}^*$) and CP-even ($a_{l,m}=(-1)^la_{l,-m}^*$) parts of the corotating-frame $\WeylScalarCorot$, as
described in the text [Eq. (\ref{eq:def:AB})].
As a concrete example, we provide leading-order post-Newtonian expressions for $h_{l,m}$ below.

Nonprecessing binaries
are \emph{even} under this transformation, radiating symmetrically above and below their orbital plane.   Equivalently,
on a mode-by-mode basis, each R-handed mode $m>0$ has a corresponding L-handed mode $m<0$ that radiates identically but in
the opposite direction.     
Precessing binaries break this symmetry, even in the corotating frame.    As noted in
\cite{gwastro-mergers-nr-Alignment-ROS-Polarization}, precessing binaries show a slight bias towards  either R- or L-handed
emission at any instant, with the sign of the bias oscillating as the orbit changes the relative orientation of the
spins to the binary separation.     This symmetry-breaking bias generally persists  in the corotating
frame  on  a mode-by-mode basis: often $b_{l,m} \ne 0$.     For this reason, a conjugation-symmetry-based diagnostic provides a powerful tool to identify and quantify the
impact of precessing spins in a corotating frame.  

To illustrate the expected functional form and symmetry properties of different multipole orders, we provide selected
terms from the $a,b$ symmetry decomposition of $h_{l,m}$, as tabulated elsewhere
\cite{gw-astro-mergers-approximations-SpinningPNHigherHarmonics}.   
To highlight their symmetry properties, we convert their notation to explicit Cartesian 3-vector operations.  
Specifically, in place of the orbital phase (including tail terms) $\Psi$, we adopt a coordinate 3-vector $\hat{r}$ for
the radial separation.   To convert between coordinate expressions and vectors, we employ a
reference frame $\hat{x},\hat{y}$ that corotates with the binary\footnote{In the notation of
  \cite{gw-astro-mergers-approximations-SpinningPNHigherHarmonics}, $\alpha$ would evolve with time.}
\begin{eqnarray}
\hat{e}_\pm \equiv \frac{\mp}{\sqrt{2}}(\hat{x}\pm i \hat{y}) \\
e^{i \Psi} \equiv - \sqrt{2} \hat{r}\cdot e_{+}^* 
\end{eqnarray}
Our expressions follow directly  from their expressions, substituting $\iota=0$ and replacing powers of $\exp i \Psi$
with $\sqrt{2}\hat{r}\cdot e_+^*$ as needed.

For example, the leading-order $l=2$ multipoles have the following form, highlighting spin dependence and working to
$v^4$ order:  %
\begin{align}
a_{2,2}^h &= \text{spin-independent to }O(v^5) \\
b_{2,2}^h &=  %
 \frac{M 16 \sqrt{\pi/5}}{d_L} v^4 \eta  {\cal S}_a r_b (e_+ e_+)^{*ab} \\
a_{2,0}^h &=  \frac{4M\sqrt{2\pi/15}}{d_L}  v^4 (-i \eta) (\hat{r}\times {\cal S})\cdot\hat{L} \\
b_{2,0}^h &= 0 \\
a_{2,1}^h &= \hat{r}\cdot e_+^* \frac{M\sqrt{2\pi/5}}{d_L} \eta
 \left[ 
   \frac{8}{3}\delta  v^3  - 4 {\cal S}\cdot \hat{L} v^4 + O(v^5)
  \right] \\
b_{2,1}^h &= O(v^5)
\end{align}
where to simplify comparisons with the literature we provide the symmetry coefficients $a,b$ derived from the complex
$h$  (i.e.,   $\WeylScalar = \partial_t^2 h = \partial_t^2(a^h +b^h)$).   
In these expressions, we employ the same notation as 
\cite{gw-astro-mergers-approximations-SpinningPNHigherHarmonics}:
\begin{eqnarray}
\eta &\equiv& \frac{m_1 m_2}{(m_1+m_2)} \\ 
\delta &\equiv& \frac{m_1 - m_2}{m_1+m_2} \\
\mathbf{\chi}_\pm &=&( {\bf S}_1/m_1^2 \pm {\bf S}_2/m_2^2)/M 
\end{eqnarray}
Additionally, following \cite{gwastro-mergers-nr-Alignment-ROS-PN} we introduce
\begin{eqnarray}
\mathbf{\cal S} &=& \mathbf{\chi}_- + \delta \mathbf{\chi}_+ \\
 &=& \frac{1}{M} \left( \frac{{\bf S}_1}{m_1} - \frac{{\bf S}_2}{m_2} \right) \nonumber
\end{eqnarray}

\section{Simulations}
\label{sec:Simulations}

\subsection{Simulations}

Table 1 in   \cite{gwastro-mergers-nr-Alignment-ROS-Polarization} enumerates many of the  simulations and groups of
simulations shown in this study; we adopt similar notation to characterize each simulation
(T,Tq,S,Sq,$\ldots$).  
 To simplify the process of identifying and distinguishing between similar simulations, we provide
each simulation with a short descriptive string.  While the specific interpretation of the string depends on the
simulation type, many of our simulations are denoted by a string of the form $X(q,a,\theta,d)$ where $q$ is the mass ratio; $a$ is the typical dimensionless spin; $\theta$ is an
angle; and $d$ is the initial binary separation in units of $M$.  
Initial data was evolved with  \texttt{Maya}, which was used in previous \bbh{} studies \cite{2007CQGra..24...33H,Herrmann:2007ex,Herrmann:2007ac,Hinder:2007qu,Healy:2008js,Hinder:2008kv,Healy:2009zm,Healy:2009ir,Bode:2009mt}.
The grid structure for each run consisted of  10 levels of refinement 
provided by \texttt{CARPET} \cite{Schnetter-etal-03b}, a 
mesh refinement package for \texttt{CACTUS} \cite{cactus-web}. 
Each successive level's resolution decreased by a factor of 2.
Sixth-order spatial finite differencing was used with the BSSN equations 
implemented with Kranc \cite{Husa:2004ip}. 
While the initial data, grid size, resolution boundaries, and peak resolution all differ between simulation, each simulation a
member of one of the following classes:
\begin{itemize}
\item S series: In this series, two equal-mass holes are positioned at  $\pm 3.1 M$, starting with $S_1/m_1^2 =
  -a \hat{x}$ and $a_2 = a \hat{n}(\theta,0)$ for several choices of $\theta$ and $a$.  
The outer boundaries are located at 317.4M.  Each simulation was performed at a resolution of $M/77$ on  the highest refinement level.
Notably, one refinement region extends between  $r=20M$ to $r=80M$.  As previously\cite{gwastro-spins-rangefit2010},  the
outermost refinement levels from $r=80M$ to $r=317M$ are too low resolution ($dx >3 M $) to safely reproduce fine,
high-frequency features in the waveform.  We therefore extract on the refinement region bounded by $[20,80]M$  (i.e., at $r=40,50,60,75$).
\item T series: In this series (also denoted the ``A series'' elsewhere), the two equal-mass holes are positioned at
  $\pm 3.1 M$, starting with $S_1/m_1^2 = a\hat{z}$ and $S_2/m_2^2 = a \hat{n}(\theta,0)$ for several choices of $\theta$ and $a$.  
The outer boundaries are at $317M$.  Each simulation was performed at a resolution of $M/77$ on the
  highest refinement level.   Refinement boundaries occur at $r=40,79 M$, with $dx=0.83M$ in that refinement region.  As previously\cite{gwastro-spins-rangefit2010}, we
  therefore extract on $r=40,50,60,75$. %

\item Sq series: Dimensionless spin vectors ${\bf S}_k/m_k^2$ are chosen as in the S series.   Each simulation was performed at a resolution of $M/140$ on the
  highest refinement level.  As above, we extract
  only in one refinement region, here  at $r=40,\ldots 90$. %
However, the grid size depended on the initial starting separation.

For simulations started with initial separation $d=9 M$, the outer boundaries are at $307M$.    Between $r=32M$ and $r=102M$, the refinement levels have $dx=0.9M$.  

For simulations started with $d=6.2 M$, the outer boundaries are at $409M$.   Between $r=25M$ and $102M$, refinement levels
have $dx=0.9 M$.

\item Tq series:  Dimensionless spin vectors ${\bf S}_k/m_k^2$ are chosen as in the T series.    The outer boundaries are at $409.6M$.  Each simulation was performed at a resolution of $M/120$ on the
  highest refinement level.   Refinement boundaries occur at $r=20,102M$, between which $dx=1M$.   We extract in this
  refinement region along   $r=40,\ldots 90$ and extrapolate to infinity in that zone.

When both spins are in the orbital plane ($\theta=\pi/2$ or $3\pi/2$), the initial data adopted for these series can be
comparable to evolutions performed elsewhere.   For example, the  Sq(4,0.6,90) and Tq(4,0.6,270) sequences are
almost identical, modulo a small shift in initial separation.

\item V series: In this series, the two equal-mass holes are positioned with a separation $d=6.2$, starting with $S_1/m_1^2=a
  \hat{n}(\theta,\phi)$ and $S_2/m_2^2 = a \hat{n}(\theta,\phi+\pi)$.  The outer boundaries are at $317M$; the highest
  resolution is $M/77$; and we extract information in the refinement region between  $20M \ldots 80 M$, in which $dx=0.8 M$.

\item  Lq  series:  Similar to the Sq series, except using a wider range
  for the smaller black hole's spin direction.  Each simulation was performed with a resolution $M/140$ on the highest refinement level;
  the simulation extends to $r=409.6 M$; 
  we extract only in one refinement region between $40 M\ldots 102 M$, where grid spacing is $dx=0.9 M$.

\item Eq series:  Similar to the Lq series, except starting at a larger separation ($d=7 M$) and
  both spins are in the $xy$ plane and antiparallel, with $S_1/m_1^2=0.6
  \hat{n}(\pi/2,\phi)$.  Resolutions and extraction radii are as described above.

 \item z and zq (aligned) series: Equal-mass (z) and unequal mass (zq)  aligned-spin binaries, with identical specific angular momentum ${\bf S}/m^2$ on
   each black hole.  The outer boundaries are at $317M$; most simulations have their finest level of refinement
   at $h=M/103$; and
   refinement boundaries occur at $r=39M$ and $r=79 M$, between which $dx=1.24 M$.

\item zU (aligned, asymmetric): Unequal-mass aligned-spin binaries, where the two black holes do not have identical
  specific angular momentum ${\bf S}/m^2$.  
The outer boundaries are at $r=409.6M$; each simulation's finest refinement level has $h=M/160$; and in the interval
$25\ldots 102.4 M$, where we extract information, the grid spacing is $dx=0.8 M$.  

\end{itemize}

Given the significant systematic errors inherent in any comparison between \texttt{IMRPhenomB} and numerical
simulations,  we did not extrapolate the waveform to infinity or in resolution.  Instead, we performed calculations on a
fixed resolution and fixed extraction radius.  We used multiple resolutions and extraction radii to assess our
(relatively small) numerical error.

\subsection{Extrapolation, resolution, and duration tests}
\label{ap:Resolution}

\begin{figure}
\ifpdf{
\includegraphics[width=\columnwidth]{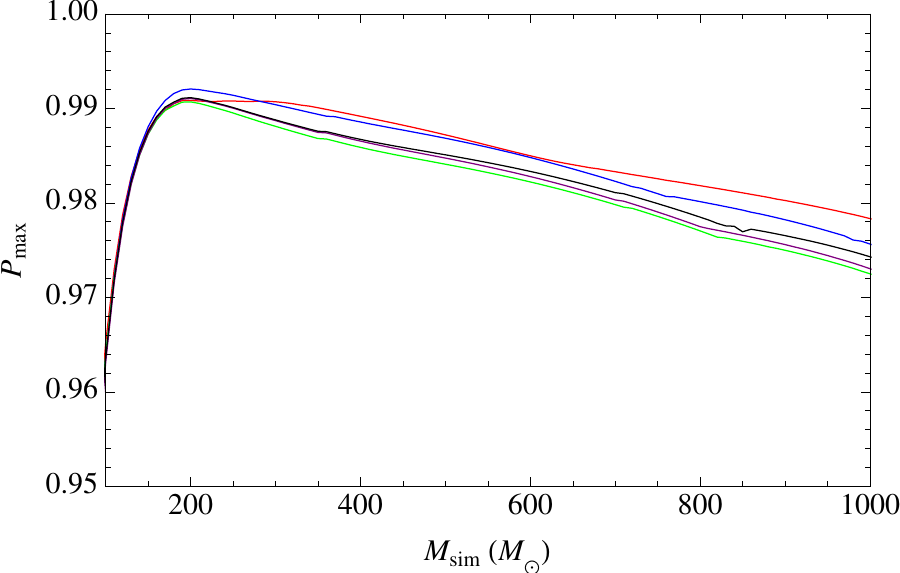}
\includegraphics[width=\columnwidth]{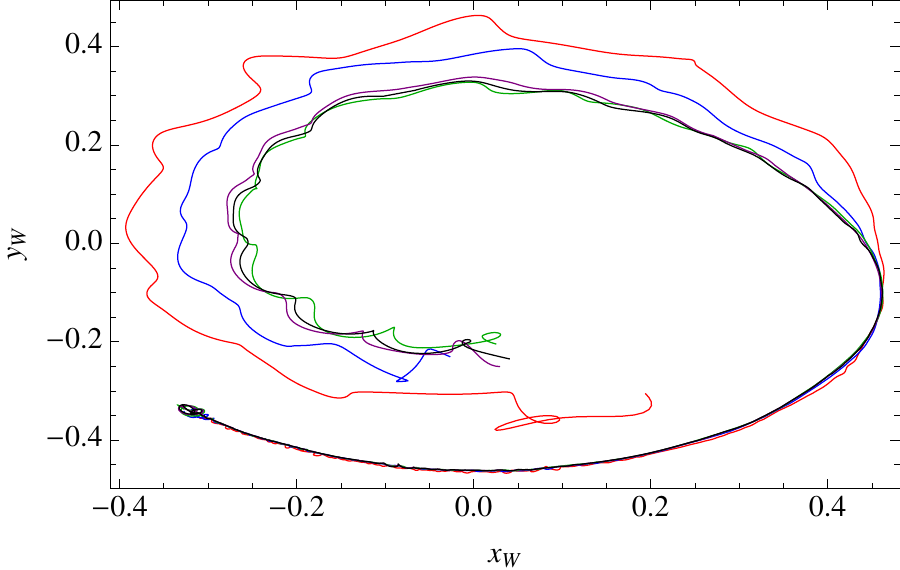}
}\fi
\caption{\label{fig:ResolutionTest:3} \textbf{Comparing resolutions}:
\emph{Top panel}: A plot of $P_{\rm max,corot}$ versus mass for the  Sq(4,0.6,270,9) simulation, performed at resolutions  $h=M/100$
(red),  $M/120$ (blue); $M/140$ (green); $M/160$ (purple); and $M/180$ (black).   Though significant, the differences
between resolutions are nonetheless smaller than the typical differences between simulations; compare to Figure
\ref{fig:ProofFitWorks}.
\emph{Bottom panel}: The path of the preferred direction derived from simulations with different resolution, viewed in
projection in a frame aligned with the precession cone (i.e., with $\hat{W}\simeq \hat{J}$; see \cite{gwastro-mergers-nr-Alignment-ROS-Polarization}).
}
\end{figure}

\begin{figure}
\includegraphics[width=\columnwidth]{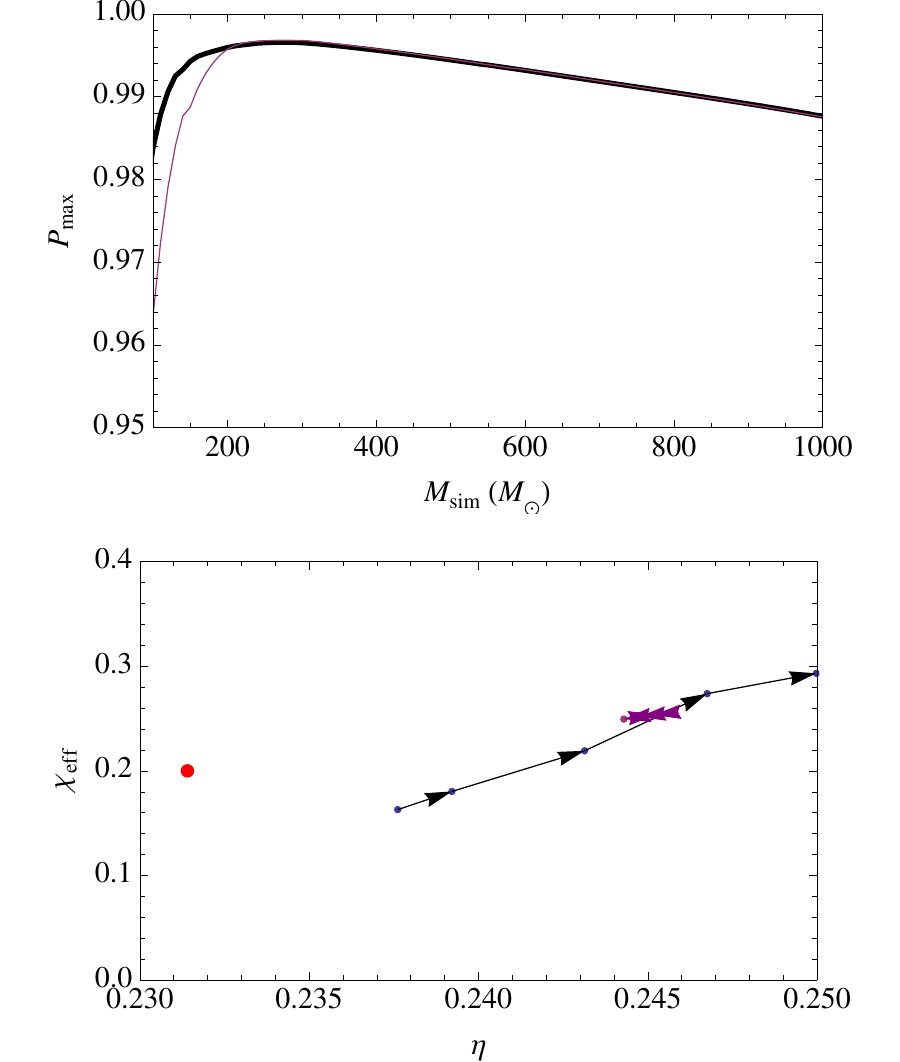}
\caption{\label{fig:DurationTest}\textbf{Signal duration and the low-mass error}: \emph{Top panel}: The best-fitting match $P_{\rm max,sim}$ between
  \texttt{IMRPhenomB} and the $(2,2)$ mode of an unequal-mass aligned-spin binary [Tq(1.75,0.2,0,10)].  Dotted curve shows
  $P_{\rm max}$ derived using the whole signal; solid curve shows $P_{\rm max}$ derived after first truncating the
  signal, to mimic the results of a shorter simulation starting at $d=6.2$.  Comparing with Figure \ref{fig:ProofFitWorks} and
  the text, this illustration shows that signal duration dominates our error at low mass and has little impact on our
  results at high mass.  
\emph{Bottom panel}: At $M=\ReferenceMassLow{} M_\odot$ (blue and black arrows) and $M=250 M_\odot$ (purple arrows), the best-fitting \texttt{IMRPhenomB}
parameters which reproduce  Tq(1.75,0.2,0,10), truncated to different lengths.   For comparison, the red point indicates
the corresponding physical properties expected from the initial data.   Arrows connect longer to shorter
signals, indicating the effect of truncating the signal: a significant error in $\eta$ and a moderate bias in $\chi$.
As illustrated by the discrepancy between the best-fit points and physical parameters,  differences between
\texttt{IMRPhenomB} model and our simulations  also contribute significantly to systematic error in parameter recovery,
particularly in mass ratio.
}
\end{figure}

Using the  $\WeylScalar$ extracted on each of the constant-simulation-radius spheres listed above, we have calculated
 preferred directions and corotating-frame waveforms, both on constant radial slices and using $\WeylScalar$ extrapolated to infinity.  
The  preferred direction  agrees almost exactly between these different options.  
Though the simulation-frame and corotating-frame $\WeylScalar$ do change slightly with extraction radius, they evolve
principally in amplitude and in common.   
To use a specific quantitative measure that is directly relevant to  our principal result,  we both evaluated normalized  overlaps
$\qmstateproduct{\WeylScalar_{2,2}(r)}{\WeylScalar_{2,2}(r')}/|\WeylScalar_{2,2}(r)|\WeylScalar_{2,2}(r')|$ between (2,2)
modes extracted from different radial slices and evaluated $P_{max}$ using different extraction radii.  For the selected
cases examined, the choice of extraction radius had minimal impact ($\Delta P_{\rm max} \lesssim 0.001$).  
We therefore limit attention to a single extraction radius ($r=75 M$).  
For selected nonprecessing and generic initial configurations, we have also performed simulations at successively higher
resolutions.  As concrete example, we have performed the $Sq(4,0.6,90,9)$ simulation at resolutions
$M/100,M/120,M/140,M/160,M/180$; see Figure \ref{fig:ResolutionTest:3}.   Each resolution produces a slightly different duration, with the dominant source of
error being a slowly-accumulating phase shift, principally accumulating prior to merger.  
After recentering our simulations, we find convergence consistent with our differencing order, for the waveform epochs most significant for our results.
Our default
resolutions (e.g., $M/140$) provide more than enough precision for our most striking results: the persistent
precession of $\hat{V}$ long after merger.    The lower resolutions used on some shorter simulations are also more than
adequate to identify principal features during merger.

Many of our simulations commence shortly prior to merger, from separations as short as $d\simeq 6.2M$.  These short
(few hundred $M$) simulations cannot produce a gravitational wave signal from an inspiral which they do not simulate.
To quantify the systematic errors introduced by our choice of signal duration, we have artificially truncated a long,
high-resolution nonprecessing signal to a duration comparable to a typical short signal length from $d\simeq 6.2 M$.  As
illustrated in the top panel of Figure \ref{fig:DurationTest}, this truncated signal matches \texttt{IMRPhenomB}
significantly less well than our simulation  below $M \simeq 500 M_\odot$ but agrees at higher masses.    Since signal
duration depends sensitively on spin orientation, we believe that most disagreement shown in Figure \ref{fig:ProofFitWorks} between our simulations'
corotating-frame signals and
\texttt{IMRPhenomB} is dominated by our limited signal duration.  In other words, we anticipate the \texttt{IMRPhenomB}
model is an adequate representation of any observationally-accessible interval of the corotating-frame signal.  
The signal duration of our shortest simulations also imposed a severe limit on our ability to reconstruct the pre-merger
binary's properties using the best-fitting low-mass parameters.  As a concrete example, the bottom panel of Figure \ref{fig:DurationTest}
compares the physical simulation properties (red point) with the parameters derived by fitting
\texttt{IMRPhenomB} to a truncated copy of a long, aligned-spin simulation, using $M=\ReferenceMassLow{} M_\odot$.   
Due to systematic differences between \texttt{IMRPhenomB} and our signal, the best-fitting
and simulation parameters are significantly offset.   
The recovered mass ratio $\eta$ at low mass depends sensitively on the simulation duration --  for the case illustrated here,
the best-fit mass ratio changed from $q\simeq 1.5$ to $q\simeq 1$.  
By contrast, not only was the recovered spin  nearly independent of the simulation duration, but also it could be
reliably estimated using a much later phase of the signal (i.e., with a much higher mass); see, for example, the purple
arrows in Figure \ref{fig:DurationTest} and the best-fit spin versus mass shown in Figure \ref{fig:BestFitCurves:Chi}.  
For this reason, in the text we use the recovered spin at $M=\ReferenceMassMiddle{} M_\odot$  to illustrate reasonable agreement between the
\texttt{IMRPhenomB} model and typical simulations.   Using this choice, we could include many more simulations in Figure
\ref{fig:ParameterRecovery:LowMass:Spins}.

\begin{acknowledgements}
The authors have benefited from conversations with Andrew Lundgren, Pablo Laguna, William East, Frans Pretorius, and the attendees of the KITP
``Chirps, bursts, and mergers'' conference.  
DS is supported by NSF  awards PHY-0955825, PHY-1212433 and TG-PHY060013N.
ROS is supported by NSF award PHY-0970074 and the UWM Research
Growth Initiative.
\end{acknowledgements}

\bibliography{nr,paperexport}

\begin{widetext}
\begin{center}
\begin{longtable}{l|rll|llllll|r|lll}
\caption{\label{tab:Simulations}\noindent \textbf{Simulations}: 
The first column is a key, encoding
  the family, mass ratio, black hole spin magnitude(s) $|S_1|/M_1^2$ and $|S_2|/M_2^2$ and alignment.   The next  column
  is the simulation resolution $h/M$.  In a handful of cases used as resolution tests in the text, multiple resolutions of the same initial data appear in
  the table.  The next 8 columns
  provide specific initial conditions: the initial separation ($r_{start}$), mass ratio $q=m_1/m_2$, and two component spins
  $S_k^2/M^2$ relative to the total initial mass.  
The column labelled $T_{\rm wave}$ provides an estimate of the duration of the well-resolved $\WeylScalar_{22}$ mode.  
The column labelled $\chi_{\rm PB}$ evaluates $\chi_{\rm PB}$ [Eq. (\ref{eq:def:AlignedSpinOptions})] using 
 the specific  mass ratio and spins provided in previous columns.  
  Finally, the last two columns provide the final black hole mass and angular momentum, derived from the late-time horizon.
}
\\  
\hline
Key & $h^{-1}$ & $r_{start}$ & $q$ & $S_{1,x}$ & $S_{1,y}$ & $S_{1,z}$ & $S_{2,x}$ & $S_{2,y}$ & $S_{2,z}$ & $T_{wave}$ & $\chi_{\rm PB,sim}$ & $M_{\rm f}$ & $J_{\rm f}$\\ 
      & $M^{-1}$ & $M$     &       & $M^{2}$   & $M^{2}$ & $M^{2}$ & $M^{2}$   &   $M^{2}$ & $M^{2}$ & $M$       & &$M$& $M_f^2$ \\
\hline
\endfirsthead
\hline
\endfoot
\hline\hline
\endlastfoot
 {Eq(2.5, 0.6, 0, 7)} & 140 & 7 & 2.5 & 0.3061 & 0 & 0 & -0.049 & 0 & 0 & 310. & 0 & 0.965 & 0.632 \\
 {Eq(2.5, 0.6, 30, 7)} & 140 & 7 & 2.5 & 0.2651 & 0.153 & 0 & -0.0424 & -0.0245 & 0 & 310. & 0 & 0.965 & 0.632 \\
 {Eq(2.5, 0.6, 60, 7)} & 140 & 7 & 2.5 & 0.153 & 0.2651 & 0 & -0.0245 & -0.0424 & 0 & 310. & 0 & 0.966 & 0.636 \\
 {Eq(2.5, 0.6, 90, 7)} & 140 & 7 & 2.5 & 0 & 0.3061 & 0 & 0 & -0.049 & 0 & 310. & 0 & 0.966 & 0.637 \\
 {Eq(2.5, 0.6, 120, 7)} & 140 & 7 & 2.5 & -0.153 & 0.2651 & 0 & 0.0245 & -0.0424 & 0 & 320. & 0 & 0.966 & 0.636 \\
 {Eq(2.5, 0.6, 150, 7)} & 140 & 7 & 2.5 & -0.2651 & 0.153 & 0 & 0.0424 & -0.0245 & 0 & 320. & 0 & 0.965 & 0.634 \\
 {Eq(2.5, 0.6, 180, 7)} & 140 & 7 & 2.5 & -0.3061 & 0 & 0 & 0.049 & 0 & 0 & 320. & 0 & 0.965 & 0.632 \\
 {Eq(2.5, 0.6, 210, 7)} & 140 & 7 & 2.5 & -0.2651 & -0.153 & 0 & 0.0424 & 0.0245 & 0 & 320. & 0 & 0.965 & 0.632 \\
 {Eq(2.5, 0.6, 240, 7)} & 140 & 7 & 2.5 & -0.153 & -0.2651 & 0 & 0.0245 & 0.0424 & 0 & 310. & 0 & 0.966 & 0.636 \\
 {Eq(2.5, 0.6, 270, 7)} & 140 & 7 & 2.5 & 0 & -0.3061 & 0 & 0 & 0.049 & 0 & 310. & 0 & 0.966 & 0.637 \\
 {Eq(2.5, 0.6, 300, 7)} & 140 & 7 & 2.5 & 0.153 & -0.2651 & 0 & -0.0245 & 0.0424 & 0 & 310. & 0 & 0.966 & 0.636 \\
 {Eq(2.5, 0.6, 330, 7)} & 140 & 7 & 2.5 & 0.2651 & -0.153 & 0 & -0.0424 & 0.0245 & 0 & 310. & 0 & 0.965 & 0.634 \\
 {Eq(3., 0.6, 0, 7)} & 140 & 7 & 3. & 0.3375 & 0 & 0 & -0.0375 & 0 & 0 & 320. & 0 & 0.97 & 0.619 \\
 {Eq(3., 0.6, 30, 7)} & 140 & 7 & 3. & 0.2923 & 0.1687 & 0 & -0.0325 & -0.0187 & 0 & 320. & 0 & 0.969 & 0.616 \\
 {Eq(3., 0.6, 60, 7)} & 140 & 7 & 3. & 0.1687 & 0.2923 & 0 & -0.0187 & -0.0325 & 0 & 310. & 0 & 0.969 & 0.616 \\
 {Eq(3., 0.6, 90, 7)} & 140 & 7 & 3. & 0 & 0.3375 & 0 & 0 & -0.0375 & 0 & 310. & 0 & 0.97 & 0.619 \\
 {Eq(3., 0.6, 120, 7)} & 140 & 7 & 3. & -0.1687 & 0.2923 & 0 & 0.0187 & -0.0325 & 0 & 320. & 0 & 0.97 & 0.62 \\
 {Eq(3., 0.6, 150, 7)} & 140 & 7 & 3. & -0.2923 & 0.1687 & 0 & 0.0325 & -0.0187 & 0 & 320. & 0 & 0.97 & 0.62 \\
 {Eq(3., 0.6, 180, 7)} & 140 & 7 & 3. & -0.3375 & 0 & 0 & 0.0375 & 0 & 0 & 330. & 0 & 0.97 & 0.619 \\
 {Eq(3., 0.6, 210, 7)} & 140 & 7 & 3. & -0.2923 & -0.1687 & 0 & 0.0325 & 0.0187 & 0 & 330. & 0 & 0.969 & 0.616 \\
 {Eq(3., 0.6, 240, 7)} & 140 & 7 & 3. & -0.1687 & -0.2923 & 0 & 0.0187 & 0.0325 & 0 & 320. & 0 & 0.969 & 0.616 \\
 {Eq(3., 0.6, 270, 7)} & 140 & 7 & 3. & 0 & -0.3375 & 0 & 0 & 0.0375 & 0 & 320. & 0 & 0.97 & 0.619 \\
 {Eq(3., 0.6, 300, 7)} & 140 & 7 & 3. & 0.1687 & -0.2923 & 0 & -0.0187 & 0.0325 & 0 & 320. & 0 & 0.97 & 0.62 \\
 {Eq(3., 0.6, 330, 7)} & 140 & 7 & 3. & 0.2923 & -0.1687 & 0 & -0.0325 & 0.0187 & 0 & 320. & 0 & 0.97 & 0.62 \\
 {Lq(2.5, 0.6, 0, 6.2)} & 140 & 6.2 & 2.5 & 0 & 0 & 0.3061 & 0 & 0 & 0.049 & 410. & 0.6 & 0.947 & 0.825 \\
 {Lq(2.5, 0.6, 15, 6.2)} & 140 & 6.2 & 2.5 & 0.0792 & 0 & 0.2957 & -0.0127 & 0 & 0.0473 & 390. & 0.58 & 0.948 & 0.82 \\
 {Lq(2.5, 0.6, 30, 6.2)} & 140 & 6.2 & 2.5 & 0.153 & 0 & 0.2651 & -0.0245 & 0 & 0.0424 & 370. & 0.52 & 0.951 & 0.804 \\
 {Lq(2.5, 0.6, 45, 6.2)} & 140 & 6.2 & 2.5 & 0.2164 & 0 & 0.2164 & -0.0346 & 0 & 0.0346 & 350. & 0.42 & 0.953 & 0.775 \\
 {Lq(2.5, 0.6, 60, 6.2)} & 140 & 6.2 & 2.5 & 0.2651 & 0 & 0.153 & -0.0424 & 0 & 0.0245 & 310. & 0.3 & 0.957 & 0.737 \\
 {Lq(2.5, 0.6, 75, 6.2)} & 140 & 6.2 & 2.5 & 0.2957 & 0 & 0.0792 & -0.0473 & 0 & 0.0127 & 280. & 0.16 & 0.962 & 0.693 \\
 {Lq(2.5, 0.6, 90, 6.2)} & 140 & 6.2 & 2.5 & 0.3061 & 0 & 0 & -0.0489 & 0 & 0 & 250. & 0 & 0.965 & 0.631 \\
 {Lq(2.5, 0.6, 120, 6.2)} & 140 & 6.2 & 2.5 & 0.2651 & 0 & -0.153 & -0.0424 & 0 & -0.0245 & 190. & -0.3 & 0.971 & 0.499 \\
 {Lq(2.5, 0.6, 150, 6.2)} & 140 & 6.2 & 2.5 & 0.153 & 0 & -0.2651 & -0.0245 & 0 & -0.0424 & 170. & -0.52 & 0.975 & 0.373 \\
 {Lq(2.5, 0.6, 210, 6.2)} & 140 & 6.2 & 2.5 & -0.153 & 0 & -0.2651 & 0.0245 & 0 & -0.0424 & 170. & -0.52 & 0.975 & 0.373 \\
 {Lq(2.5, 0.6, 240, 6.2)} & 140 & 6.2 & 2.5 & -0.2651 & 0 & -0.153 & 0.0424 & 0 & -0.0245 & 200. & -0.3 & 0.971 & 0.499 \\
 {Lq(2.5, 0.6, 270, 6.2)} & 140 & 6.2 & 2.5 & -0.3061 & 0 & 0 & 0.0489 & 0 & 0 & 240. & 0 & 0.965 & 0.631 \\
 {Lq(2.5, 0.6, 300, 6.2)} & 140 & 6.2 & 2.5 & -0.2651 & 0 & 0.153 & 0.0424 & 0 & 0.0245 & 320. & 0.3 & 0.957 & 0.737 \\
 {Lq(2.5, 0.6, 315, 6.2)} & 140 & 6.2 & 2.5 & -0.2164 & 0 & 0.2164 & 0.0346 & 0 & 0.0346 & 350. & 0.42 & 0.954 & 0.775 \\
 {Lq(2.5, 0.6, 330, 6.2)} & 140 & 6.2 & 2.5 & -0.153 & 0 & 0.2651 & 0.0245 & 0 & 0.0424 & 370. & 0.52 & 0.951 & 0.804 \\
 {Lq(2.5, 0.6, 345, 6.2)} & 140 & 6.2 & 2.5 & -0.0792 & 0 & 0.2957 & 0.0127 & 0 & 0.0473 & 400. & 0.58 & 0.948 & 0.82 \\
 {Lq(3., 0.6, 30, 6.2)} & 140 & 6.2 & 3. & 0.1687 & 0 & 0.2923 & -0.0187 & 0 & 0.0325 & 390. & 0.52 & 0.957 & 0.791 \\
 {Lq(3., 0.6, 45, 6.2)} & 140 & 6.2 & 3. & 0.2386 & 0 & 0.2386 & -0.0265 & 0 & 0.0265 & 360. & 0.42 & 0.959 & 0.763 \\
 {Lq(3., 0.6, 60, 6.2)} & 140 & 6.2 & 3. & 0.2923 & 0 & 0.1687 & -0.0325 & 0 & 0.0187 & 330. & 0.3 & 0.963 & 0.727 \\
 {Lq(3., 0.6, 210, 6.2)} & 140 & 6.2 & 3. & -0.1687 & 0 & -0.2922 & 0.0187 & 0 & -0.0324 & 160. & -0.52 & 0.978 & 0.324 \\
 {Lq(3., 0.6, 240, 6.2)} & 140 & 6.2 & 3. & -0.2923 & 0 & -0.1687 & 0.0325 & 0 & -0.0187 & 200. & -0.3 & 0.975 & 0.472 \\
 {Lq(3., 0.6, 270, 6.2)} & 140 & 6.2 & 3. & -0.3375 & 0 & 0 & 0.0375 & 0 & 0 & 240. & 0 & 0.969 & 0.615 \\
 {S(1, 0.2, 0)} & 77 & 6.2 & 1 & 0 & 0 & 0.05 & -0.05 & 0 & 0 & 250. & 0.10 & 0.95 & 0.802 \\
 {S(1, 0.2, 45)} & 77 & 6.2 & 1 & 0.0354 & 0 & 0.0354 & -0.05 & 0 & 0 & 250. & 0.071 & 0.951 & 0.788 \\
 {S(1, 0.2, 90)} & 77 & 6.2 & 1 & 0.05 & 0 & 0 & -0.05 & 0 & 0 & 220. & 0 & 0.953 & 0.76 \\
 {S(1, 0.2, 135)} & 77 & 6.2 & 1 & 0.0354 & 0 & -0.0354 & -0.05 & 0 & 0 & 210. & -0.071 & 0.954 & 0.734 \\
 {S(1, 0.2, 180)} & 77 & 6.2 & 1 & 0 & 0 & -0.05 & -0.05 & 0 & 0 & 210. & -0.10 & 0.955 & 0.725 \\
 {S(1, 0.2, 225)} & 77 & 6.2 & 1 & -0.0354 & 0 & -0.0354 & -0.05 & 0 & 0 & 210. & -0.071 & 0.954 & 0.743 \\
 {S(1, 0.2, 270)} & 77 & 6.2 & 1 & -0.05 & 0 & 0 & -0.05 & 0 & 0 & 220. & 0 & 0.952 & 0.774 \\
 {S(1, 0.2, 315)} & 77 & 6.2 & 1 & -0.0354 & 0 & 0.0354 & -0.05 & 0 & 0 & 250. & 0.071 & 0.951 & 0.798 \\
 {S(1, 0.4, 0)} & 77 & 6.2 & 1 & 0 & 0 & 0.1 & -0.1 & 0 & 0 & 280. & 0.20 & 0.946 & 0.849 \\
 {S(1, 0.4, 45)} & 77 & 6.2 & 1 & 0.0707 & 0 & 0.0707 & -0.1 & 0 & 0 & 260. & 0.14 & 0.948 & 0.814 \\
 {S(1, 0.4, 90)} & 77 & 6.2 & 1 & 0.1 & 0 & 0 & -0.1 & 0 & 0 & 230. & 0 & 0.953 & 0.76 \\
 {S(1, 0.4, 135)} & 77 & 6.2 & 1 & 0.0707 & 0 & -0.0707 & -0.1 & 0 & 0 & 210. & -0.14 & 0.956 & 0.706 \\
 {S(1, 0.4, 180)} & 77 & 6.2 & 1 & 0 & 0 & -0.1 & -0.1 & 0 & 0 & 210. & -0.20 & 0.957 & 0.697 \\
 {S(1, 0.4, 225)} & 77 & 6.2 & 1 & -0.0707 & 0 & -0.0707 & -0.1 & 0 & 0 & 210. & -0.14 & 0.955 & 0.744 \\
 {S(1, 0.4, 270)} & 77 & 6.2 & 1 & -0.1 & 0 & 0 & -0.1 & 0 & 0 & 240. & 0 & 0.952 & 0.813 \\
 {S(1, 0.4, 315)} & 77 & 6.2 & 1 & -0.0707 & 0 & 0.0707 & -0.1 & 0 & 0 & 270. & 0.14 & 0.948 & 0.853 \\
 {S(1, 0.6, 0)} & 77 & 6.2 & 1 & 0 & 0 & 0.15 & -0.15 & 0 & 0 & 290. & 0.30 & 0.942 & 0.904 \\
 {S(1, 0.6, 15)} & 77 & 6.2 & 1 & 0.0388 & 0 & 0.1449 & -0.15 & 0 & 0 & 290. & 0.29 & 0.943 & 0.887 \\
 {S(1, 0.6, 30)} & 77 & 6.2 & 1 & 0.075 & 0 & 0.1299 & -0.15 & 0 & 0 & 280. & 0.26 & 0.944 & 0.866 \\
 {S(1, 0.6, 45)} & 77 & 6.2 & 1 & 0.1061 & 0 & 0.1061 & -0.15 & 0 & 0 & 270. & 0.21 & 0.945 & 0.841 \\
 {S(1, 0.6, 60)} & 77 & 6.2 & 1 & 0.1299 & 0 & 0.075 & -0.15 & 0 & 0 & 260. & 0.15 & 0.946 & 0.814 \\
 {S(1, 0.6, 75)} & 77 & 6.2 & 1 & 0.1449 & 0 & 0.0388 & -0.15 & 0 & 0 & 250. & 0.078 & 0.949 & 0.788 \\
 {S(1, 0.6, 90)} & 77 & 6.2 & 1 & 0.15 & 0 & 0 & -0.15 & 0 & 0 & 230. & 0 & 0.953 & 0.759 \\
 {S(1, 0.6, 105)} & 77 & 6.2 & 1 & 0.1449 & 0 & -0.0388 & -0.15 & 0 & 0 & 200. & -0.078 & 0.954 & 0.728 \\
 {S(1, 0.6, 120)} & 77 & 6.2 & 1 & 0.1299 & 0 & -0.075 & -0.15 & 0 & 0 & 210. & -0.15 & 0.955 & 0.699 \\
 {S(1, 0.6, 135)} & 77 & 6.2 & 1 & 0.1061 & 0 & -0.1061 & -0.15 & 0 & 0 & 210. & -0.21 & 0.957 & 0.679 \\
 {S(1, 0.6, 150)} & 77 & 6.2 & 1 & 0.075 & 0 & -0.1299 & -0.15 & 0 & 0 & 200. & -0.26 & 0.958 & 0.667 \\
 {S(1, 0.6, 165)} & 77 & 6.2 & 1 & 0.0388 & 0 & -0.1449 & -0.15 & 0 & 0 & 200. & -0.29 & 0.959 & 0.665 \\
 {S(1, 0.6, 180)} & 77 & 6.2 & 1 & 0 & 0 & -0.15 & -0.15 & 0 & 0 & 200. & -0.30 & 0.959 & 0.675 \\
 {S(1, 0.6, 195)} & 77 & 6.2 & 1 & -0.0388 & 0 & -0.1449 & -0.15 & 0 & 0 & 200. & -0.29 & 0.959 & 0.696 \\
 {S(1, 0.6, 210)} & 77 & 6.2 & 1 & -0.075 & 0 & -0.1299 & -0.15 & 0 & 0 & 200. & -0.26 & 0.957 & 0.726 \\
 {S(1, 0.6, 225)} & 77 & 6.2 & 1 & -0.1061 & 0 & -0.1061 & -0.15 & 0 & 0 & 210. & -0.21 & 0.955 & 0.761 \\
 {S(1, 0.6, 240)} & 77 & 6.2 & 1 & -0.1299 & 0 & -0.075 & -0.15 & 0 & 0 & 200. & -0.15 & 0.954 & 0.8 \\
 {S(1, 0.6, 255)} & 77 & 6.2 & 1 & -0.1449 & 0 & -0.0388 & -0.15 & 0 & 0 & 230. & -0.078 & 0.953 & 0.839 \\
 {S(1, 0.6, 260)} & 77 & 6.2 & 1 & -0.1477 & 0 & -0.026 & -0.15 & 0 & 0 & 250. & -0.052 & 0.953 & 0.851 \\
 {S(1, 0.6, 265)} & 77 & 6.2 & 1 & -0.1494 & 0 & -0.0131 & -0.15 & 0 & 0 & 250. & -0.026 & 0.952 & 0.862 \\
 {S(1, 0.6, 270)} & 77 & 6.2 & 1 & -0.15 & 0 & 0 & -0.15 & 0 & 0 & 250. & 0 & 0.951 & 0.873 \\
 {S(1, 0.6, 285)} & 77 & 6.2 & 1 & -0.1449 & 0 & 0.0388 & -0.15 & 0 & 0 & 270. & 0.078 & 0.948 & 0.899 \\
 {S(1, 0.6, 300)} & 77 & 6.2 & 1 & -0.1299 & 0 & 0.075 & -0.15 & 0 & 0 & 280. & 0.15 & 0.945 & 0.917 \\
 {S(1, 0.6, 315)} & 77 & 6.2 & 1 & -0.1061 & 0 & 0.1061 & -0.15 & 0 & 0 & 290. & 0.21 & 0.943 & 0.925 \\
 {S(1, 0.6, 330)} & 77 & 6.2 & 1 & -0.075 & 0 & 0.1299 & -0.15 & 0 & 0 & 290. & 0.26 & 0.943 & 0.925 \\
 {S(1, 0.6, 345)} & 77 & 6.2 & 1 & -0.0388 & 0 & 0.1449 & -0.15 & 0 & 0 & 290. & 0.29 & 0.942 & 0.917 \\
 {S(1, 0.8, 0)} & 77 & 6.2 & 1 & 0 & 0 & 0.2 & -0.2 & 0 & 0 & 310. & 0.40 & 0.938 & 0.936 \\
 {S(1, 0.8, 120)} & 77 & 6.2 & 1 & 0.1732 & 0 & -0.1 & -0.2 & 0 & 0 & 210. & -0.20 & 0.955 & 0.676 \\
 {S(1, 0.8, 150)} & 77 & 6.2 & 1 & 0.1 & 0 & -0.1732 & -0.2 & 0 & 0 & 200. & -0.35 & 0.96 & 0.634 \\
 {S(1, 0.8, 180)} & 77 & 6.2 & 1 & 0 & 0 & -0.2 & -0.2 & 0 & 0 & 190. & -0.40 & 0.961 & 0.641 \\
 {S(1, 0.8, 210)} & 77 & 6.2 & 1 & -0.1 & 0 & -0.1732 & -0.2 & 0 & 0 & 200. & -0.35 & 0.958 & 0.698 \\
 {S(1, 0.8, 240)} & 77 & 6.2 & 1 & -0.1732 & 0 & -0.1 & -0.2 & 0 & 0 & 230. & -0.2 & 0.954 & 0.783 \\
 {S(1, 0.8, 255)} & 77 & 6.2 & 1 & -0.1932 & 0 & -0.0518 & -0.2 & 0 & 0 & 260. & -0.10 & 0.954 & 0.83 \\
 {S(1, 0.8, 270)} & 77 & 6.2 & 1 & -0.2 & 0 & 0 & -0.2 & 0 & 0 & 280. & 0 & 0.947 & 0.866 \\
 {S(1, 0.8, 30)} & 77 & 6.2 & 1 & 0.1 & 0 & 0.1732 & -0.2 & 0 & 0 & 290. & 0.35 & 0.94 & 0.898 \\
 {S(1, 0.8, 300)} & 77 & 6.2 & 1 & -0.1732 & 0 & 0.1 & -0.2 & 0 & 0 & 290. & 0.20 & 0.943 & 0.923 \\
 {S(1, 0.8, 330)} & 77 & 6.2 & 1 & -0.1 & 0 & 0.1732 & -0.2 & 0 & 0 & 310. & 0.35 & 0.94 & 0.948 \\
 {S(1, 0.8, 60)} & 77 & 6.2 & 1 & 0.1732 & 0 & 0.1 & -0.2 & 0 & 0 & 270. & 0.20 & 0.944 & 0.83 \\
 {S(1, 0.8, 90)} & 77 & 6.2 & 1 & 0.2 & 0 & 0 & -0.2 & 0 & 0 & 230. & 0. & 0.952 & 0.758 \\
 {S(1, 0.6, 0,8)} & 77 & 8 & 1 & 0 & 0 & 0.15 & -0.15 & 0 & 0 & {} & 0.30 & {} & {} \\
 {S(1, 0.6, 135, 8)} & 77 & 8 & 1 & 0.1061 & 0 & -0.1061 & -0.15 & 0 & 0 & {} & -0.21 & {} & {} \\
 {S(1, 0.6, 180, 8)} & 77 & 8 & 1 & 0 & 0 & -0.15 & -0.15 & 0 & 0 & {} & -0.30 & {} & {} \\
 {S(1, 0.6, 225, 8)} & 77 & 8 & 1 & -0.1061 & 0 & -0.1061 & -0.15 & 0 & 0 & {} & -0.21 & {} & {} \\
 {S(1, 0.6, 240, 8)} & 77 & 8 & 1 & -0.1299 & 0 & -0.075 & -0.15 & 0 & 0 & {} & -0.15 & {} & {} \\
 {S(1, 0.6, 255, 8)} & 77 & 8 & 1 & -0.1449 & 0 & -0.0388 & -0.15 & 0 & 0 & {} & -0.078 & {} & {} \\
 {S(1, 0.6, 270, 8)} & 77 & 8 & 1 & -0.15 & 0 & 0 & -0.15 & 0 & 0 & {} & 0 & {} & {} \\
 {S(1, 0.6, 315, 8)} & 77 & 8 & 1 & -0.1061 & 0 & 0.1061 & -0.15 & 0 & 0 & {} & 0.21 & {} & {} \\
 {S(1, 0.6, 45, 8)} & 77 & 8 & 1 & 0.1061 & 0 & 0.1061 & -0.15 & 0 & 0 & {} & 0.21 & {} & {} \\
 {S(1, 0.6, 90, 8)} & 77 & 8 & 1 & 0.15 & 0 & 0 & -0.15 & 0 & 0 & {} & 0 & {} & {} \\
 {Sq(2, 0.6, 0, 6.2)} & 120 & 6.2 & 2 & 0 & 0 & 0.2666 & -0.0666 & 0 & 0 & 340. & 0.40 & 0.946 & 0.81 \\
 {Sq(2, 0.6, 150, 6.2)} & 120 & 6.2 & 2 & 0.1333 & 0 & -0.2309 & -0.0666 & 0 & 0 & 200. & -0.35 & 0.969 & 0.457 \\
 {Sq(2, 0.6, 180, 6.2)} & 120 & 6.2 & 2 & 0 & 0 & -0.2666 & -0.0666 & 0 & 0 & 190. & -0.40 & 0.97 & 0.429 \\
 {Sq(2, 0.6, 90, 6.2)} & 120 & 6.2 & 2 & 0.2666 & 0 & 0 & -0.0666 & 0 & 0 & 260. & 0 & 0.96 & 0.652 \\
 {Sq(2.5, 0.6, 0, 6.2)} & 120 & 6.2 & 2.5 & 0 & 0 & 0.3061 & -0.0489 & 0 & 0 & 370. & 0.43 & 0.952 & 0.805 \\
 {Sq(2.5, 0.6, 120, 6.2)} & 120 & 6.2 & 2.5 & 0.2651 & 0 & -0.153 & -0.0489 & 0 & 0 & 210. & -0.21 & 0.97 & 0.51 \\
 {Sq(2.5, 0.6, 15, 6.2)} & 120 & 6.2 & 2.5 & 0.0792 & 0 & 0.2957 & -0.0489 & 0 & 0 & 360. & 0.41 & 0.952 & 0.797 \\
 {Sq(2.5, 0.6, 150, 6.2)} & 120 & 6.2 & 2.5 & 0.153 & 0 & -0.2651 & -0.0489 & 0 & 0 & 180. & -0.37 & 0.974 & 0.389 \\
 {Sq(2.5, 0.6, 180, 6.2)} & 120 & 6.2 & 2.5 & 0 & 0 & -0.3061 & -0.0489 & 0 & 0 & 180. & -0.43 & 0.975 & 0.344 \\
 {Sq(2.5, 0.6, 210, 6.2)} & 120 & 6.2 & 2.5 & -0.153 & 0 & -0.2651 & -0.0489 & 0 & 0 & 190. & -0.37 & 0.973 & 0.421 \\
 {Sq(2.5, 0.6, 240, 6.2)} & 120 & 6.2 & 2.5 & -0.2651 & 0 & -0.153 & -0.0489 & 0 & 0 & 230. & -0.21 & 0.97 & 0.552 \\
 {Sq(2.5, 0.6, 270, 6.2)} & 120 & 6.2 & 2.5 & -0.3061 & 0 & 0 & -0.0489 & 0 & 0 & 270. & 0 & 0.964 & 0.668 \\
 {Sq(2.5, 0.6, 30, 6.2)} & 120 & 6.2 & 2.5 & 0.153 & 0 & 0.2651 & -0.0489 & 0 & 0 & 350. & 0.37 & 0.954 & 0.783 \\
 {Sq(2.5, 0.6, 300, 6.2)} & 120 & 6.2 & 2.5 & -0.2651 & 0 & 0.153 & -0.0489 & 0 & 0 & 330. & 0.21 & 0.958 & 0.75 \\
 {Sq(2.5, 0.6, 315, 6.2)} & 120 & 6.2 & 2.5 & -0.2164 & 0 & 0.2164 & -0.0489 & 0 & 0 & 340. & 0.3 & 0.955 & 0.776 \\
 {Sq(2.5, 0.6, 330, 6.2)} & 120 & 6.2 & 2.5 & -0.153 & 0 & 0.2651 & -0.0489 & 0 & 0 & 360. & 0.37 & 0.953 & 0.795 \\
 {Sq(2.5, 0.6, 345, 6.2)} & 120 & 6.2 & 2.5 & -0.0792 & 0 & 0.2957 & -0.0489 & 0 & 0 & 370. & 0.41 & 0.952 & 0.804 \\
 {Sq(2.5, 0.6, 45, 6.2)} & 120 & 6.2 & 2.5 & 0.2164 & 0 & 0.2164 & -0.0489 & 0 & 0 & 330. & 0.3 & 0.957 & 0.759 \\
 {Sq(2.5, 0.6, 60, 6.2)} & 120 & 6.2 & 2.5 & 0.2651 & 0 & 0.153 & -0.0489 & 0 & 0 & 300. & 0.21 & 0.958 & 0.724 \\
 {Sq(2.5, 0.6, 75, 6.2)} & 120 & 6.2 & 2.5 & 0.2957 & 0 & 0.0792 & -0.0489 & 0 & 0 & 280. & 0.11 & 0.963 & 0.686 \\
 {Sq(2.5, 0.6, 90, 6.2)} & 120 & 6.2 & 2.5 & 0.3061 & 0 & 0 & -0.0489 & 0 & 0 & 260. & 0 & 0.965 & 0.631 \\
 {Sq(3, 0.6, 0, 6.2)} & 120 & 6.2 & 3 & 0 & 0 & 0.3375 & -0.0375 & 0 & 0 & 390. & 0.45 & 0.957 & 0.798 \\
 {Sq(3, 0.6, 120, 6.2)} & 120 & 6.2 & 3 & 0.2923 & 0 & -0.1687 & -0.0375 & 0 & 0 & 210. & -0.22 & 0.975 & 0.479 \\
 {Sq(3, 0.6, 15, 6.2)} & 120 & 6.2 & 3 & 0.0873 & 0 & 0.326 & -0.0375 & 0 & 0 & 380. & 0.43 & 0.958 & 0.791 \\
 {Sq(3, 0.6, 30, 6.2)} & 120 & 6.2 & 3 & 0.1687 & 0 & 0.2923 & -0.0375 & 0 & 0 & 370. & 0.39 & 0.959 & 0.774 \\
 {Sq(3, 0.6, 45, 6.2)} & 120 & 6.2 & 3 & 0.2386 & 0 & 0.2386 & -0.0375 & 0 & 0 & 350. & 0.32 & 0.962 & 0.752 \\
 {Sq(3, 0.6, 60, 6.2)} & 120 & 6.2 & 3 & 0.2923 & 0 & 0.1687 & -0.0375 & 0 & 0 & 320. & 0.22 & 0.964 & 0.715 \\
 {Sq(3, 0.6, 75, 6.2)} & 120 & 6.2 & 3 & 0.326 & 0 & 0.0873 & -0.0375 & 0 & 0 & 290. & 0.12 & 0.967 & 0.674 \\
 {Sq(3, 0.6, 90, 6.2)} & 120 & 6.2 & 3 & 0.3375 & 0 & 0 & -0.0375 & 0 & 0 & 260. & 0 & 0.969 & 0.615 \\
 {Sq(3, 0.6, 150, 6.2)} & 120 & 6.2 & 3 & 0.1687 & 0 & -0.2922 & -0.0375 & 0 & 0 & 170. & -0.39 & 0.978 & 0.333 \\
 {Sq(3, 0.6, 180, 6.2)} & 120 & 6.2 & 3 & 0 & 0 & -0.3374 & -0.0375 & 0 & 0 & 180. & -0.45 & 0.978 & 0.269 \\
 {Sq(3, 0.6, 210, 6.2)} & 120 & 6.2 & 3 & -0.1687 & 0 & -0.2922 & -0.0375 & 0 & 0 & 180. & -0.39 & 0.977 & 0.364 \\
 {Sq(3, 0.6, 240, 6.2)} & 120 & 6.2 & 3 & -0.2923 & 0 & -0.1687 & -0.0375 & 0 & 0 & 220. & -0.22 & 0.974 & 0.517 \\
 {Sq(3, 0.6, 270, 6.2)} & 120 & 6.2 & 3 & -0.3375 & 0 & 0 & -0.0375 & 0 & 0 & 270. & 0 & 0.969 & 0.648 \\
 {Sq(3, 0.6, 300, 6.2)} & 120 & 6.2 & 3 & -0.2923 & 0 & 0.1687 & -0.0375 & 0 & 0 & 340. & 0.22 & 0.963 & 0.736 \\
 {Sq(3, 0.6, 315, 6.2)} & 120 & 6.2 & 3 & -0.2386 & 0 & 0.2386 & -0.0375 & 0 & 0 & 360. & 0.32 & 0.96 & 0.765 \\
 {Sq(3, 0.6, 330, 6.2)} & 120 & 6.2 & 3 & -0.1687 & 0 & 0.2923 & -0.0375 & 0 & 0 & 380. & 0.39 & 0.959 & 0.786 \\
 {Sq(3, 0.6, 345, 6.2)} & 120 & 6.2 & 3 & -0.0873 & 0 & 0.326 & -0.0375 & 0 & 0 & 390. & 0.43 & 0.957 & 0.796 \\
 {Sq(4, 0.6, 0, 6.2)} & 120 & 6.2 & 4 & 0 & 0 & 0.384 & -0.024 & 0 & 0 & 430. & 0.48 & 0.966 & 0.781 \\
 {Sq(4, 0.6, 0, 9, 6.2)} & 120 & 9 & 4 & 0 & 0 & 0.384 & -0.024 & 0 & 0 & 1300. & 0.48 & 0.965 & 0.779 \\
 {Sq(4, 0.6, 30, 6.2)} & 120 & 6.2 & 4 & 0.192 & 0 & 0.3325 & -0.024 & 0 & 0 & 430. & 0.42 & 0.967 & 0.759 \\
 {Sq(4, 0.6, 90, 6.2)} & 120 & 6.2 & 4 & 0.384 & 0 & 0 & -0.024 & 0 & 0 & 260. & 0 & 0.977 & 0.594 \\
 {Sq(4, 0.6, 90, 9)} & 120 & 9& 4 & 0.384 & 0 & 0 & -0.024 & 0 & 0 & 870. & 0 & 0.976 & 0.593 \\
 {Sq(4, 0.6, 150, 6.2)} & 120 & 6.2 & 4 & 0.192 & 0 & -0.3325 & -0.024 & 0 & 0 & 170. & -0.42 & 0.983 & 0.255 \\
 {Sq(4, 0.6, 150, 9)} & 120 & 9 & 4 & 0.192 & 0 & -0.3325 & -0.024 & 0 & 0 & 460. & -0.42 & 0.983 & 0.258 \\
 {Sq(4, 0.6, 180, 6.2)} & 120 & 6.2 & 4 & 0 & 0 & -0.384 & -0.024 & 0 & 0 & 170. & -0.48 & 0.983 & 0.146 \\
 {Sq(4, 0.6, 180, 9)} & 120 & 9 & 4 & 0 & 0 & -0.384 & -0.024 & 0 & 0 & 430. & -0.48 & 0.983 & 0.15 \\
 {Sq(4, 0.6, 210, 9)} & 120 & 9 & 4 & -0.192 & 0 & -0.3325 & -0.024 & 0 & 0 & 490. & -0.42 & 0.982 & 0.285 \\
 {Sq(4, 0.6, 270, 6.2)} & 120 & 6.2 & 4 & -0.384 & 0 & 0 & -0.024 & 0 & 0 & 280. & 0 & 0.976 & 0.616 \\
 {Sq(4, 0.6, 270, 9)} & 120 & 9 & 4 & -0.384 & 0 & 0 & -0.024 & 0 & 0 & 890. & 0 & 0.975 & 0.612 \\
 {Sq(4, 0.6, 270, 9)} & 140 & 9 & 4 & -0.384 & 0 & 0 & -0.024 & 0 & 0 & 890. & 0 & 0.975 & 0.612 \\
 {Sq(4, 0.6, 270, 9)} & 160 & 9 & 4 & -0.384 & 0 & 0 & -0.024 & 0 & 0 & 890. & 0 & 0.975 & 0.612 \\
 {Sq(4, 0.6, 270, 9)} & 180 & 9 & 4 & -0.384 & 0 & 0 & -0.024 & 0 & 0 & 890. & 0 & 0.975 & 0.613 \\
 {T(1, 0., 0)} & 77 & 6.2 & 1 & 0 & 0 & 0 & 0 & 0 & 0 & 890. & 0 & 0.951 & 0.685 \\
 {T(1, 0.2, 0)} & 77 & 6.2 & 1 & 0 & 0 & 0.05 & 0 & 0 & 0.05 & 910. & 0.2 & {} & {} \\
 {T(1, 0.2, 45)} & 77 & 6.2 & 1 & 0 & 0 & 0.05 & 0.0354 & 0 & 0.0354 & 850. & 0.17 & {} & {} \\
 {T(1, 0.2, 60)} & 77 & 6.2 & 1 & 0 & 0 & 0.05 & 0.0433 & 0 & 0.025 & 840. & 0.15 & {} & {} \\
 {T(1, 0.2, 90)} & 77 & 6.2 & 1 & 0 & 0 & 0.05 & 0.05 & 0 & 0 & 840. & 0.1 & {} & {} \\
 {T(1, 0.4, 0)} & 90 & 6.2 & 1 & 0 & 0 & 0.1 & 0 & 0 & 0.1 & 1000. & 0.4 & {} & {} \\
 {T(1, 0.4, 45)} & 77 & 6.2 & 1 & 0 & 0 & 0.1 & 0.0707 & 0 & 0.0707 & 990. & 0.34 & {} & {} \\
 {T(1, 0.4, 60)} & 77 & 6.2 & 1 & 0 & 0 & 0.1 & 0.0866 & 0 & 0.05 & 970. & 0.3 & {} & {} \\
 {T(1, 0.4, 90)} & 77 & 6.2 & 1 & 0 & 0 & 0.1 & 0.1 & 0 & 0 & 910. & 0.2 & {} & {} \\
 {T(1, 0.6, 0)} & 77 & 6.2 & 1 & 0 & 0 & 0.15 & 0 & 0 & 0.15 & 1100. & 0.6 & {} & {} \\
 {T(1, 0.6, 45)} & 77 & 6.2 & 1 & 0 & 0 & 0.15 & 0.1061 & 0 & 0.1061 & 1100. & 0.51 & {} & {} \\
 {T(1, 0.6, 60)} & 77 & 6.2 & 1 & 0 & 0 & 0.15 & 0.1299 & 0 & 0.075 & 990. & 0.45 & {} & {} \\
 {T(1, 0.6, 90)} & 77 & 6.2 & 1 & 0 & 0 & 0.15 & 0.15 & 0 & 0 & 920. & 0.3 & {} & {} \\
 {T(1, 0.8, 0)} & 90 & 6.2 & 1 & 0 & 0 & 0.2 & 0 & 0 & 0.2 & 1200. & 0.8 & 0.906 & 0.9 \\
 {Tq(1.5, 0.4, 60, 10)} & 120 & 10 & 1.5 & 0.1247 & 0 & 0.072 & 0 & 0 & 0.064 & 1200. & 0.28 & 0.947 & 0.753 \\
 {Tq(1.5, 0.6, 45, 10)} & 120 & 10 & 1.5 & 0.1527 & 0 & 0.1527 & 0 & 0 & 0.096 & 1300. & 0.49 & 0.937 & 0.822 \\
 {Tq(2, 0.4, 60, 10)} & 120 & 10 & 2 & 0.1539 & 0 & 0.0889 & 0 & 0 & 0.0444 & 1200. & 0.27 & 0.954 & 0.722 \\
 {Tq(2, 0.6, 135, 10)} & 120 & 10 & 2 & 0.1885 & 0 & -0.1885 & 0 & 0 & 0.0666 & 930. & -0.083 & 0.964 & 0.549 \\
 {Tq(2, 0.6, 180, 10)} & 120 & 10 & 2 & 0 & 0 & -0.2666 & 0 & 0 & 0.0666 & 810. & -0.20 & 0.967 & 0.465 \\
 {Tq(2, 0.6, 270, 10)} & 120 & 10 & 2 & -0.2666 & 0 & 0 & 0 & 0 & 0.0666 & 1200. & 0.20 & 0.955 & 0.698 \\
 {Tq(2, 0.6, 45, 10)} & 120 & 10 & 2 & 0.1885 & 0 & 0.1885 & 0 & 0 & 0.0666 & 1400. & 0.48 & 0.945 & 0.804 \\
 {Tq(2, 0.6, 60, 10)} & 120 & 10 & 2 & 0.2309 & 0 & 0.1333 & 0 & 0 & 0.0666 & 1300. & 0.40 & 0.948 & 0.777 \\
 {Tq(2, 0.6, 90, 10)} & 120 & 10 & 2 & 0.2666 & 0 & 0 & 0 & 0 & 0.0666 & 1200. & 0.20 & 0.955 & 0.698 \\
 {Tq(2.5, 0.4, 45, 10)} & 120 & 10 & 2.5 & 0.1443 & 0 & 0.1443 & 0 & 0 & 0.0326 & 1400. & 0.32 & 0.958 & 0.715 \\
 {Tq(2.5, 0.4, 60, 10)} & 120 & 10 & 2.5 & 0.1767 & 0 & 0.102 & 0 & 0 & 0.0326 & 1300. & 0.26 & 0.96 & 0.693 \\
 {Tq(2.5, 0.4, 90, 10)} & 120 & 10 & 2.5 & 0.2041 & 0 & 0 & 0 & 0 & 0.0326 & 1200. & 0.11 & 0.964 & 0.626 \\
 {Tq(2.5, 0.6, 45, 10)} & 120 & 10 & 2.5 & 0.2164 & 0 & 0.2165 & 0 & 0 & 0.049 & 1500. & 0.47 & 0.952 & 0.787 \\
 {Tq(2.5, 0.6, 60, 10)} & 120 & 10 & 2.5 & 0.2651 & 0 & 0.1531 & 0 & 0 & 0.049 & 1400. & 0.39 & 0.955 & 0.756 \\
 {Tq(2.5, 0.6, 90, 10)} & 120 & 10 & 2.5 & 0.3061 & 0 & 0 & 0 & 0 & 0.049 & 1200. & 0.17 & 0.962 & 0.671 \\
 {Tq(4, 0.6, 45, 10)} & 120 & 10 & 4 & 0.2715 & 0 & 0.2715 & 0 & 0 & 0.024 & 1800. & 0.46 & 0.968 & 0.749 \\
 {Tq(4, 0.6, 60, 10)} & 120 & 10 & 4 & 0.3325 & 0 & 0.192 & 0 & 0 & 0.024 & 1700. & 0.36 & 0.97 & 0.715 \\
 {Tq(4, 0.6, 90, 10)} & 120 & 10 & 4 & 0.384 & 0 & 0 & 0 & 0 & 0.024 & 1400. & 0.12 & 0.975 & 0.61 \\
{V(1, 0.6, 34, 0)} & 77 & 6.2 & 1 & 0.0839 & 0 & 0.1243 & -0.0839 & 0 & 0.1243 & 290. & 0.50 & {} & {} \\
 {V(1, 0.6, 34, 30)} & 77 & 6.2 & 1 & 0.0726 & 0.0419 & 0.1243 & -0.0726 & -0.0419 & 0.1243 & 290. & 0.50 & {} & {} \\
 {V(1, 0.6, 34, 45)} & 77 & 6.2 & 1 & 0.0593 & 0.0593 & 0.1243 & -0.0593 & -0.0593 & 0.1243 & 290. & 0.50 & {} & {} \\
 {V(1, 0.6, 34, 60)} & 77 & 6.2 & 1 & 0.0419 & 0.0726 & 0.1243 & -0.0419 & -0.0726 & 0.1243 & 290. & 0.50 & {} & {} \\
 {V(1, 0.6, 34, 90)} & 77 & 6.2 & 1 & 0 & 0.0839 & 0.1243 & 0 & -0.0839 & 0.1243 & 300. & 0.50 & 0.932 & 0.827 \\
 {V(1, 0.6, 34, 120)} & 77 & 6.2 & 1 & -0.0419 & 0.0726 & 0.1243 & 0.0419 & -0.0726 & 0.1243 & 300. & 0.50 & 0.933 & 0.828 \\
 {V(1, 0.6, 34, 150)} & 77 & 6.2 & 1 & -0.0726 & 0.0419 & 0.1243 & 0.0726 & -0.0419 & 0.1243 & 300. & 0.50 & 0.933 & 0.829 \\
 {V(1, 0.6, 34, 180)} & 77 & 6.2 & 1 & -0.0839 & 0 & 0.1243 & 0.0839 & 0 & 0.1243 & 300. & 0.50 & {} & {} \\
 {V(1, 0.6, 34, 210)} & 77 & 6.2 & 1 & -0.0726 & -0.0419 & 0.1243 & 0.0726 & 0.0419 & 0.1243 & 320. & 0.5 & {} & {} \\
 {V(1, 0.6, 34, 240)} & 77 & 6.2 & 1 & -0.0419 & -0.0726 & 0.1243 & 0.0419 & 0.0726 & 0.1243 & 320. & 0.50 & {} & {} \\
 {V(1, 0.6, 34, 270)} & 77 & 6.2 & 1 & 0 & -0.0839 & 0.1243 & 0 & 0.0839 & 0.1243 & 310. & 0.5 & 0.932 & 0.827 \\
 {V(1, 0.6, 34, 300)} & 77 & 6.2 & 1 & 0.0419 & -0.0726 & 0.1243 & -0.0419 & 0.0726 & 0.1243 & 310. & 0.50 & 0.933 & 0.828 \\
 {V(1, 0.6, 34, 330)} & 77 & 6.2 & 1 & 0.0726 & -0.0419 & 0.1243 & -0.0726 & 0.0419 & 0.1243 & 310. & 0.50 & 0.933 & 0.829 \\
 {V(1, 0.6, 66, 0)} & 77 & 6.2 & 1 & 0.137 & 0 & 0.061 & -0.137 & 0 & 0.061 & 270. & 0.24 & {} & {} \\
 {V(1, 0.6, 66, 30)} & 77 & 6.2 & 1 & 0.1186 & 0.0685 & 0.061 & -0.1186 & -0.0685 & 0.061 & 270. & 0.24 & {} & {} \\
 {V(1, 0.6, 66, 60)} & 77 & 6.2 & 1 & 0.0685 & 0.1186 & 0.061 & -0.0685 & -0.1186 & 0.061 & 270. & 0.24 & {} & {} \\
 {V(1, 0.6, 66, 90)} & 77 & 6.2 & 1 & 0 & 0.137 & 0.061 & 0 & -0.137 & 0.061 & 270. & 0.24 & {} & {} \\
 {V(1, 0.6, 66, 120)} & 77 & 6.2 & 1 & -0.0685 & 0.1186 & 0.061 & 0.0685 & -0.1186 & 0.061 & 270. & 0.24 & {} & {} \\
 {V(1, 0.6, 66, 150)} & 77 & 6.2 & 1 & -0.1186 & 0.0685 & 0.061 & 0.1186 & -0.0685 & 0.061 & 270. & 0.24 & {} & {} \\
 {V(1, 0.6, 90, 0)} & 77 & 6.2 & 1 & 0.1499 & 0 & 0 & -0.1499 & 0 & 0 & 220. & 0 & 0.951 & 0.686 \\
 {V(1, 0.6, 90, 30)} & 77 & 6.2 & 1 & 0.1299 & 0.075 & 0 & -0.1299 & -0.075 & 0 & 220. & 0 & 0.951 & 0.684 \\
 {V(1, 0.6, 90, 45)} & 77 & 6.2 & 1 & 0.106 & 0.106 & 0 & -0.106 & -0.106 & 0 & 210. & 0. & 0.95 & 0.683 \\
 {V(1, 0.6, 90, 60)} & 77 & 6.2 & 1 & 0.075 & 0.1299 & 0 & -0.075 & -0.1299 & 0 & 200. & 0. & 0.95 & 0.681 \\
 {V(1, 0.6, 90, 90)} & 77 & 6.2 & 1 & 0 & 0.1499 & 0 & 0 & -0.1499 & 0 & 200. & 0 & 0.949 & 0.68 \\
 {V(1, 0.6, 90, 120)} & 77 & 6.2 & 1 & -0.075 & 0.1299 & 0 & 0.075 & -0.1299 & 0 & 210. & 0 & 0.95 & 0.682 \\
 {V(1, 0.6, 90, 150)} & 77 & 6.2 & 1 & -0.1299 & 0.075 & 0 & 0.1299 & -0.075 & 0 & 210. & 0 & 0.951 & 0.685 \\
 {V(1, 0.6, 90, 180)} & 77 & 6.2 & 1 & -0.1499 & 0 & 0 & 0.1499 & 0 & 0 & 220. & 0 & 0.951 & 0.686 \\
 {V(1, 0.6, 90, 210)} & 77 & 6.2 & 1 & -0.1299 & -0.075 & 0 & 0.1299 & 0.075 & 0 & 220. & 0 & 0.951 & 0.684 \\
 {V(1, 0.6, 90, 240)} & 77 & 6.2 & 1 & -0.075 & -0.1299 & 0 & 0.075 & 0.1299 & 0 & 220. & 0. & 0.95 & 0.681 \\
 {V(1, 0.6, 90, 270)} & 77 & 6.2 & 1 & 0 & -0.1499 & 0 & 0 & 0.1499 & 0 & 220. & 0 & 0.949 & 0.68 \\
 {V(1, 0.6, 90, 300)} & 77 & 6.2 & 1 & 0.075 & -0.1299 & 0 & -0.075 & 0.1299 & 0 & 220. & 0 & 0.95 & 0.682 \\
 {V(1, 0.6, 90, 330)} & 77 & 6.2 & 1 & 0.1299 & -0.075 & 0 & -0.1299 & 0.075 & 0 & 220. & 0 & 0.951 & 0.685 \\
 {z(1,0)} & 103 & 10 & 1. & 0 & 0 & 0 & 0 & 0 & 0 & 960. & 0. & 0.952 & 0.686 \\
 {z(1,0)} & 100 & 10 & 1 & 0 & 0 & 0 & 0 & 0 & 0 & 930. & 0. & 0.951 & 0.687 \\
 {z(1,0, 11,a)} & 120 & 11 & 1 & 0 & 0 & 0 & 0 & 0 & 0 & 1300. & 0. & 0.951 & 0.686 \\
 {z(1, 0, 11,b)} & 160 & 11 & 1 & 0 & 0 & 0 & 0 & 0 & 0 & 1400. & 0. & 0.952 & 0.686 \\
 {z(1, 0, 11,c)} & 200 & 11 & 1 & 0 & 0 & 0 & 0 & 0 & 0 & 1400. & 0. & 0.952 & 0.686 \\
 {z(1, 0, 11,d)} & 240 & 11 & 1 & 0 & 0 & 0 & 0 & 0 & 0 & 1400. & 0. & 0.952 & 0.686 \\
 {zq(1.15,0 )} & 103 & 10 & 1.15 & 0 & 0 & 0 & 0 & 0 & 0 & 960. & 0. & 0.952 & 0.684 \\
 {zq(1.25,0)} & 100 & 10 & 1.25 & 0 & 0 & 0 & 0 & 0 & 0 & 960. & 0. & 0.952 & 0.68 \\
 {zq(1.3,0 )} & 103 & 10 & 1.3 & 0 & 0 & 0 & 0 & 0 & 0 & 970. & 0. & 0.953 & 0.677 \\
 {zq(1.45,0)} & 103 & 10 & 1.45 & 0 & 0 & 0 & 0 & 0 & 0 & 980. & 0. & 0.955 & 0.667 \\
 {zq(1.5,0)} & 103 & 10 & 1.5 & 0 & 0 & 0 & 0 & 0 & 0 & 990. & 0. & 0.955 & 0.664 \\
 {zq(1.5,0)} & 100 & 10 & 1.5 & 0 & 0 & 0 & 0 & 0 & 0 & 1000. & 0. & 0.955 & 0.664 \\
 {zq(1.5,0, 11)} & 200 & 11 & 1.5 & 0 & 0 & 0 & 0 & 0 & 0 & 1400. & 0. & 0.955 & 0.664 \\
 {zq(1.6,0)} & 103 & 10 & 1.6 & 0 & 0 & 0 & 0 & 0 & 0 & 990. & 0. & 0.956 & 0.656 \\
 {zq(1.75,0)} & 103 & 10 & 1.75 & 0 & 0 & 0 & 0 & 0 & 0 & 1000. & 0. & 0.958 & 0.644 \\
 {zq(1.9,0)} & 103 & 10 & 1.9 & 0 & 0 & 0 & 0 & 0 & 0 & 1000. & 0. & 0.96 & 0.632 \\
 {zq(2,0)} & 103 & 10 & 2. & 0 & 0 & 0 & 0 & 0 & 0 & 1000. & 0. & 0.961 & 0.623 \\
 {zq(2,0)} & 100 & 10 & 2. & 0 & 0 & 0 & 0 & 0 & 0 & 1000. & 0. & 0.961 & 0.623 \\
 {zq(2.05,0)} & 103 & 10 & 2.05 & 0 & 0 & 0 & 0 & 0 & 0 & 1000. & 0. & 0.962 & 0.619 \\
 {zq(2.2,0)} & 103 & 10 & 2.2 & 0 & 0 & 0 & 0 & 0 & 0 & 1100. & 0. & 0.963 & 0.606 \\
 {zq(2.35,0)} & 103 & 10 & 2.35 & 0 & 0 & 0 & 0 & 0 & 0 & 1100. & 0. & 0.965 & 0.593 \\
 {zq(2.5,0)} & 103 & 10 & 2.5 & 0 & 0 & 0 & 0 & 0 & 0 & 1100. & 0. & 0.967 & 0.581 \\
 {zq(2.5,0, 11)} & 200 & 11 & 2.5 & 0 & 0 & 0 & 0 & 0 & 0 & 1600. & 0. & 0.967 & 0.581 \\
 {zq(3,0)} & 200 & 11 & 3 & 0 & 0 & 0 & 0 & 0 & 0 & 1700. & 0.& 0.971 & 0.54 \\
 {zq(4, 0,11)} & 200 & 11 & 4 & 0 & 0 & 0 & 0 & 0 & 0 & 2000. & 0. & 0.978 & 0.472 \\
 {z(1,-0.4,   11)} & 200 & 11 & 1 & 0 & 0 & -0.1 & 0 & 0 & -0.1 & 1100. & -0.40 & 0.96 & 0.56 \\
 {z(1, -0.2,  11)} & 200 & 11 & 1 & 0 & 0 & -0.05 & 0 & 0 & -0.05 & 1200. & -0.20 & 0.956 & 0.624 \\
 {z(1, 0.2)} & 100 & 10 & 1 & 0 & 0 & 0.05 & 0 & 0 & 0.05 & 1100. & 0.20 & 0.945 & 0.746 \\
 {z(1, 0.2,  11)} & 200 & 11 & 1 & 0 & 0 & 0.05 & 0 & 0 & 0.05 & 1500. & 0.20 & 0.945 & 0.746 \\
 {zq(1.25 , 0.2)} & 100 & 10 & 1.25 & 0 & 0 & 0.0617 & 0 & 0 & 0.0395 & 1100. & 0.2 & 0.946 & 0.74 \\
 {zq(1.5, 0.2)} & 100 & 10 & 1.5 & 0 & 0 & 0.072 & 0 & 0 & 0.032 & 1100. & 0.2 & 0.949 & 0.729 \\
 {zq(1.5, 0.2, 11)} & 200 & 11 & 1.5 & 0 & 0 & 0.072 & 0 & 0 & 0.032 & 1600. & 0.2 & 0.949 & 0.729 \\
 {zq(1.75, 0.2)} & 100 & 10 & 1.75 & 0 & 0 & 0.081 & 0 & 0 & 0.0264 & 1200. & 0.2 & 0.952 & 0.714 \\
 {zq(2, 0.2)} & 100 & 10 & 2. & 0 & 0 & 0.0889 & 0 & 0 & 0.0222 & 1200. & 0.2 & 0.956 & 0.698 \\
 {zq(2, 0.2, 11)} & 160 & 11 & 2. & 0 & 0 & 0.0889 & 0 & 0 & 0. & 1700. & 0.2 & 0.956 & 0.698 \\
 {z(1, 0.4,  11)} & 200 & 11 & 1 & 0 & 0 & 0.1 & 0 & 0 & 0.1 & 1700. & 0.40 & 0.938 & 0.804 \\
 {zq(1.25, 0.4)} & 100 & 10 & 1.25 & 0 & 0 & 0.1234 & 0 & 0 & 0.079 & 1200. & 0.4 & 0.939 & 0.8 \\
 {zq(1.5, 0.4)} & 100 & 10 & 1.5 & 0 & 0 & 0.144 & 0 & 0 & 0.064 & 1300. & 0.4 & 0.941 & 0.792 \\
 {zq(1.5, 0.4)} & 200 & 11 & 1.5 & 0 & 0 & 0.144 & 0 & 0 & 0.064 & 1700. & 0.4 & 0.942 & 0.792 \\
 {zq(1.75, 0.4)} & 100 & 10 & 1.75 & 0 & 0 & 0.162 & 0 & 0 & 0.0529 & 1300. & 0.4 & 0.945 & 0.781 \\
 {zq(2,, 0.4)} & 100 & 10 & 2. & 0 & 0 & 0.1778 & 0 & 0 & 0.0444 & 1300. & 0.4 & 0.949 & 0.768 \\
 {zq(2, 0.4, 11)} & 160 & 11 & 2. & 0 & 0 & 0.1778 & 0 & 0 & 0.0444 & 1900. & 0.4 & 0.949 & 0.77 \\
 {z(1, 0.6)} & 100 & 10 & 1 & 0 & 0 & 0.15 & 0 & 0 & 0.15 & 1400. & 0.60 & 0.926 & 0.858 \\
 {z(1, 0.6,11)} & 200 & 11 & 1 & 0 & 0 & 0.15 & 0 & 0 & 0.15 & 1900. & 0.60 & 0.927 & 0.858 \\
 {zq(1.25, 0.6)} & 100 & 10 & 1.25 & 0 & 0 & 0.1852 & 0 & 0 & 0.1185 & 1400. & 0.6 & 0.928 & 0.855 \\
 {zq(1.5, 0.6)} & 100 & 10 & 1.5 & 0 & 0 & 0.216 & 0 & 0 & 0.096 & 1400. & 0.6 & 0.931 & 0.851 \\
 {zq(1.75, 0.6)} & 100 & 10 & 1.75 & 0 & 0 & 0.243 & 0 & 0 & 0.0793 & 1400. & 0.6 & 0.935 & 0.845 \\
 {zq(2, 0.6)} & 100 & 10 & 2. & 0 & 0 & 0.2666 & 0 & 0 & 0.0666 & 1500. & 0.6 & 0.939 & 0.839 \\
 {zq(2, 0.6,11)} & 160 & 11 & 2. & 0 & 0 & 0.2666 & 0 & 0 & 0.0666 & 2000. & 0.6 & 0.94 & 0.839 \\
 {z(1, 0.8)} & 100 & 10 & 1 & 0 & 0 & 0.2 & 0 & 0 & 0.2 & 1500. & 0.80 & 0.912 & 0.908 \\
 {z(1, 0.8, 11)} & 200 & 11 & 1 & 0 & 0 & 0.2 & 0 & 0 & 0.2 & 1900. & 0.8 & 0.912 & 0.909 \\
 {zU(1, 0., 0.2,  11)} & 160 & 11 & 1 & 0 & 0 & 0 & 0 & 0 & 0.05 & 1400. & 0.10 & 0.949 & 0.716 \\
 {zU(1, 0., 0.4, 11)} & 160 & 11 & 1 & 0 & 0 & 0 & 0 & 0 & 0.1 & 1500. & 0.20 & 0.945 & 0.746 \\
 {zU(1, 0., 0.6, 11)} & 160 & 11 & 1 & 0 & 0 & 0 & 0 & 0 & 0.15 & 1600. & 0.30 & 0.942 & 0.775 \\
 {zU(1, 0., 0.8,  11)} & 160 & 11 & 1 & 0 & 0 & 0 & 0 & 0 & 0.2 & 1700. & 0.40 & 0.937 & 0.802 \\
 {zU(1, 0.2, 0.4,  11)} & 160 & 11 & 1 & 0 & 0 & 0.05 & 0 & 0 & 0.1 & 1600. & 0.30 & 0.942 & 0.775 \\
 {zU(1, 0.2, 0.6, 11)} & 160 & 11 & 1 & 0 & 0 & 0.05 & 0 & 0 & 0.15 & 1700. & 0.40 & 0.937 & 0.803 \\
 {zU(1, 0.2, 0.8,  11)} & 160 & 11 & 1 & 0 & 0 & 0.05 & 0 & 0 & 0.2 & 1700. & 0.50 & 0.932 & 0.83
\end{longtable}

\end{center}
\end{widetext}

\end{document}